\documentclass[%
reprint,
superscriptaddress,
amsmath,amssymb,
aps,
floatfix,
nofootinbib
]{revtex4-1}

\usepackage{graphicx}% Include figure files
\usepackage{subcaption}
\usepackage{rotating}
\usepackage{dcolumn}% Align table columns on decimal point
\usepackage{bm}% bold math
\usepackage{braket}
\usepackage{amsmath}
\usepackage{mathtools}
\usepackage[colorlinks,allcolors=blue]{hyperref} 
\usepackage{hyperref}% add hypertext capabilities
\newcommand{\expect}[1]{\left\langle #1 \right\rangle}

\begin{document}
	
	\raggedbottom
	
	\preprint{APS/123-QED}
	
	\title{Ab-Initio Calculations of Nonlinear Susceptibility and Multi-Phonon Mixing Processes in a 2DEG-Piezoelectric Heterostructure}
	
	\author{Eric Chatterjee}
	\affiliation{Wyant College of Optical Sciences, The University of Arizona, Tucson, Arizona 85721, USA}
    \author{Alexander Wendt}
    \affiliation{Wyant College of Optical Sciences, The University of Arizona, Tucson, Arizona 85721, USA}
    \author{Daniel Soh}
    \email{danielsoh@arizona.edu}
    \affiliation{Wyant College of Optical Sciences, The University of Arizona, Tucson, Arizona 85721, USA}
	\author{Matt Eichenfield}
    \email{eichenfield@arizona.edu}
    \affiliation{Wyant College of Optical Sciences, The University of Arizona, Tucson, Arizona 85721, USA}
    \affiliation{Sandia National Laboratories, Albuquerque, New Mexico 87123, USA}

	\date{\today}% It is always \today, today,
	
	\begin{abstract}
		Solid-state elastic-wave phonons are a promising platform for a wide range of quantum information applications including facilitating quantum transduction from microwave to optical electromagnetic fields, long-lived quantum memories, in addition to potentially acting as qubits themselves. An outstanding challenge and enabling capability in harnessing phonons for quantum information processing is achieving sufficiently strong nonlinear interactions between them. To this end, we propose a general architecture for generating strong quantum phononic nonlinearity using piezoelectric-semiconductor heterostructures consisting of a piezoelectric acoustic material hosting phononic modes that is in direct proximity to a two-dimensional electron gas (2DEG). Each phonon in the piezoelectric material carries an electric field, which extends into the 2DEG. The fields induce polarization of 2DEG electrons, which in turn interact with other piezoelectric phononic electric fields. The net result is coupling between the electric fields associated with the various phonon modes. We derive, from first principles, the nonlinear phononic susceptibility of a piezo-2DEG system and provide a prescription to calculate all orders of the susceptibility in a perturbative expansion. We show that many nonlinear processes are strongly favored at high electron mobility, motivating the use of the 2DEG to mediate the nonlinearities. We derive the first, second, and third-order susceptibilities and calculate them for the case of a lithium niobate surface acoustic wave interacting with a GaAs-AlGaAs heterostructure 2DEG. We show that, for this system, the strong third-order phononic nonlinearities generated could enable single-phonon Kerr shift in an acoustic cavity that exceeds realistic cavity linewidths, potentially leading to a new class of acoustic qubit. We further show that the strong second-order nonlinearity could be used to produce a high-gain, traveling-wave parametric amplifier to amplify--and ultimately detect--the outputs of the acoustic cavity qubits. Assuming favorable losses in such a system, the combination of these capabilities, combined with the ability to efficiently transduce phonons from microwave electromagnetic fields in transmission lines, thus hold promise for creating all-acoustic quantum information processors.
	\end{abstract}
	
	\pacs{Valid PACS appear here}% PACS, the Physics and Astronomy
	\maketitle

\footnote{Distribution Statement A: Approved for Public Release, Distribution Unlimited}

\section{Introduction}

Quantized elastic waves in solids, or phonons, have become increasingly ubiquitous in quantum information processing applications. For example, they can act as a nearly ``universal'' quantum bus \cite{raniwala2023piezoelectricnanocavity, neuman2021phononicinterface}, allowing coupling to many useful solid state qubit modalities. Phonons naturally couple to optomechanical cavities \cite{eichenfield2009optomechanicalcrystals, aspelmeyer2014cavityoptomechanics, kippenberg2007cavityoptomechanics, lemonde2013nonlinearinteraction,  metcalfe2014applicationscavity}, which provide a means for optical generation of nonclassical phonon states \cite{riedinger2016nonclassical, hong2017hanburybrown, li2018generationdetection, zivari2022nonclassical} and optical teleportation of quantum information \cite{fiaschi2021optomechanicalquantum, soh2021highfidelity}. Moreover, piezoelectric elastic wave phonons can be used to efficiently couple to traveling wave \cite{hackett2023aluminumscandium, safavinaeini2011proposaloptomechanical, massel2011microwavemplification, wu2020microwaveoptical, arnold2020convertingmicrowave} and standing wave \cite{teufel2008dynamicalbackaction, safavinaeini2019controllingphonons} microwave fields in important quantum phononic materials, couple to superconducting circuits \cite{chamberland2022faulttolerant, maccabe2020nanoacoustic, arrangoizarriola2018coupling, barzanjeh2022optomechanicsquantum, viennot2018phononnumber, satzinger2018quantumcontrol}, or potentially act as quantum memories to store quantum information from superconducting circuits due to their long lifetime \cite{taylor2022reconfigurablequantum, kilina2011theoreticalstudy, hann2019hardwareefficient, wallucks2020quantummemory, chamberland2022faulttolerant, maccabe2020nanoacoustic, chan2011lasercooling}. If those piezoelectric phonons are stored within optomechanical cavities, then they provide a means for optical addressing of quantum information in superconducting qubits \cite{bochmann2013nanomechanicalcoupling, mirhosseini2020superconductingqubit, forsch2020microwaveoptics, jiang2020efficientbidirectional, balram2022piezoelectricoptomechanical}. Phonons also can couple to solid state color centers and their spin degrees of freedom \cite{kepesidis2013phononcooling, albrecht2013couplingnitrogen, albrecht2013couplingsingle, bennett2013phononinduced, teissier2014straincoupling, goldman2015phononinduced, meesala2016enhancedstrain, li2016hybridquantum, oeckinghaus2020spinphonon, liu2020spectroscopicsignatures, neuman2021phononicinterface}.

\begin{figure*}[t]
	\centering
	\includegraphics[width=\textwidth]{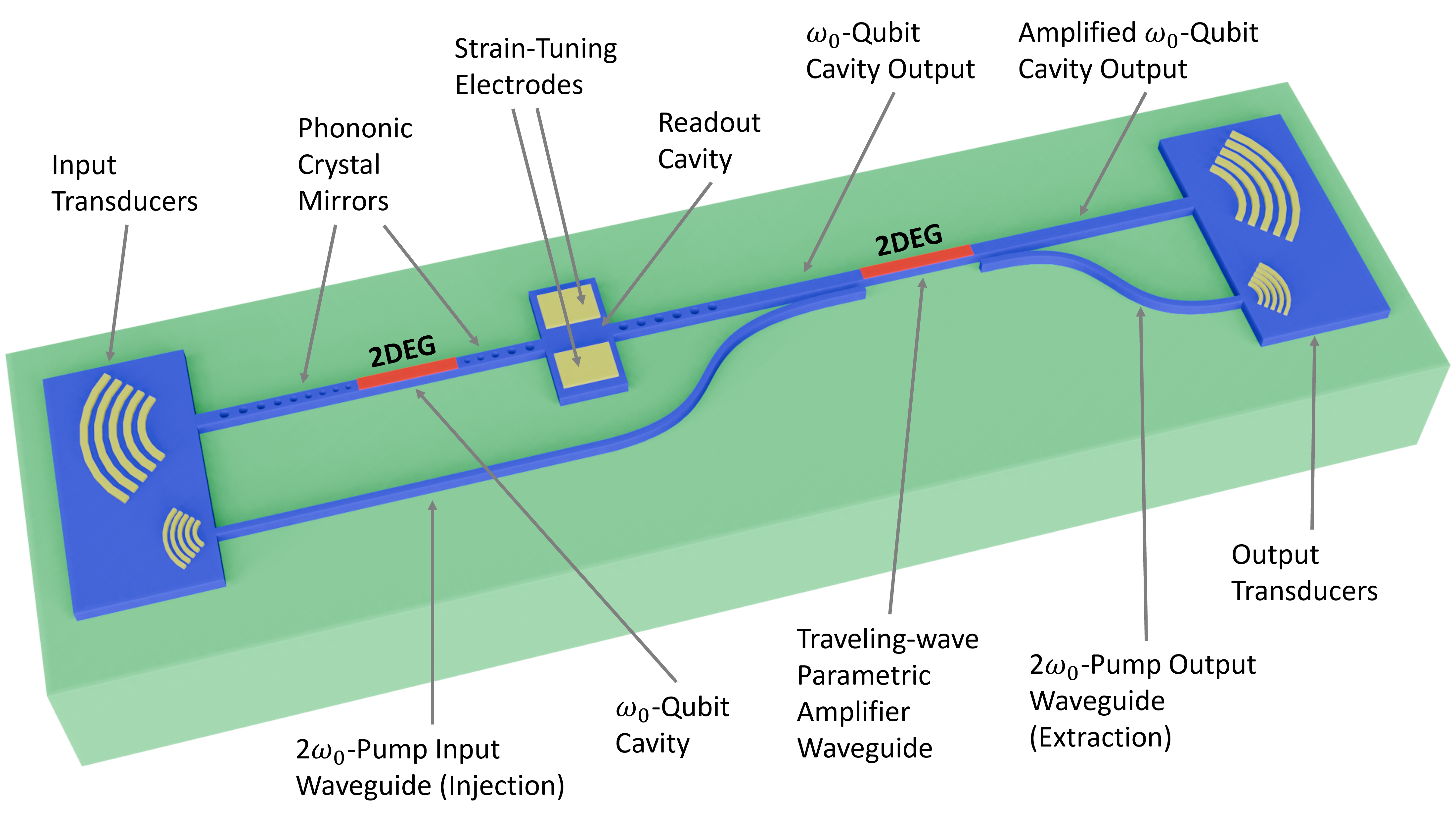}
	\caption{Diagram of a system consisting of an acoustic Kerr cavity (left) coupled to a degenerate traveling-wave parametric amplifier (right). The coupling in the acoustic cavity between the 2DEG (red) and the piezoelectric material (blue) gives rise to a Kerr nonlinearity and in turn a qubit. The qubit is then amplified by the 3-wave-mixing in the parametric amplifier due to the coupling between the 2DEG there (red) and the piezoelectric material (blue), before being transduced into a microwave field at the system output. Note that the thin waveguide carries the pump phonons.}
	\label{fig:systemimage}
\end{figure*}

%Phonons have increasingly been recognized as an essential supplement to electromagnetic fields for quantum information applications. One key application is in optomechanical oscillators \cite{eichenfield2009optomechanicalcrystals}, which, when coupled with piezoelectric transducers, facilitate quantum transduction between microwave electromagnetic fields and optical fields (which are optimal for long-range communication). Furthermore, phonon modes hold great promise as quantum memories due to their long lifetime \cite{galliou2013lowloss, kharel2018ultrahighq}, as well as the newly-discovered ability to tunably couple phonon cavities to input/output acoustic channels \cite{taylor2022reconfigurablequantum}. At the cutting edge, phonons can serve as artificial atoms that provide a platform for generating qubits. As a result, phonon-based systems have become increasingly popular in quantum computing \cite{ruskov2012coherentphonons, neuman2021phononicinterface, qiao2023splittingphonons}. 

The ability of phonons to act as a universal quantum bus or even as a qubit in universal linear acoustics are undoubtedly useful, but these abilities would be far more useful if one could perform quantum logic operations on the phonons, themselves. This, of course, requires single-phonon nonlinearities that are large compared to system losses. No material currently possesses such strong nonlinearities and low losses; thus, it is imperative to explore other means for generating these properties. Recently, we have demonstrated that phononic nonlinearities can be increased by orders of magnitude by mediating them with electronic nonlinearities in a piezoelectric-semiconductor heterostructure \cite{hackett2023mixing}. Classical theories of such nonlinearities \cite{conwellganguly1971mixing, wuspector1972harmonic, johrispector1975boltzmann} show that the nonlinearity depends directly on the mobility. Thus, one must wonder what would be possible in materials with ultra-high mobilities, such as two-dimensional electron gasses (2DEGs), which can attain  mobilities multiple orders of magnitude higher than bulk semiconductors \cite{pfeiffer1989electronmobilities, umansky2009mbegrowth}.

To this end, we study a heterostructure consisting of a 2DEG  bonded to a strongly piezoelectric elastic slab waveguide  substrate (e.g. thin-film LiNbO\textsubscript{3} on Si, sapphire, or SiC or thin-film Al\textsubscript{1-x}Sc\textsubscript{x}N on SiC). Only the critical layers of the 2DEG remain on the surface of the piezoelectric substrate, which can be achieved with modern heteroepitaxy and chemical etching (e.g. in a GaAs/AlGaAs heteroepitaxial stack). Similar piezoelectric-semiconductor phononic heterostructure systems have been realized in the classical regime utilizing bulk semiconductors to achieve phonon amplification \cite{hackett2021singlechip, hackett2023nonreciprocal, hackett2023aluminumscandium, hackett2023sband}, phononic switches \cite{storey2021acoustoelectricsurface}, phononic circulators \cite{hackett2021singlechip}, phononic mixing \cite{hackett2023mixing}, and acoustoelectrically Brillouin optomechanical processes \cite{otterstrom2023modulationbrillouin}. A 2DEG-piezoelectric heterostructure has itself been experimentally realized in the context of measuring a large acoustoelectric effect \cite{rotter1998giantacoustoelectric}, while more recently, a quantum-dot-piezoelectric heterostructure has been realized for optomechanical applications \cite{nysten2020hybridsurface}. In all these cases, including the one at-hand, the acoustic field in the piezoelectric material takes the form of a slab waveguide mode or surface acoustic wave. As in the case of the classical systems, an evanescent electric field associated with the elastic wave phonons extends into the 2DEG, where it induces the polarization of 2DEG electrons. The polarized electrons, in turn, interact with other electric fields stemming from other elastic wave phonon excitations in the system, yielding mixing of multiple phonons mediated by the 2DEG electrons.

%(itself typically a heterostructure; e.g. GaAs/AlGaAs)

Here, we theoretically analyze the electron-phonon interactions in the above system and perform ab-initio quantum calculations of the resulting acoustic nonlinearities and multi-phonon mixing processes that result from them. This allows for the exact calculation of phononic mixing processes such as second harmonic generation, parametric amplification, Kerr nonlinearity, etc. We find, for example, that these kinds of systems could produce high-gain and low-loss quantum-limited degenerate parametric amplification via three-wave mixing with only nanowatts of acoustic pump power; we also find that the effective Kerr nonlinearity in the system could be sufficient to produce an effective two-level system in a wavelength-scale cavity with sufficiently low losses, given a high-Q acoustic cavity, which has recently been achieved \cite{shao2019phononicband}. Thus, our results allow one to imagine a new class of quantum information processing tools for phonons based on electronically mediated nonlinearity in heterostructures and calculate the exact details of these processes. 

Such a system is illustrated in Fig.~\ref{fig:systemimage}. Here, a 2DEG in a nearly single-mode cavity--the qubit cavity--creates a qubit at frequency (wavevector) $\omega_0$ ($k_0$) via four-wave phononic Kerr nonlinearity. The qubit cavity is coupled to a tunable readout cavity that allows adiabatic transfer of the qubit state into a purely phononic state of the readout cavity. The readout cavity is strongly coupled to an output phononic bus waveguide that leads to a degenerate, traveling wave parametric three-wave mixing phononic amplifier. A pair of pump waveguides inject and extract the pump field at frequency (wavevector) $w\omega_0$ ($2k_0)$ via acoustic evanescent coupling. The amplified phononic field at $\omega_0$ can then be transduced into a microwave field at the system output, to be measured by room temperature electronics after a series of progressively higher temperature amplifiers. Note that elastic waves with highly restricted transverse dimensions can be efficiently excited (low microwave insertion loss and low elastic wave loss) by focusing transducers \cite{siddiqui2018lambwave, liu2019electromechanicalbrillouin, eichenfield2013designfabrication}, phononic negative refraction in phononic crystals \cite{zhang2004negativerefraction, page2016focusingultrasonic}, and phononic crystal gradient-index (GRIN) lenses \cite{wu2011focusinglowest}. Essentially, with the inclusion of 2DEGs playing the role of spatially selective sources of single-phonon nonlinearities, one can imagine circuit QED for phonons that mimic the functionalities of those of superconducting circuit QED.

%We build on the techniques developed to analyze optical nonlinearities in atoms and gases \cite{boyd}, deriving the acoustic nonlinearities poses a unique challenge. Specifically, due to the low propagation speed of the acoustic waves, phonons possess far greater momentum than a photon at the same frequency. The electronic transitions upon interaction with phonons are therefore intraband (rather than the interband transitions experienced upon interaction with photons). Thus, our result is the first of this kind, clearly presenting the acoustic susceptibility calculation through intricate intraband electronic transitions from the 2DEG in a heterostructure. 

The manuscript is organized as follows. In Sec.~\ref{sec: Deriving the Electric Field Per Phonon}, we derive the electric field per phonon experienced by the 2DEG electrons based on the piezoelectric material's properties, as well as the 2DEG electric field screening. In Sec.~\ref{sec: General Derivation of Higher-Order Susceptibilities}, we derive the  susceptibility of the 2DEG to the electric fields associated with the elastic wave phonons. In Sec.~\ref{sec: Susceptibility Results}, we present numerical and analytical results for the 2DEG susceptibility in certain relevant limits. In Sec.~\ref{sec: Calculation of Mixing Dynamics}, we combine the susceptibility values from Sec.~\ref{sec: Susceptibility Results} with the electric-field-per-phonon values from Sec.~\ref{sec: Deriving the Electric Field Per Phonon} to calculate the resulting phononic susceptibility of the system and multi-phonon mixing dynamics. As a check on validity, in Sec.~\ref{sec: Comparison of Quantum and Classical Results in Fluid Limit} we derive the classical susceptibilities up to second order based on our quantum susceptibility model and compare to existing classical results in the literature.

\section{Deriving the Electric Field Per Phonon}
\label{sec: Deriving the Electric Field Per Phonon}

We start with the critical task of deriving the electric field per phonon from the elasticity, dielectric coupling, and piezoelectric coupling tensors given a shear-wave surface acoustic field. In general, the following relationships must hold for any piezoelectric material \cite{schuetz2015universalquantum}:
\begin{align}
\label{eq: T_{ij}}
T_{ij} &= \sum_{\substack{k,l \\ l \ge k}} \Big(c_{ijkl} S_{kl} - e_{kij} E_k\Big), \\
\label{eq: D_i}
D_i &= \sum_{\substack{j,k \\ k \ge j}} \Big(\epsilon_{ij} E_j + e_{ijk} S_{jk}\Big),
\end{align}
where $c$, $\epsilon$, and $e$ are the elasticity, dielectric, and piezoelectric coupling tensors, respectively, while $T$, $S$, $E$, and $D$ represent the stress, strain, electric field, and electric displacement field, respectively.

Our first goal is to determine the electric field for a given strain. In the absence of an external electric field, the electric displacement field $\bm{D}$ goes to 0 regardless of how the coordinate system is defined, yielding the following relationship between the electric field and the strain field based on Eq.~\eqref{eq: D_i}:
\begin{equation}
\label{eq: electric field vs strain generic}
E_i = -\sum_{\substack{j,k,l \\ l \ge k}} (\epsilon^{-1})_{ij} e_{jkl} S_{kl}.
\end{equation}
The electric field per phonon can thus be calculated from the strain amplitude per phonon. 

We thus turn to solving the strain amplitude for a given mode as a function of phonon number in that mode. In general, for a surface acoustic wave propagating along the plane of the surface, the displacement amplitude will vary along the axis perpendicular to the surface. Defining the $x$-axis as the direction of propagation and the $z$-axis as the axis perpendicular to the surface, we can express the displacement field as a superposition of modes as follows:
\begin{align}
\begin{split}
\bm{u}(\bm{r}) &= \bm{u}_{n,q} f_{n,q}(z) \Big(e^{i(q x - \omega t)} + e^{-i(q x - \omega t)}\Big) \\
&= 2 \bm{u}_{n,q} f_{n,q}(z) \cos{(q x - \omega t)}.
\end{split}
\end{align}
where the functions $f_{n,q}$ are orthogonal for a given wavevector $q$:
\begin{equation}
\int dz f_{n,q}(z) f_{m,q}(z) = L_z \delta_{m,n}.
\end{equation}
The strain components corresponding to the displacement $\bm{u}(\bm{r})$ are calculated as follows:
\begin{align}
\begin{split}
S_{ij}(\bm{r}) &= \frac{\partial u_i}{\partial r_j}(\bm{r}) + (1 - \delta_{i,j}) \frac{\partial u_j}{\partial r_i}(\bm{r}) \\
&= u_{n,q} \Big(g_{ij,n,q}(z) e^{i(q x - \omega t)} + g_{ij,n,q}^* e^{-i(q x - \omega t)}\Big),
\end{split}
\end{align}
where $g_{ij,n,q}$ is defined as follows:
\begin{align}
\begin{split}
&g_{ij,n,q}(z) = \\
&\quad i q f_{n,q}(z) \Big(\cos{\theta_{n,q,i}} \delta_{j,x} + \cos{\theta_{n,q,j}} (1 - \delta_{i,j}) \delta_{i,x}\Big) \\
&\quad + f'_{n,q}(z) \Big(\cos{\theta_{n,q,i}} \delta_{j,z} + \cos{\theta_{n,q,j}} (1 - \delta_{i,j}) \delta_{i,z}\Big),
\end{split}
\end{align}
where $\theta_{n,q,m}$ is defined as the angle between the displacement vector $\bm{u}_{n,q}$ and the $m$-axis.

The goal is to determine the global amplitude $u_{n,q}$ as a function of the number of phonons $N$ in the mode $(n,q)$. To that end, we relate the rate of change of the displacement $\bm{\dot{u}}$ to the kinetic energy $T$ and use the fact that the total mode energy $N\hbar\omega$ is twice the kinetic energy for a simple harmonic oscillator:
\begin{align}
\begin{split}
&\bigg(N + \frac{1}{2}\bigg) \hbar \omega \\
&= 2 T \\
&= 2 \int d^3r \frac{1}{2} \rho \sum_i \dot{u_i}^2(\bm{r}) \\
&= 4 \rho \omega^2 u_{n,q}^2 \int dy \int dx \sin^2{(q x - \omega t)} \int dz f_{n,q}^2(z) \\
&= 4 \rho \omega^2 u_{n,q}^2 L_y \frac{L_x}{2} L_z, \\
u_{n,q} &= \sqrt{\frac{(N + 1/2) \hbar}{2 \rho \omega V}},
\end{split}
\end{align}
where $V = L_x L_y L_z$ is the mode volume in the piezoelectric material. Note that if the displacement consists of a superposition of modes, then there will be no direct coupling between the modes, due to two factors. First, if the modes are of different wavelengths, then the integral of the product of the amplitudes over the $x$-axis (i.e., the propagation axis) will go to zero. Second, even for two modes of the same wavelength, the fact that the $z$-varying functions $f$ are orthogonal to one another ensures that the integral of the product of the amplitudes over the $z$-axis will go to zero.

Finally, having solved for the strain amplitude per phonon, we calculate the electric field per phonon at the surface. In general, the electric field amplitude corresponding to a plane wave of frequency $\omega$ and wavevector $q$ propagating in the $\hat{m}$-direction can be expanded in a manner analogous to the displacement and strain amplitudes:
\begin{equation}
\bm{E}(\bm{r}) = \bm{E}_{n,q}(z) e^{i(q x_m - \omega t)} + \bm{E^*}_{n,q}(z) e^{-i(q x_m - \omega t)}.
\end{equation}
We solve for the $z$-dependent electric field amplitude $\bm{E}_{n,q}$ as a function of the phonon numbers in the respective modes using Eq.~\eqref{eq: electric field vs strain generic} as follows:
\begin{widetext}
\begin{align}
\begin{split} \label{eq: electric field piezo generic}
&E_{n,q,i}(z) \\
&= -\sum_{\substack{j,k,l \\ l \ge k}} (\epsilon^{-1})_{ij} e_{jkl} u_{n,q} g_{kl,n,q}(z) \\
&= -\sqrt{\frac{(N + 1/2) \hbar}{2 \rho \omega V}} \sum_{\substack{j,k,l \\ l \ge k}} (\epsilon^{-1})_{ij} e_{jkl} \bigg(i q f_{n,q}(z) \Big(\cos{\theta_{n,q,k}} \delta_{k,x} + \cos{\theta_{n,q,l}} (1 - \delta_{k,l}) \delta_{k,x}\Big) \\
&\quad\quad + f'_{n,q}(z) \Big(\cos{\theta_{n,q,k}} \delta_{l,z} + \cos{\theta_{n,q,l}} (1 - \delta_{k,l}) \delta_{k,z}\Big)\bigg).
\end{split}
\end{align}
\end{widetext}
We can assume that a surface acoustic wave exponentially decays into the bulk with a decay length similar to the wavelength, yielding $f_{n,q}(z) \propto e^{-qz}$. Then, given a diagonal dielectric tensor (i.e., $\epsilon_{ij} = \epsilon_{ii} \delta_{i,j}$), the electric field amplitude at the surface (which we define as $z = 0$) reduces to the following:
\begin{equation} \label{eq: electric field diagonal dielectric exponential envelope}
E = -\frac{C}{\epsilon} \sqrt{\frac{\hbar \omega}{v_s}} A_\mathrm{ph},
\end{equation}
where $C$ is the effective ratio between the piezoelectric coupling coefficient and the square-root of the elasticity (i.e., in a linear piezoelectric material, $C = e/\sqrt{\kappa}$, where $e$ and $\kappa$ are the piezoelectric coupling and elasticity coefficients, respectively), $\epsilon$ is the effective dielectric constant, and $A_\mathrm{ph}$ is a measurement of the square-root of the propagating strain field power density (i.e., power per unit area) in units of $\hbar \omega$. Each of these parameters is defined in the following manner:
\begin{widetext}
\begin{align}
C &= \frac{f_{n,q}(0)}{\sqrt{2 v_s^2 \rho}} \sum_{\substack{k,l \\ l \ge k}} e_{ikl} \bigg(i\Big(\cos{\theta_{n,q,k}} \delta_{k,x} + \cos{\theta_{n,q,l}} (1 - \delta_{k,l}) \delta_{k,x}\Big) - \Big(\cos{\theta_{n,q,k}} \delta_{l,z} + \cos{\theta_{n,q,l}} (1 - \delta_{k,l}) \delta_{k,z}\Big)\bigg), \\
\epsilon &= \epsilon_{xx}, \\
A_\mathrm{ph} &= \sqrt{v_s \frac{(N + 1/2)}{V}}.
\end{align}
\end{widetext}
For the electron-phonon interaction, we use the dielectric tensor for the 2DEG as $\epsilon$, instead of the dielectric tensor for the piezoelectric material. In Sec.~\ref{sec: Calculation of Mixing Dynamics}, we will use $A_\mathrm{ph}$ as a measurement of the overall field amplitude for each mode.

\section{General Derivation of Higher-Order Susceptibilities}
\label{sec: General Derivation of Higher-Order Susceptibilities}

Having determined the electric field amplitude per phonon, we now derive the general $N^{\mathrm{th}}$-order susceptibility of the 2DEG electrons as a function of the input electric field amplitudes. By definition, the $N^{\mathrm{th}}$-order susceptibility corresponds to the average electron dipole moment (normalized to material volume and amplitude of each input field) induced by $N$ field modes simultaneously interacting with the electrons. We specifically consider the interaction at a quantum level: i.e., how $N$ phonons simultaneously polarize a single electron via the electric field amplitudes associated with the respective phonons. First, we calculate the amplitude for a single electron-phonon interaction, which is proportional to the strength of an electron's dipole moment if it successfully interacts with $N$ phonons. Then, we derive the probability (as a function of the field amplitudes) that the electron successfully interacts with $N$ phonons, which yields the $N^{\mathrm{th}}$-order susceptibility.

\subsection{Deriving the electron-phonon interaction amplitude}
\label{sec:: Deriving the electron-phonon interaction amplitude}

Here, we derive the electron-phonon interaction Hamiltonian from first principles, ultimately yielding the effective electron dipole moment as a function of the acoustic field wavelength. As previously discussed, the phonon acts on the electron through the piezoelectric field associated with the phonon. Generically defining the effective electron dipole moment along the axis of field polarization as $d_\mathrm{eff}$ and the electric field amplitude as $E$, we apply the standard dipole interaction Hamiltonian:
\begin{equation}
H_\mathrm{int} = -d_\mathrm{eff} E.
\end{equation}
To solve for the dipole matrix elements, we thus need to calculate the Hamiltonian matrix elements. The reason for using this indirect method is that given the longitudinal nature of the electric field associated with the acoustic field, the dipole moment cannot be separately promoted to operator form in the basis of 2DEG electronic states, since the electric field also spatially varies along the axis of propagation. We therefore start with the fundamental piezo-2DEG interaction Hamiltonian, given an electric field polarized in the $x$-direction:
\begin{equation} \label{eq: piezo-2DEG Hamiltonian}
H_\mathrm{int} = -\int d^3r E_x P_x,
\end{equation}
where $P$ is the 2DEG polarization. For an acoustic field (and thus an associated piezoelectric field) propagating in the $x$-direction with a wavevector $q$, the electric field $E_x$ takes the following form in terms of the acoustic field latter operators $b^{(\dag)}$:
\begin{equation}
E_x = E_\mathrm{zpf} be^{i q x} + E^*_\mathrm{zpf} b^{\dag} e^{-i q x},
\end{equation}
where $E_\mathrm{zpf}$ is the zero-point piezoelectric field amplitude. Note that the spatial dependence of the field is entirely isolated to the phases $e^{\pm i q x}$. 

Next, we express the polarization $P_x$ in the basis of 2DEG electronic states. Here, we assume that the spacing between consecutive impurities is significantly smaller than the $y$-direction span of the 2DEG, yet much larger than the $z$-direction span (which represents the axis perpendicular to the piezo-2DEG interface). Consequently, the Fermi level cuts through several subbands in the $k_y$-direction, while cutting through only the lowest subband in the $k_z$-direction (with the next $k_z$-subband being located a vast distance from the lowest subband in phase space). We can therefore ignore $k_z$ and index all states as $(k_x,k_y)$. The polarization $P_x$ can thus be expanded in terms of the carrier transitions between any initially occupied ground state $(k_i,k_y)$ and an initially unoccupied excited state $(k_i + \Delta k,k_y)$ \cite{todorov2015electrongas}:
\begin{align}
\begin{split}
P_x &= \frac{|\phi(z)|^2}{L_y} \sum_{k_i,\Delta k,k_y} \frac{q_e \hbar}{2 m \omega_{k_i + \Delta k,k_i}} \\
&\quad \times \Big(\xi_{k_i + \Delta k,k_i}(x) c_{k_i + \Delta k,k_y}^{\dag} c_{k_i,k_y} \\
&\quad\quad + \xi^*_{k_i + \Delta k,k_i}(x) c_{k_i,k_y}^{\dag} c_{k_i + \Delta k,k_y}\Big),
\end{split}
\end{align}
where $c^{(\dag)}_{k_x,k_y}$ is the fermionic annihilation (creation) operator for the electronic state at $(k_x,k_y)$, $\omega_{k_i + \Delta k,k_i} = \omega_{k_i + \Delta k} - \omega_{k_i}$ is the change in carrier energy (this is independent of $k_y$), $\phi(z)$ is the wavefunction along the $z$ axis, $L_y$ is the span of the 2DEG along the $y$-axis, and $\xi$ is defined as follows:
\begin{align}
\begin{split}
&\xi_{k_i + \Delta k,k_i}(x) \\
&= \frac{1}{L} \Big(e^{-i (k_i + \Delta k) x} \partial_x e^{i k_i x} - (\partial_x e^{-i (k_i + \Delta k) x}) e^{i k_i x}\Big) \\
&= i\frac{2k_i + \Delta k}{L} e^{-i (\Delta k) x},
\end{split}
\end{align}
where $L$ is the waveguide length. It is worth noting that $\omega_{k_i + \Delta k,k_i}$ can be solved as follows:
\begin{align} 
\begin{split} \label{eq: omegagap}
\omega_{k_i + \Delta k,k_i} &= \omega_{k_i + \Delta k} - \omega_{k_i} \\
&= \frac{\hbar}{2m} \Big((k_i + \Delta k)^2 - k_i^2\Big) \\
&= \frac{\hbar \Delta k (2k_i + \Delta k)}{2m}.
\end{split}
\end{align}
Substituting the above expressions into the Hamiltonian in Eq.~\eqref{eq: piezo-2DEG Hamiltonian}, we express the Hamiltonian in electron-phonon interaction form as follows:
\begin{widetext}
\begin{align}
\begin{split} \label{eq: solution to interaction Hamiltonian}
&H_\mathrm{int} \\
&= \int dx \Big(E_\mathrm{zpf} be^{i q x} + E^*_\mathrm{zpf} b^{\dag} e^{-i q x}\Big) \sum_{k_i,\Delta k,k_y} i \frac{q_e}{L \Delta k} \Big(e^{-i (\Delta k) x} c_{k_i + \Delta k,k_y}^{\dag} c_{k_i,k_y} - e^{i (\Delta k) x} c_{k_i,k_y}^{\dag} c_{k_i + \Delta k,k_y}\Big) \int dz |\phi(z)|^2 \int \frac{dy}{L_y}  \\
&= i \sum_{k_i,\Delta k,k_y} \frac{q_e}{L \Delta k} \Big(E_\mathrm{zpf} b c_{k_i + \Delta k,k_y}^{\dag} c_{k_i,k_y} \int dx e^{i (q - \Delta k) x} + E^*_\mathrm{zpf} b^{\dag} c_{k_i,k_y}^{\dag} c_{k_i + \Delta k,k_y} \int dx e^{-i (q + \Delta k) x}\Big)  \\
&= \frac{q_e E_\mathrm{zpf}}{q} (c_{k_i + q,k_i}^{\dag} b \delta_{q,\Delta k} + b^{\dag} c_{k_i + q,k_i}  \delta_{q,-\Delta k}).
\end{split}
\end{align}
\end{widetext}
As desired, the spatially-varying phases in the field and the polarization combine to enforce momentum conservation. The effective dipole matrix element thus becomes a function of the phonon wavevector amplitude $q$:
\begin{equation} \label{eq: dipole matrix element delocalized}
d_{{\mathrm{eff}},k_x + \Delta k,k_x}(q) = -\frac{q_e}{q} \delta_{q,\Delta k},
\end{equation}
for any initial wavevector $k_x$. This implies that the matrix element amplitude is inversely proportional to the phonon momentum, and hence directly proportional to the acoustic wavelength.

\begin{figure*}[!tb]
	\centering
	\includegraphics[width=\textwidth]{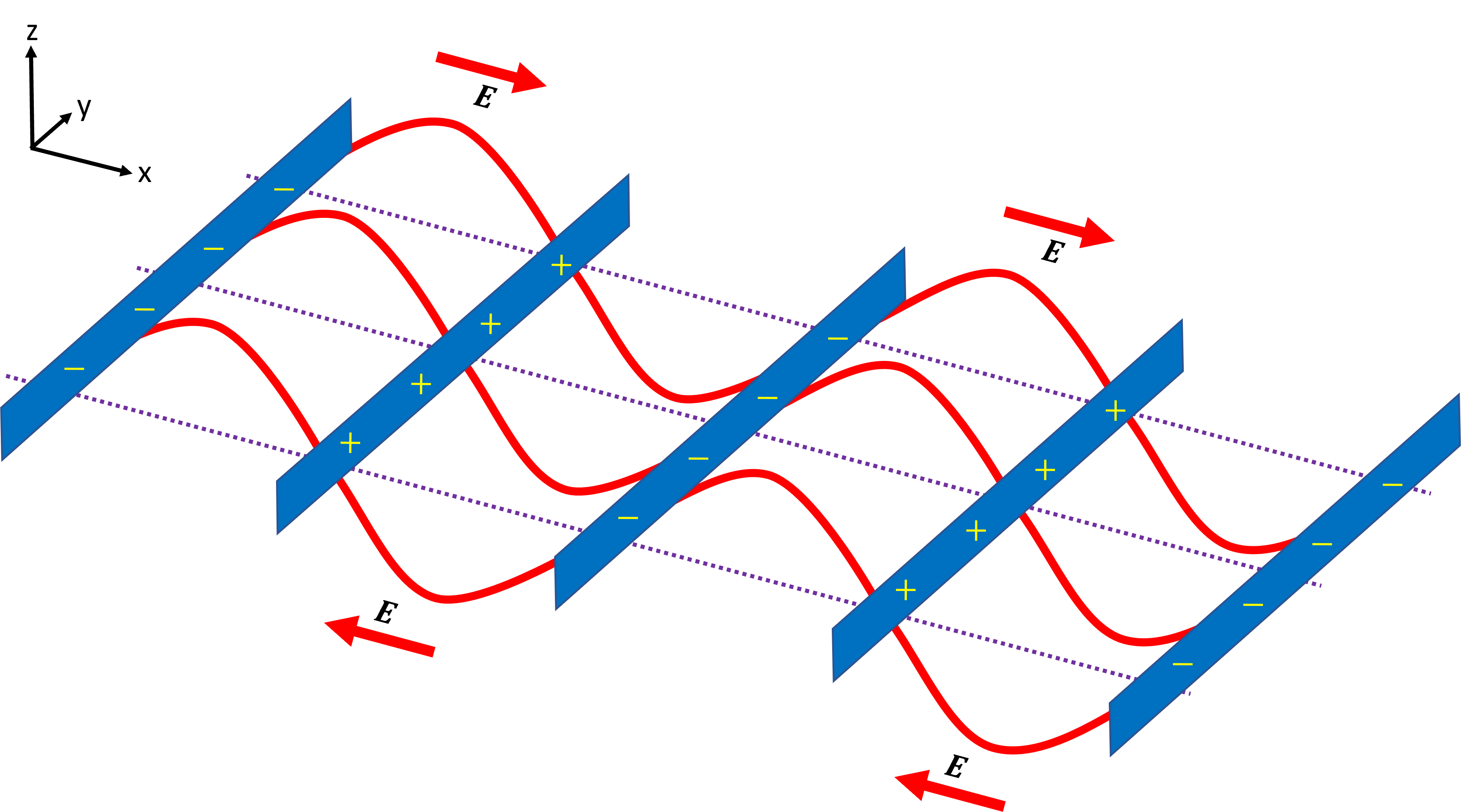}
	\caption{Conceptual diagram of how the alternating electric field $\bm{E}$ associated with the acoustic field induces dipole formation in the 2DEG. Note that the dipole size is limited to the order of a half-wavelength.} 
	\label{fig:chargeseparation}
\end{figure*}

It is worth analyzing the reason why the dipole moment is proportional to the acoustic wavelength but independent of the material length. Intuitively, this results from the alternating nature of the piezoelectric field (as shown in Fig.~\ref{fig:chargeseparation}), which limits the separation of charges to the order of a half-wavelength, regardless of the material dimensions.
As we increase the number of phonons (corresponding to an increase in field intensity), the dipole moment per polarized electron will stay constant, but the number of polarized electrons will increase, as will be shown in the next subsection.

\subsection{Deriving the higher-order interaction probabilities}
\label{sec: Deriving the higher-order interaction probabilities}

We now seek to determine the higher-order electron-phonon interaction probabilities, which yields the polarization field as a function of the acoustic (and thus piezoelectric) field strength, by deriving the generic $N^\textrm{th}$-order susceptibility in terms of the general electronic decay rate $\gamma = q_e/(m \mu)$, and from this deriving the conversion efficiency for specific nonlinear processes. As in the case of an atom interacting with an optical field \cite{boyd}, the energy conservation requirement for the intermediate states is relaxed due to the momentary nature of these intermediate states (per the energy-time uncertainty principle). However, unlike the atom-optical interaction, the much slower propagation speed of the acoustic signal ensures that each phonon features a significant momentum relative to its energy. The need for momentum conservation thus collapses the range of possible intermediate states, as the results in this section will show.

We start by considering the self-Hamiltonian of each excited 2DEG state. We incorporate the decay rate $\gamma$ as an anti-Hermitian effective Hamiltonian term \cite{chatterjee2022inputresonator}, yielding the following unperturbed Hamiltonian for the 2DEG states:
\begin{equation} \label{eq: 2DEG self-energy}
H_0 = \sum_{l > g} \hbar \bigg(\omega_{lg} - i \frac{\gamma_l}{2}\bigg) \ket{l} \bra{l},
\end{equation}
where $\omega_{lg} = \omega_l - \omega_g$ represents the energy of the excited state $\ket{l}$ (with the energy of the ground state $\ket{g}$ set as the baseline), and $\gamma_l = \gamma$ for all excited states $\ket{l}$ in our system (though we express the decay rate independently for each state for the sake of generality). Turning to the piezoelectric field, we specifically consider a set of waves of varying frequencies polarized along the $x$-axis and traveling in the $+\hat{x}$-direction. The fields interact with the electron dipole moment $\bm{d}$ via the following time-dependent interaction Hamiltonian:
\begin{align}
\begin{split} \label{eq: interaction Hamiltonian multiple modes}
&H_\mathrm{int}(t) \\
&= -d_x \sum_{p (\omega_p > 0)} \Big(E(\omega_p) e^{i(q_p x - \omega_p t)} + E^*(\omega_p) e^{-i(q_p x - \omega_p t)}\Big) \\
&= -\sum_p d_\mathrm{eff}(q_p) E(\omega_p) e^{-i \omega_p t},
\end{split}
\end{align}
where $q_p = \omega_p/v_s$, and the spatial variation of the field is incorporated into the effective dipole moment $d_\mathrm{eff}$ (which is defined as in Eq.~\eqref{eq: dipole matrix element delocalized}). Note that $E(-\omega_p) = E^*(\omega_p)$, ensuring the realness of the composite field at all points. We can also express a generic field of frequency $\omega_b > 0$ in terms of the corresponding ladder operators $a_b^{(\dag)}$ as $E(\omega_b) = E_\mathrm{zpf}(\omega_b) a_b$ and $E(-\omega_b) = E^*(\omega_b) = E_\mathrm{zpf}^*(\omega_b) a_b^{\dag}$. This will play an important role later in this section when we convert the semiclassical fields to quantum form for the purpose of deriving the frequency conversion rate from the Heisenberg equation of motion.

The interaction Hamiltonian enables us to derive the expectation value of the effective dipole moment, $\expect{d_\mathrm{eff}}$, as a series expansion in the field amplitudes $E$. Reducing the expectation value of the interaction Hamiltonian using the rotating-wave approximation (such that we only consider the time-independent $\expect{H_\mathrm{int}}$ terms), we express this expectation value as follows:
\begin{align}
\begin{split} \label{eq: dipole moment expectation value}
&\expect{d_\mathrm{eff}(q_p)} \\
&= \epsilon_0 V_\mathrm{2DEG} \sum_n \sum_{\substack{p_1,...,p_N \\ \omega_{p_1} + ... + \omega_{p_N} = -\omega_p}} \chi^{(N)}(\omega_{p_1},...,\omega_{p_N}) \\
&\quad\quad \times E(\omega_{p_1})...E(\omega_{p_N}) e^{i \omega_p t},
\end{split}
\end{align}
where the $N^\textrm{th}$-order susceptibility $\chi^{(n)}$ is defined as a function proportional to the $N^\textrm{th}$-order expansion of $\expect{d_\mathrm{eff}}$, in the case where a unique set of frequencies $\omega_{p_1},...,\omega_{p_n}$ add up to $-\omega_p$:
\begin{widetext}
\begin{align}
\begin{split} \label{eq: chi(N)}
\chi^{(N)}(\omega_{p_1},...,\omega_{p_n}) 
&= \frac{\int_{-\infty}^{\infty} \expect{d_\mathrm{eff}^{(N)}(-q_{p_1} - ... - q_{p_N})}_t e^{i (\omega_{p_1} + ... + \omega_{p_N}) t}}{\epsilon_0 V_\mathrm{2DEG} E(\omega_{p_1})...E(\omega_{p_N})} \\
&=  \frac{\int_{-\infty}^{\infty} e^{i (\omega_{p_1} + ... + \omega_{p_N}) t} \sum_{j = 0}^N \braket{\Psi^{(j)}(t)|d_\mathrm{eff}(-q_{p_1} - ... - q_{p_N})|\Psi^{(N-j)}(t)}}{\epsilon_0 V_\mathrm{2DEG} E(\omega_{p_1})...E(\omega_{p_N})},
\end{split}
\end{align}
\end{widetext}
where $\ket{\psi^{(n)}(t)}$ is the $n^\textrm{th}$-order expansion of the composite 2DEG wavefunction $\ket{\Psi(t)}$. Note that the variation of $\expect{d_\mathrm{eff}}$ with the field amplitudes in Eq.~\eqref{eq: dipole moment expectation value} is entirely due to the variation of the density of polarized electrons rather than the dipole strength corresponding to a given polarized electron (which is fixed by the acoustic wavelength, as discussed in the previous subsection). Intuitively, $\chi^{(N)}(\omega_{p_1},...,\omega_{p_n})$ corresponds to the probability that an electron becomes polarized upon interacting with $N$ fields of wavevectors $q_{p_1},...,q_{p_N}$ respectively (normalized to the amplitude of each field and to the material volume), multiplied by the strength of the dipole moment itself.

\begin{figure}[!tb]
	\centering
	\includegraphics[width=\columnwidth]{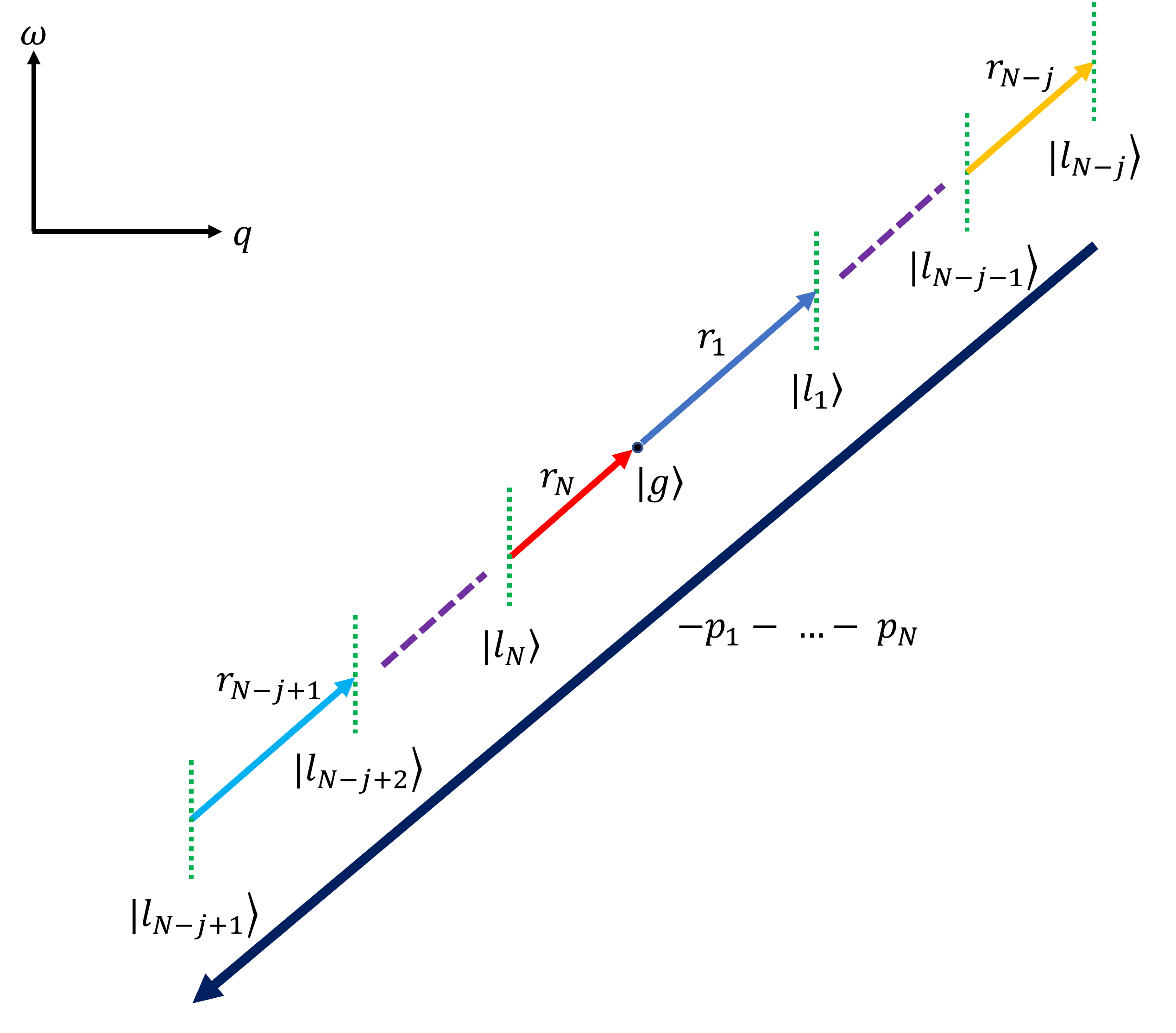}
	\caption{Depiction of the ladder of intermediate states for the $N$-step transition described by $\chi^{(N)}(\omega_{p_1},...,\omega_{p_N})$, with the horizontal (vertical) axis representing the wavevector $q$ (frequency $\omega$). Note that the set of generic modes $(r_1,...,r_N)$ can be composed of any permutation of the set of input modes $(p_1,...,p_N)$, with the requirement that $\omega_{r_1} + ... + \omega_{r_N} = \omega_{p_1} + ... + \omega_{p_N}$. Also note that each intermediate state is broadened in energy but fixed in momentum.}
	\label{fig:transitionladder}
\end{figure}

Appendix~\ref{sec: Calculation of Higher-Order Susceptibilities from Time-Dependent Perturbation Theory} shows how $\chi^{(N)}$ is calculated for a set of $N$ propagating acoustic waves with wavevectors $q_{p_1},...,q_{p_N}$. As for the 2DEG states, the well-defined momentum for each state ensures strict momentum conservation for each transition. However, due to the spectral broadening encapsulated in the decay rate $\gamma$, the energy conservation requirement is somewhat relaxed. This dynamic is depicted in Fig.~\ref{fig:transitionladder}.
Due to the availability of permutations within the set of modes $(p_1,...,p_N)$, we have left the mode indices in the generic form $(r_1,...,r_N)$, where $\omega_{r_1} + ... + \omega_{r_N} = \omega_{p_1} + ... + \omega_{p_N}$, and hence $q_{r_1} + ... + q_{r_N} = q_{p_1} + ... + q_{p_N}$. The transition to (from) $\ket{l_m}$ is mediated by the mode $r_m$ for $m \leq N - j$ ($m > N - j$). The transition from $\ket{l_{N - j}}$ to $\ket{l_{N - j +1}}$ is unique in that it is not mediated by any individual mode in the set $(p_1,...,p_N)$, but rather by the composite mode $-p_1 - ... - p_N$. Quantitatively, this mode will become relevant when considering the interaction Hamiltonian term associated with $\chi^{(N)}(\omega_{p_1},...,\omega_{p_N})$.

The most critical takeaway from Fig.~\ref{fig:transitionladder} is that while the intermediate 2DEG states are broadened in frequency $\omega$ (corresponding to a wide range of possible real states), each intermediate state is fixed in wavevector $q$, since the effective dipole moment $d_\mathrm{eff}$ conserves momentum. As a result, for a given initial state, each intermediate state reduces to a \textit{single} real state (rather than a superposition of multiple states). The summation over intermediate states in Eq.~\eqref{eq: chi(N) general} thus reduces to a sum over initial states (which we index with the 2D wavevector $(k_x,k_y)$):
\begin{widetext}
\begin{align}
\begin{split} \label{eq: chi(N) reduced}
&\chi^{(N)}(\omega_{p_1},...,\omega_{p_N}) \\
&= \frac{1}{\hbar^N \epsilon_0 V_\mathrm{2DEG}} \frac{q_e^{N+1}}{|q_{p_1}...q_{p_n} (q_{p_1} + ... + q_{p_n})|} \sum_{j = 0}^N \sum_{k_x,k_y} \mathcal{P} \sum_{p_1,...,p_N}   \\
&\quad \frac{1}
{(\omega_{k_x - q_{p_N},k_x} + \omega_{p_N} + i\frac{\gamma}{2}) ... (\omega_{k_x - q_{p_N} - ... - q_{p_{N - j + 1}},k_x} + (\omega_{p_{N - j + 1}} + ... + \omega_{p_N}) + i\frac{\gamma}{2})} \\
&\quad \times \frac{1}{(\omega_{k_x + q_{p_1} + ... + q_{p_{N-j}},k_x} - (\omega_{p_1} + ... + \omega_{p_{N - n}}) - i\frac{\gamma}{2}) ... (\omega_{k_x + q_{p_1},k_x} - \omega_{p_1} - i\frac{\gamma}{2})},
\end{split}
\end{align}
\end{widetext}
where we have used the fact that $d_{\mathrm{eff},k_f,k_i}(q) = (q_e/|q|) \delta_{q,k_f-k_i}$, and the symbol $\mathcal{P}$ preceding the summation over $p_1,...,p_N$ denotes a permutation of the mode indices. The expression thus includes $N!(N+1) = (N+1)!$ terms, due to the $N$ possible values of $j$ and the $N!$ permutations of the field mode indices.

\section{Susceptibility Results}
\label{sec: Susceptibility Results}

Here, we specifically consider the susceptibility results up to third order. In calculating the nonlinearity produced by the third-order Kerr susceptibility, it is also important to compare its value to the spectral broadening. For a high-Q acoustic cavity (which has recently been achieved \cite{shao2019phononicband}), the spectral broadening is dominated by the phonon absorption by the 2DEG electrons, for which we need to determine the imaginary part of $\chi^{(1)}$. In performing these calculations, we make the assumption that the average spacing between consecutive charge carriers is much smaller than the acoustic wavelength, corresponding to the limit $q \ll k_F$ (where $q$ and $k_F$ are the acoustic and Fermi wavevectors, respectively).

The general procedure for calculating the susceptibilities from phase-space integrals is derived in Appendix~\ref{sec: Calculating Susceptibility Through Phase-Space Integrals}. We start by considering the third-order Kerr susceptibility. To this end, the degenerate four-wave-mixing susceptibility $\chi^{(3)}(\omega,\omega,-\omega_1)$ (corresponding to the processes where 2 phonons of frequency $\omega$ each are absorbed, and 2 phonons of frequency $\omega_1$ and $\omega_2 = 2\omega- \omega_1$ are emitted) reduces to the following form:
\begin{align}
\begin{split} \label{eq: chi(3) integral}
&\chi^{(3)}(\omega,\omega,-\omega_1) = \\
&\quad \frac{q_e^4}{\pi^2 \hbar^3 \epsilon_0 t_\mathrm{2DEG}} \frac{v_s^4}{\omega^2 \omega_1 (2\omega - \omega_1)} \\
&\quad \times \Bigg(\int_{-q/2}^{k_F} dk_x 2 \sqrt{k_F^2 - k_x^2} f_{k_x}^{(3)}(\omega,\omega,-\omega_1) \\
&\quad - \int_{-q/2}^{k_F - q} dk_x 2 \sqrt{k_F^2 - (q + k_x)^2} f_{k_x}^{(3)}(\omega,\omega,-\omega_1)\Bigg),
\end{split}
\end{align}
where $f_{k_x}^{(3)}(\omega,\omega,-\omega_1)$ is approximately the following:
\begin{widetext}
\begin{align}
\begin{split} \label{eq: chi Kerr denominator terms}
f_{k_x}^{(3)}(\omega,\omega,-\omega_1) 
&\approx \frac{1}{(\omega_{k_x + q_2,k_x} - \omega_2 - i\frac{\gamma}{2}) (\omega_{k_x + 2q,k_x} - 2\omega - i\frac{\gamma}{2}) (\omega_{k_x + q,k_x} - \omega - i\frac{\gamma}{2})} \\
&\quad + \frac{1}{(\omega_{k_x + q_1,k_x} - \omega_1 + i\frac{\gamma}{2}) (\omega_{k_x + 2q,k_x} - 2\omega - i\frac{\gamma}{2}) (\omega_{k_x + q,k_x} - \omega - i\frac{\gamma}{2})} \\
&\quad + \frac{1}{(\omega_{k_x + q,k_x} + \omega + i\frac{\gamma}{2}) (\omega_{k_x + 2q,k_x} + 2\omega + i\frac{\gamma}{2}) (\omega_{k_x + q_1,k_x} + \omega_1 - i\frac{\gamma}{2})} \\
&\quad + \frac{1}{(\omega_{k_x + q,k_x} + \omega + i\frac{\gamma}{2}) (\omega_{k_x + 2q,k_x} + 2\omega + i\frac{\gamma}{2}) (\omega_{k_x + q_2,k_x} + \omega_2 + i\frac{\gamma}{2})},
\end{split}
\end{align}
\end{widetext}
where we have defined $q = \omega/v_s$ and $q_1 = \omega_1/v_s$. The first two terms are resonant, corresponding to the absorption of 2 phonons of frequency $\omega$ each, followed by the emission of a phonon of frequency $\omega_1$ and another of frequency $\omega_2 = 2\omega - \omega_1$. On the other hand, the latter two terms are counter-resonant, corresponding to the emission of $\omega_1$ and $\omega_2$ phonons followed by absorption of two $\omega$ phonons. The process described by each term is depicted in Fig.~\ref{fig:chi3(omega,omega,-omega1)}.
\begin{figure*}[!tb]
	\centering
	\begin{subfigure}{\columnwidth}
		\centering
	    \includegraphics[width=\linewidth]{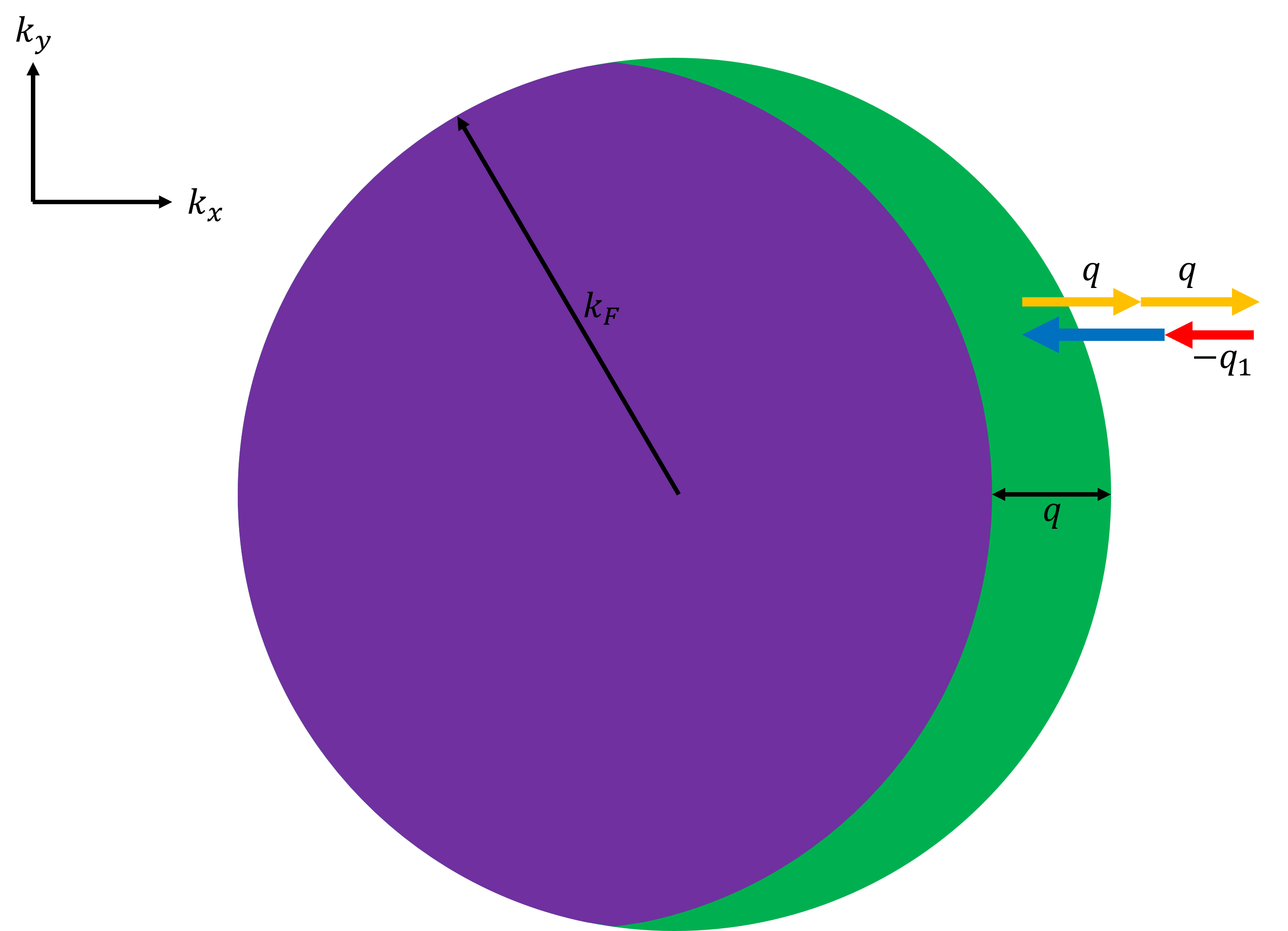}
	    \caption{}
	    \label{fig:chi3term1}
	\end{subfigure}
	\begin{subfigure}{\columnwidth}
		\centering
	    \includegraphics[width=\linewidth]{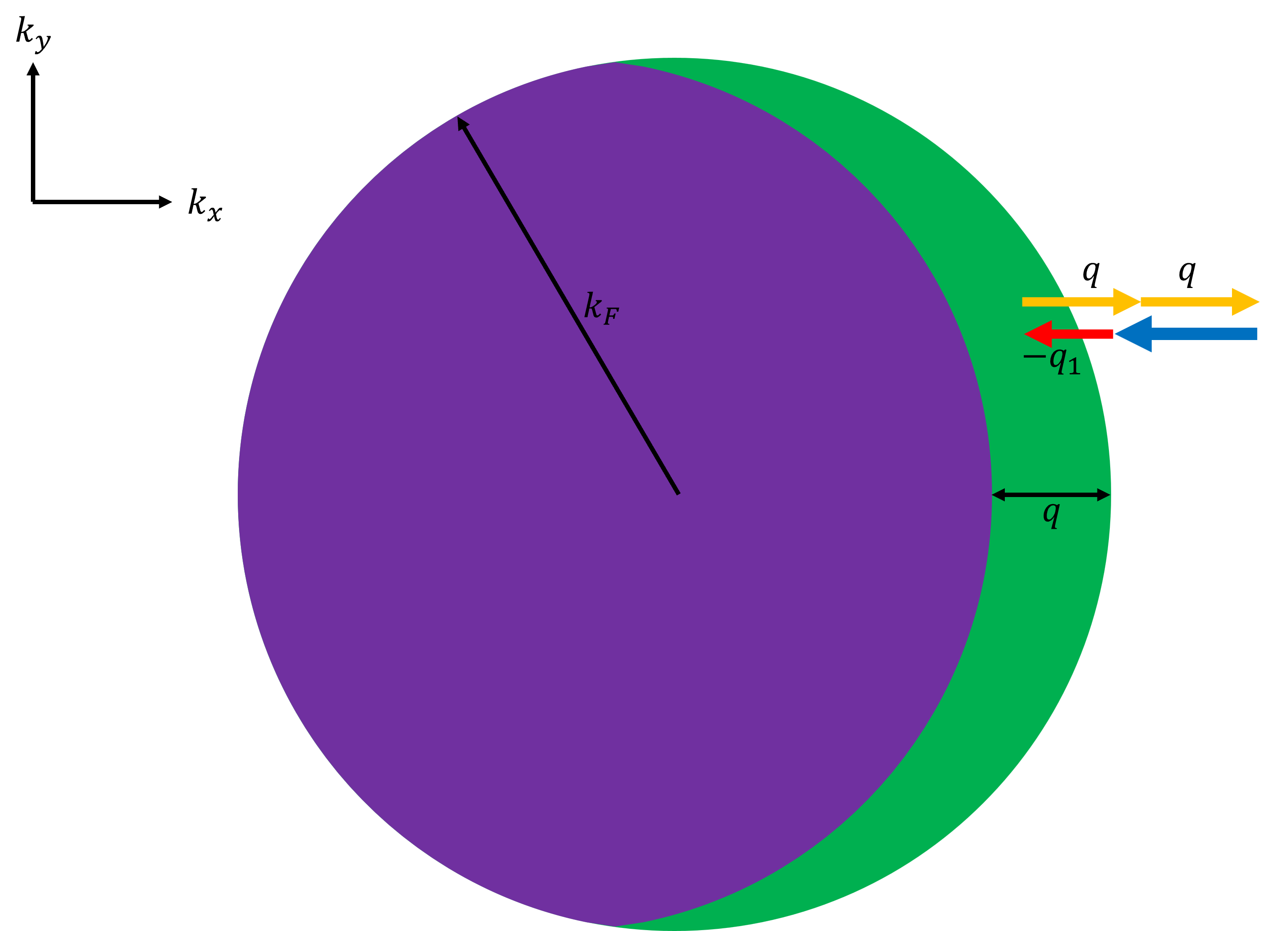}
		\caption{}
		\label{fig:chi3term2}
	\end{subfigure}
\end{figure*}
\begin{figure*}[!tb]\ContinuedFloat
    \begin{subfigure}{\columnwidth}
		\centering
	    \includegraphics[width=\linewidth]{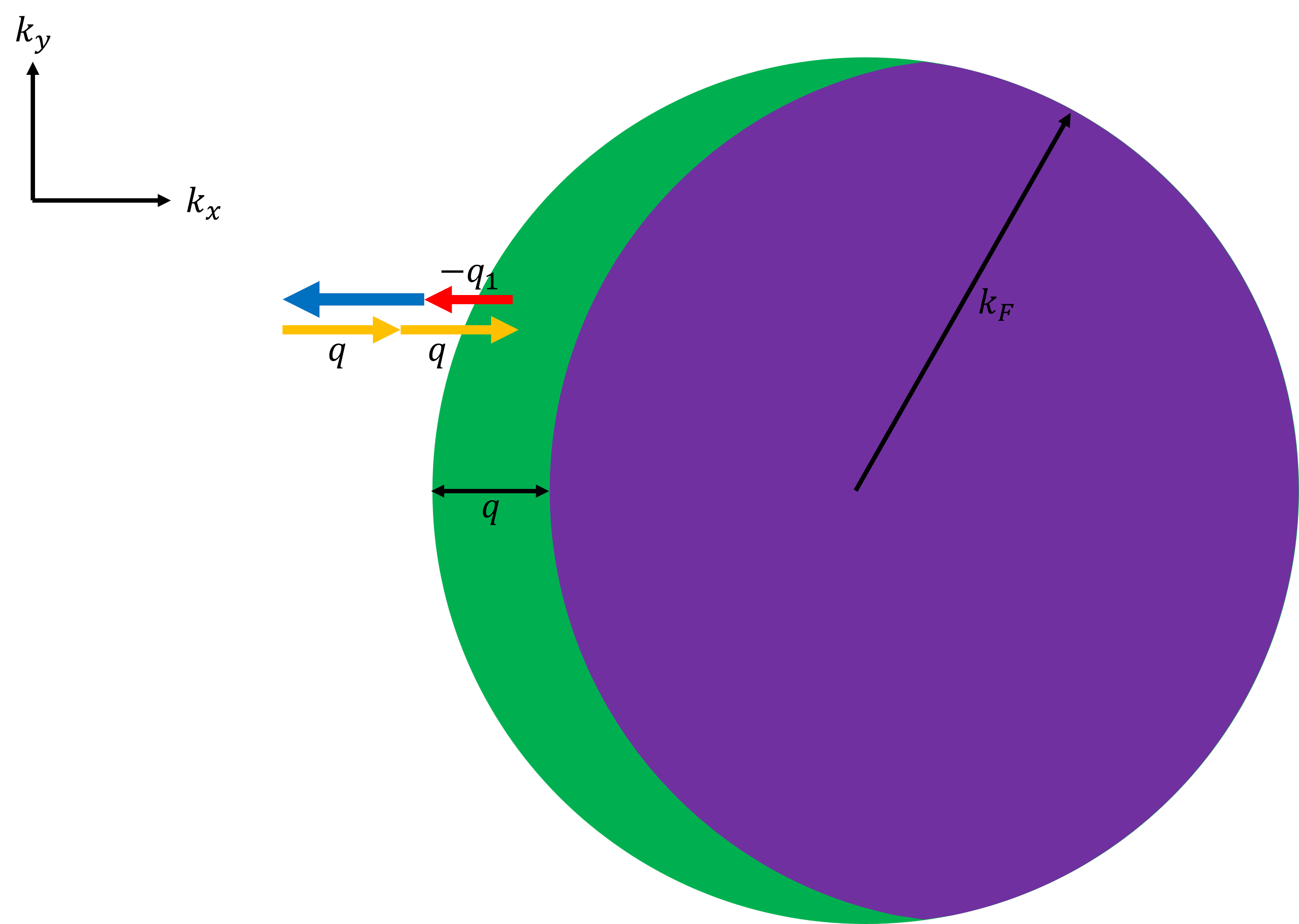}
		\caption{}
		\label{fig:chi3term3}
	\end{subfigure}
    \begin{subfigure}{\columnwidth}
		\centering
	    \includegraphics[width=\linewidth]{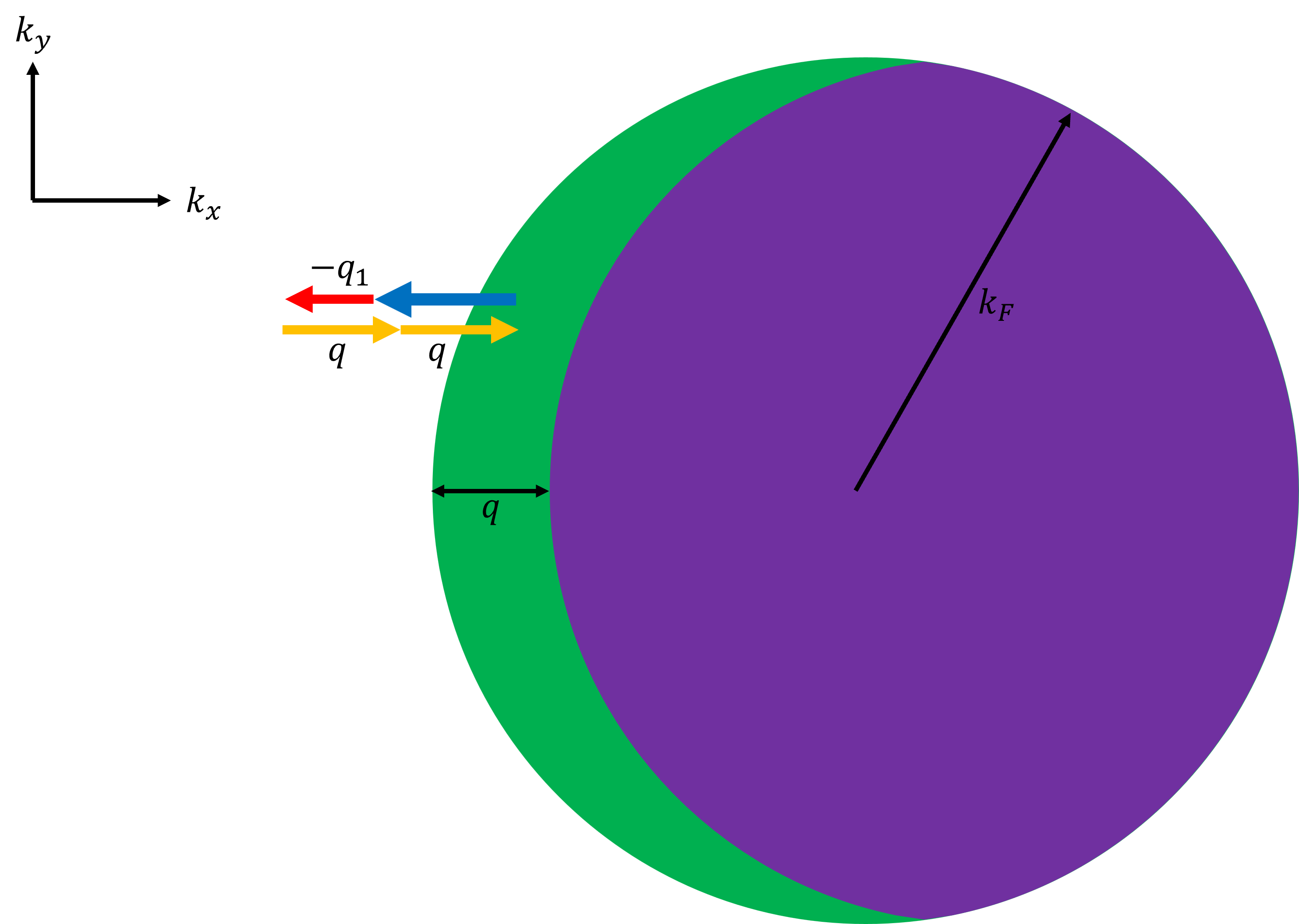}
		\caption{}
		\label{fig:chi3term4}
	\end{subfigure}
	\caption{Phase-space depiction of processes contributing to $\chi^{(3)}(\omega,\omega,-\omega_1)$, with (a), (b), (c), and (d) corresponding respectively to the first, second, third, and fourth terms of Eq.~\eqref{eq: chi Kerr denominator terms}. The shaded (unshaded) regions represent initially occupied (unoccupied) electronic states. Note that the crescent region represents the range of allowed initial states, while electrons in the rest of the Fermi circle are prohibited from interacting with the given phonons.}
	\label{fig:chi3(omega,omega,-omega1)}
\end{figure*}
It is worth noting that if we ignored the initial occupation of the states, then there would be 12 separate processes contributing to $\chi^{(3)}(\omega,\omega,-\omega_1)$. However, assuming that $q << k_F$, and given the fact that all states within the Fermi circle $|k| < k_F$ are occupied, we can make the approximation that the only valid processes are either two absorptions followed by two emissions (resonant) or two emissions followed by two absorptions (counter-resonant). The third-order Kerr susceptibility equals the degenerate four-wave-mixing susceptibility in the limit $\omega_1 \rightarrow \omega$.

Next, we examine the second-order susceptibility, focusing first on the case of sum-frequency generation encapsulated by $\chi^{(2)}(\omega_1,\omega_2)$, where two phonons of frequency $\omega_1$ and $\omega_2$ are absorbed and a phonon of frequency $2\omega = \omega_1 + \omega_2$ is emitted:
\begin{align}
\begin{split} \label{eq: chi(2) integral}
&\chi^{(2)}(\omega_1,\omega_2) = \\
&\quad \frac{q_e^3}{2 \pi^2 \hbar^2 \epsilon_0 t_\mathrm{2DEG}} \frac{v_s^3}{\omega_1 \omega_2 (\omega_1 + \omega_2)} \\
&\quad \times \Bigg(\int_{-q/2}^{k_F} dk_x 2 \sqrt{k_F^2 - k_x^2} f_{k_x}^{(2)}(\omega_1,\omega_2) \\
&\quad - \int_{-q/2}^{k_F - q} dk_x 2 \sqrt{k_F^2 - (q + k_x)^2} f_{k_x}^{(2)}(\omega_1,\omega_2)\Bigg),
\end{split}
\end{align}
where $f_{k_x}^{(2)}(\omega_1,\omega_2)$ is approximately the following:
\begin{widetext}
\begin{align}
\begin{split} \label{eq: chi(2) denominator terms}
f_{k_x}^{(2)}(\omega_1,\omega_2) &\approx 
\bigg(\frac{1}{(\omega_{k_x + 2q,k_x} - 2\omega - i\frac{\gamma}{2}) (\omega_{k_x + q_1,k_x} - \omega_1 - i\frac{\gamma}{2})} + \frac{1}{(\omega_{k_x + 2q,k_x} - 2\omega - i\frac{\gamma}{2}) (\omega_{k_x + q_2,k_x} - \omega_2 - i\frac{\gamma}{2})} \\
&\quad\quad + \frac{1}{(\omega_{k_x + 2q,k_x} + 2\omega + i\frac{\gamma}{2}) (\omega_{k_x + q_2,k_x} + \omega_2 + i\frac{\gamma}{2})} + \frac{1}{(\omega_{k_x + 2q,k_x} + 2\omega + i\frac{\gamma}{2}) (\omega_{k_x + q_1,k_x} + \omega_1 + i\frac{\gamma}{2})}\bigg),
\end{split}
\end{align}
\end{widetext}
where $q_1 = \omega_1/v_s$ and $q_2 = \omega_2/v_s$. The first two terms are resonant and the latter two terms are counter-resonant, with the processes depicted in Fig.~\ref{fig:chi2(omega1,omega2)}.
\begin{figure*}[!tb]
	\centering
	\begin{subfigure}{\columnwidth}
		\centering
	    \includegraphics[width=\linewidth]{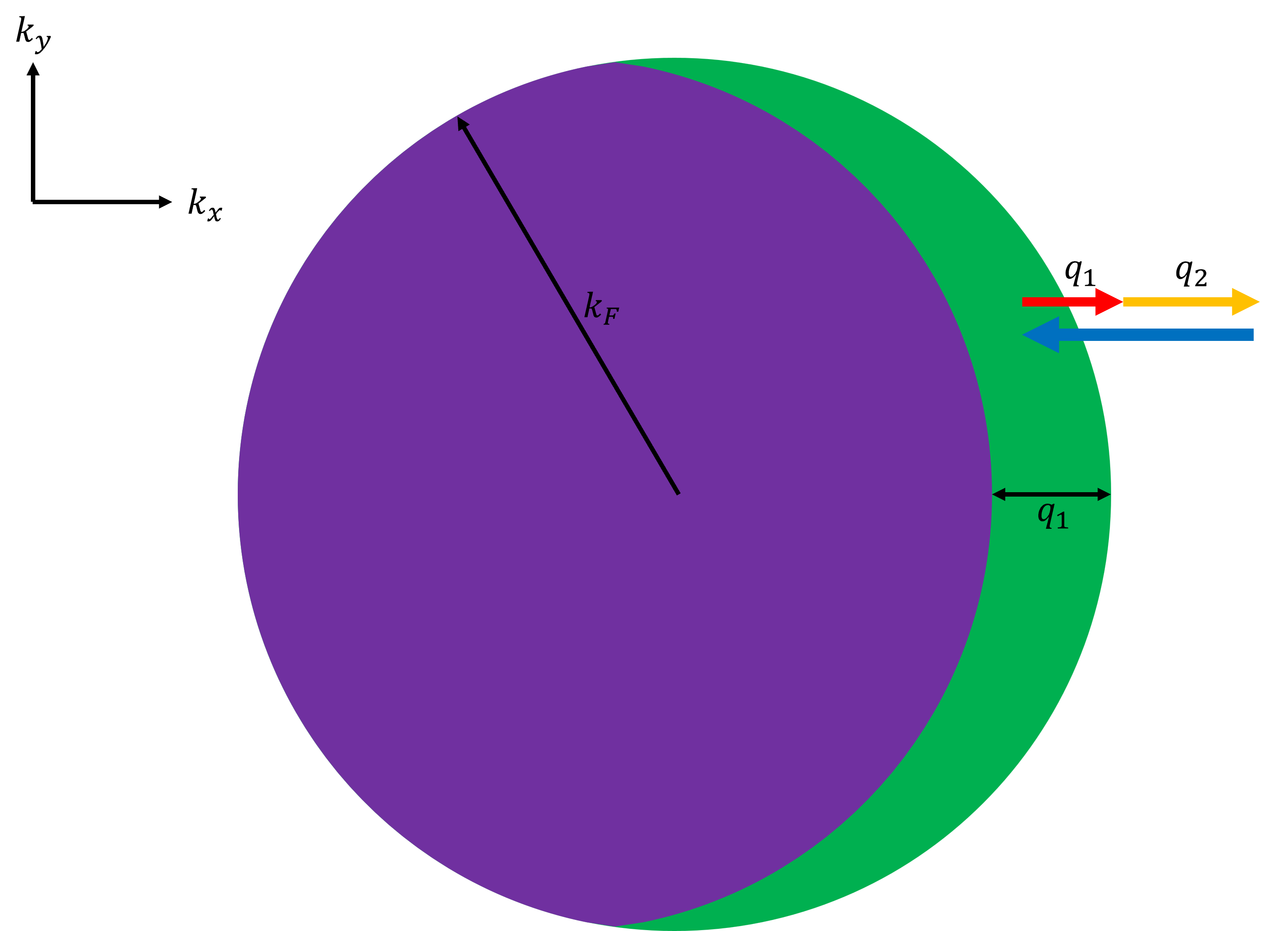}
	    \caption{}
	    \label{fig:chi2term1}
	\end{subfigure}
	\begin{subfigure}{\columnwidth}
		\centering
	    \includegraphics[width=\linewidth]{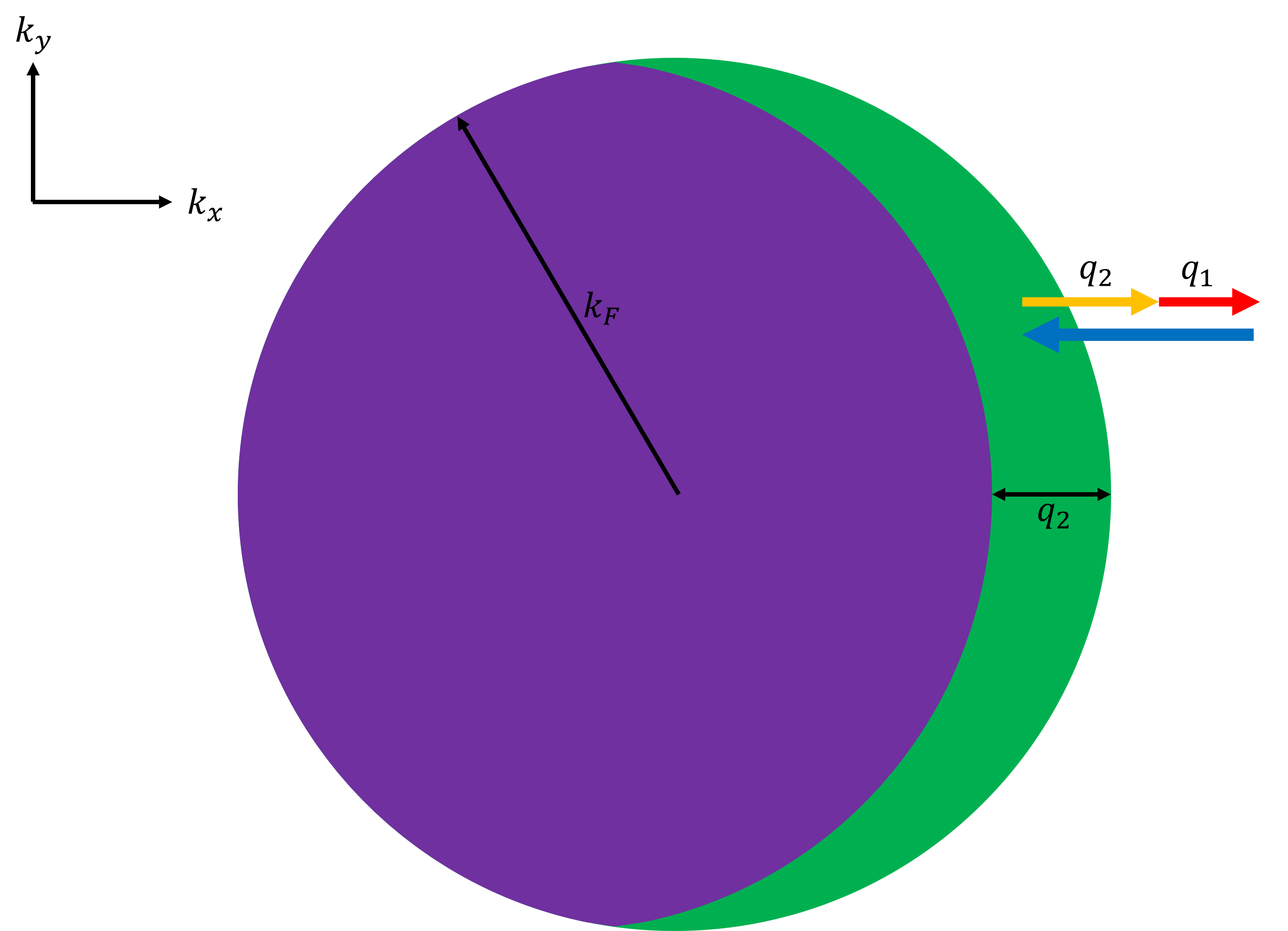}
		\caption{}
		\label{fig:chi2term2}
	\end{subfigure}
\end{figure*}
\begin{figure*}[!tb]\ContinuedFloat
    \begin{subfigure}{\columnwidth}
		\centering
	    \includegraphics[width=\linewidth]{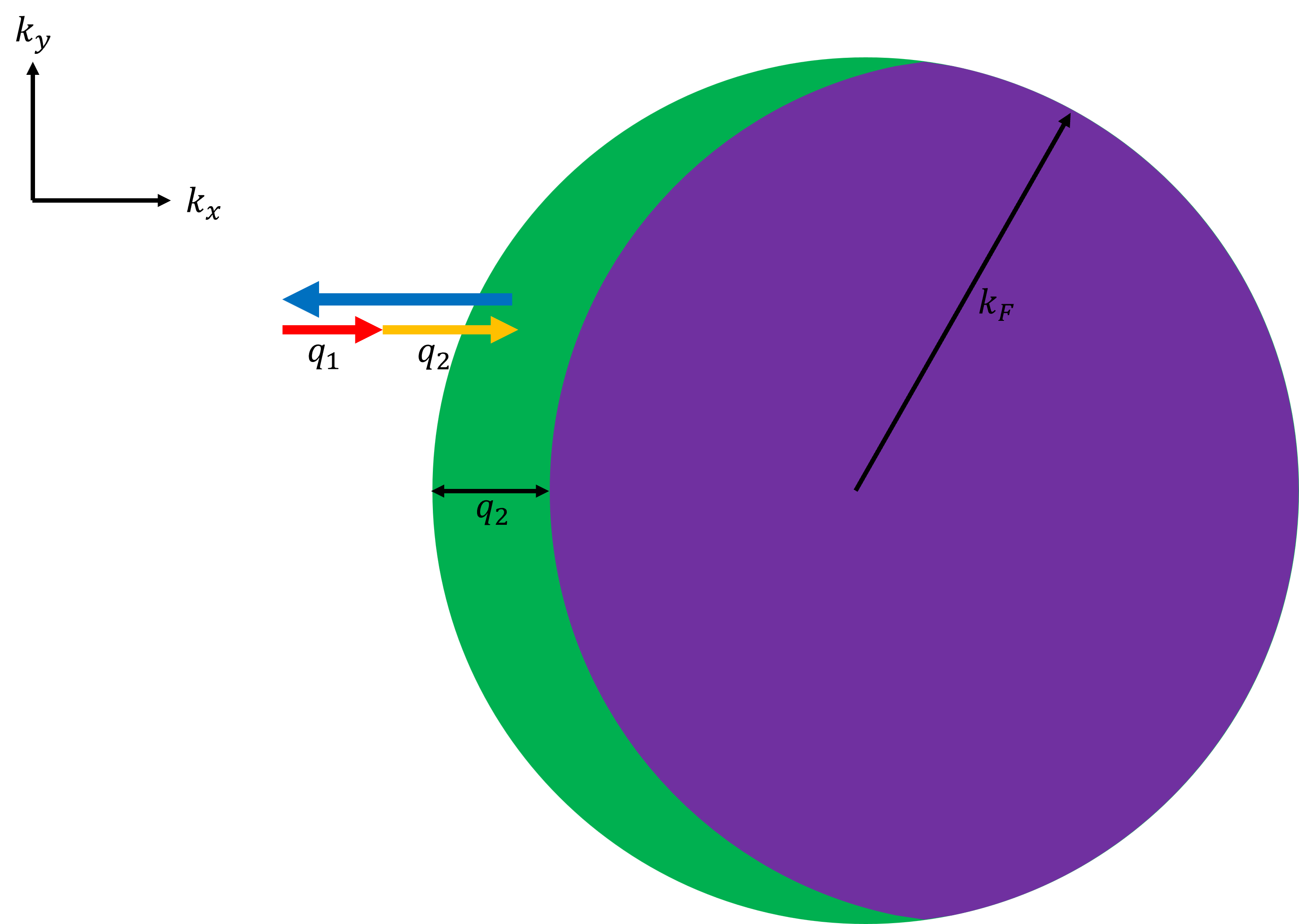}
		\caption{}
		\label{fig:chi2term3}
	\end{subfigure}
    \begin{subfigure}{\columnwidth}
		\centering
	    \includegraphics[width=\linewidth]{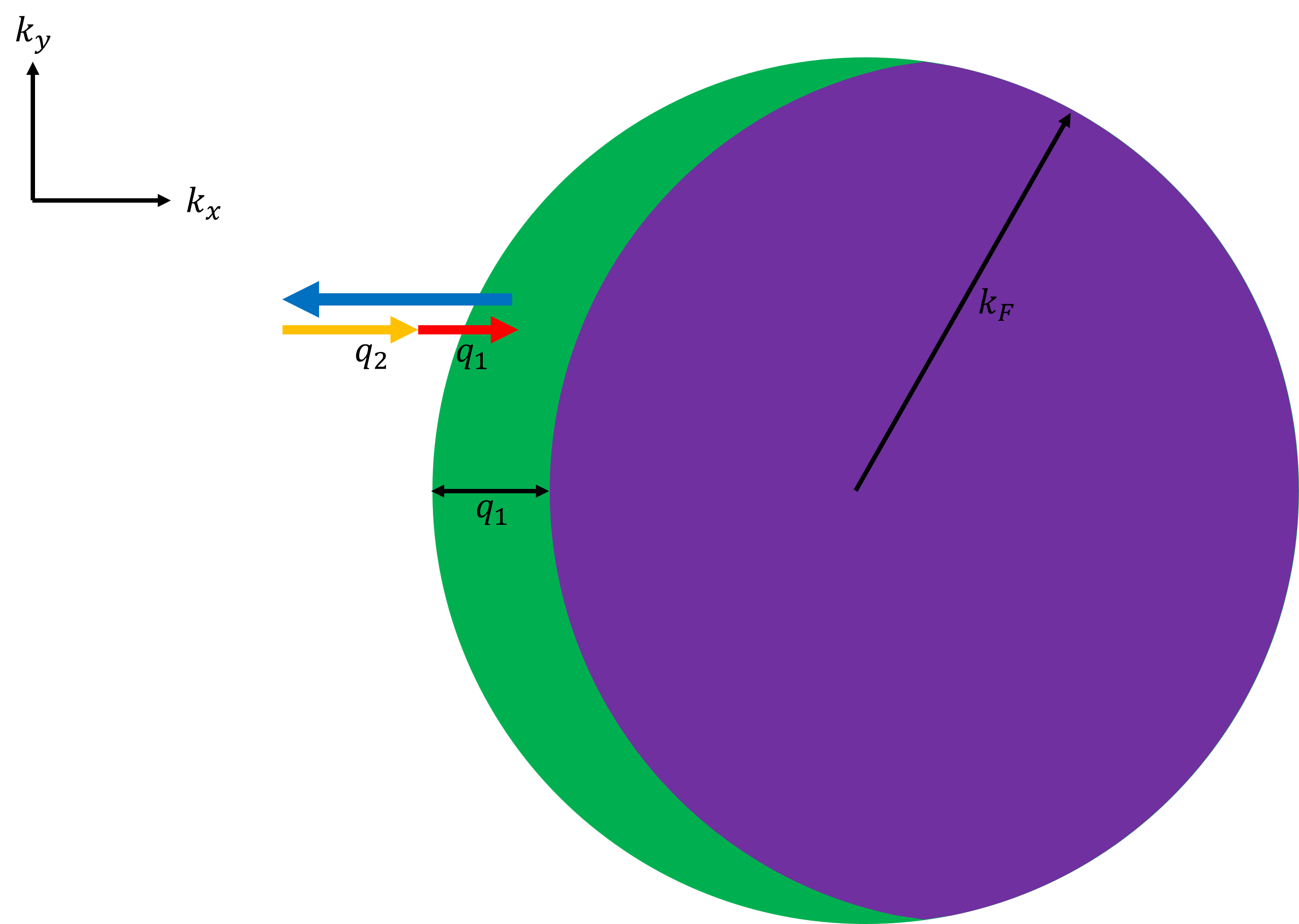}
		\caption{}
		\label{fig:chi2term4}
	\end{subfigure}
	\caption{Phase-space depiction of processes contributing to $\chi^{(2)}(\omega_1,\omega_2)$, with (a), (b), (c), and (d) corresponding respectively to the first, second, third, and fourth terms of Eq.~\eqref{eq: chi(2) denominator terms}. The shading scheme is identical to that in Fig.~\ref{fig:chi3(omega,omega,-omega1)}.}
	\label{fig:chi2(omega1,omega2)}
\end{figure*}
As with the case of $\chi^{(3)}$, we apply the high-carrier-density limit $q \ll k_F$, causing the requirement for unoccupied intermediate states to approximately reduce the 6 processes that would feed into a generic second-order process to 4 processes. The susceptibility governing the second-harmonic generation process equals the above susceptibility for the general sum-frequency generation in the limit $\omega_1 = \omega_2$.

In the context of second-order susceptibility, it is also worth examining the case of parametric amplification encapsulatd by $\chi^{(2)}(2\omega,-\omega_1)$, where a pump phonon of frequency $2\omega$ is absorbed, causing one phonon each of the signal frequency $\omega_1$ and the idler frequency $\omega_2$ to be emitted. Since the frequencies involved are the same as the sum-frequency generation, the coefficients for $\chi^{(2)}(2\omega,-\omega_1)$ are the same as those for $\chi^{(2)}(\omega_1,\omega_2)$ in Eq.~\eqref{eq: chi(2) integral}. However, $f_{k_x}^{(2)}(\omega_1,\omega_2)$ in the integral is replaced by $f_{k_x}^{(2)}(2\omega,-\omega_1)$, defined as:
\begin{widetext}
\begin{align}
\begin{split} \label{eq: chi(2) parametric denominator terms}
f_{k_x}^{(2)}(2\omega,-\omega_1) &\approx 
\bigg(\frac{1}{(\omega_{k_x + 2q,k_x} - 2\omega - i\frac{\gamma}{2}) (\omega_{k_x + q_1,k_x} - \omega_1 + i\frac{\gamma}{2})} + \frac{1}{(\omega_{k_x + 2q,k_x} - 2\omega - i\frac{\gamma}{2}) (\omega_{k_x + q_2,k_x} - \omega_2 - i\frac{\gamma}{2})} \\
&\quad\quad + \frac{1}{(\omega_{k_x + 2q,k_x} + 2\omega + i\frac{\gamma}{2}) (\omega_{k_x + q_2,k_x} + \omega_2 + i\frac{\gamma}{2})} + \frac{1}{(\omega_{k_x + 2q,k_x} + 2\omega + i\frac{\gamma}{2}) (\omega_{k_x + q_1,k_x} + \omega_1 - i\frac{\gamma}{2})}\bigg).
\end{split}
\end{align}
\end{widetext}
Note the sign flip (relative to Eq.~\eqref{eq: chi(2) denominator terms}) in the second imaginary term of the first and fourth processes. This has important implications for the relationship between the parametric-amplification susceptibility and the mobility in the case of degenerate parametric amplification ($\omega_1 = \omega_2$), as we will show in Sec.~\ref{sec: Second-Order Susceptibility}.

\begin{figure*}[!tb]
    \begin{subfigure}{\columnwidth}
		\centering
	    \includegraphics[width=0.9\linewidth]{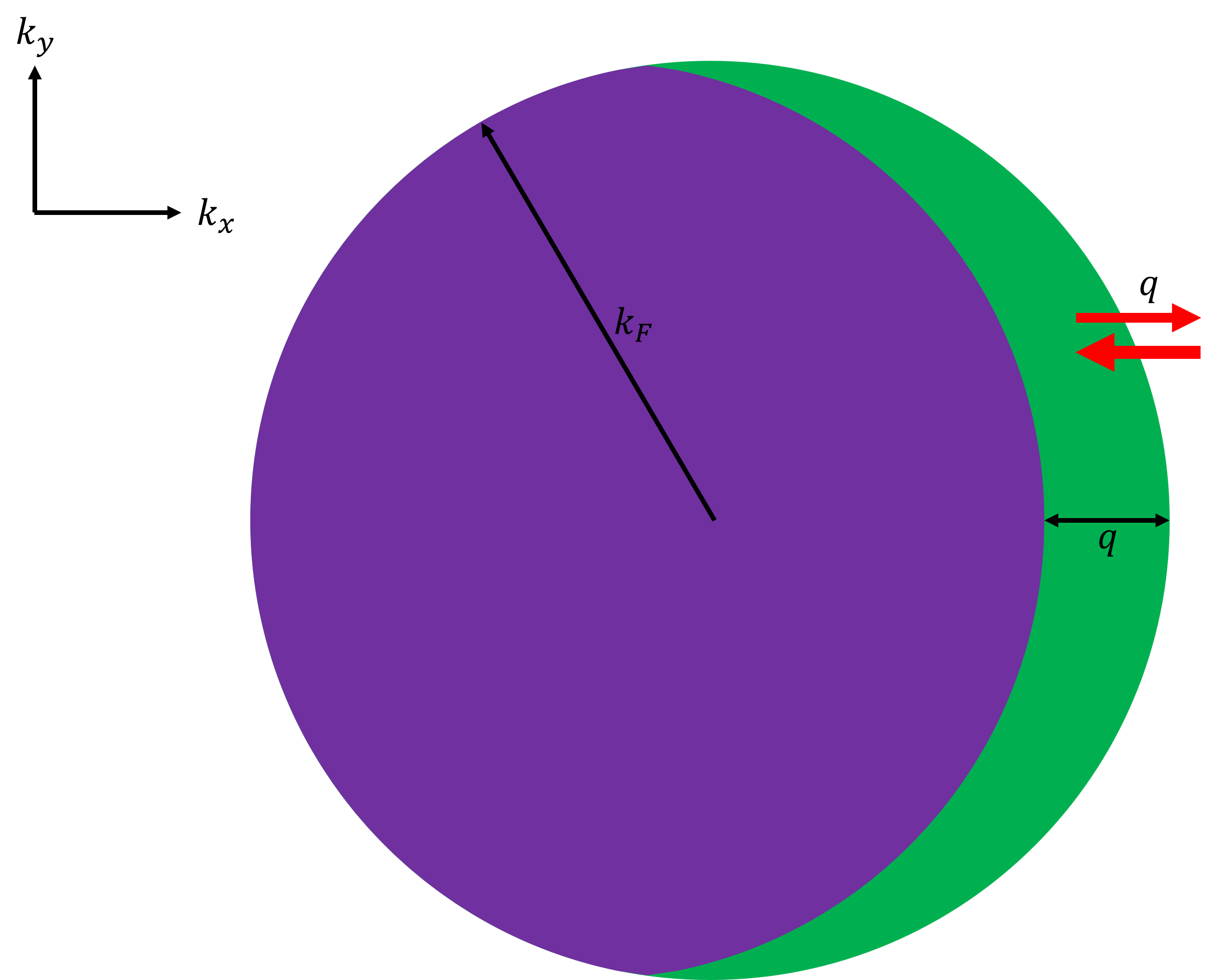}
		\caption{}
		\label{fig:chi1term1}
	\end{subfigure}
    \begin{subfigure}{\columnwidth}
		\centering
	    \includegraphics[width=\linewidth]{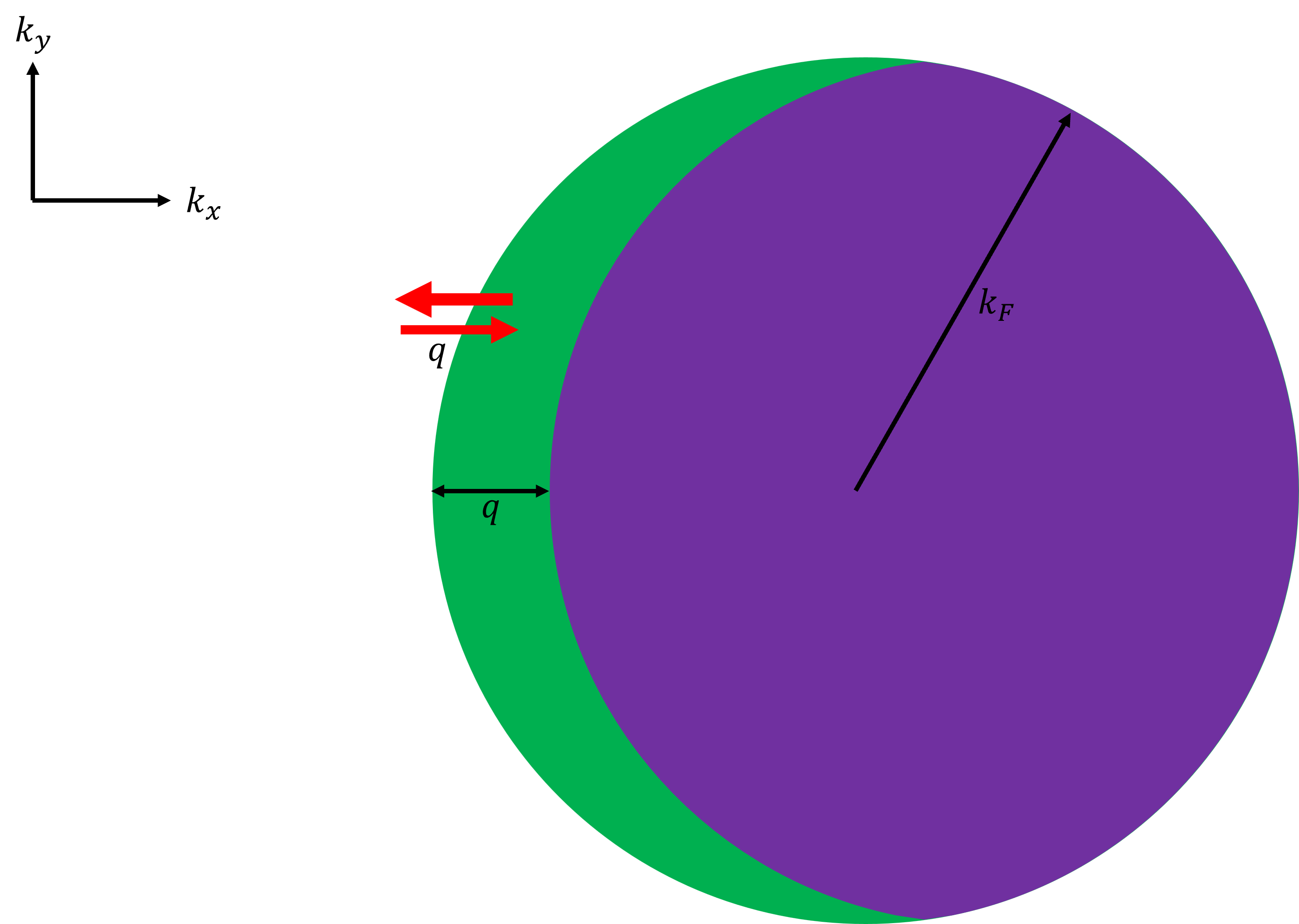}
		\caption{}
		\label{fig:chi1term2}
	\end{subfigure}
	\caption{Phase-space depiction of processes contributing to $\chi^{(1)}(\omega)$, with (a) and (b) corresponding respectively to the first and second terms of Eq.~\eqref{eq: chi(1) denominator terms}. The shading scheme is identical to that in Fig.~\ref{fig:chi3(omega,omega,-omega1)}.}
	\label{fig:chi1(omega)}
\end{figure*}

Finally, we calculate the linear susceptibility $\chi^{(1)}(\omega)$ corresponding to the absorption and emission of a phonon of frequency $\omega$:
\begin{align}
\begin{split} \label{eq: chi(1) integral}
&\chi^{(1)}(\omega) = \\
&\quad \frac{q_e^2}{2 \pi^2 \hbar \epsilon_0 t_\mathrm{2DEG}} \frac{v_s^2}{\omega^2} \Bigg(\int_{-q/2}^{k_F} dk_x 2 \sqrt{k_F^2 - k_x^2} f_{k_x}^{(1)}(\omega) \\
&\quad - \int_{-q/2}^{k_F - q} dk_x 2 \sqrt{k_F^2 - (q + k_x)^2} f_{k_x}^{(1)}(\omega)\Bigg),
\end{split}
\end{align}
where $f_{k_x}^{(1)}(\omega)$ is defined as follows:
\begin{equation} \label{eq: chi(1) denominator terms}
f_{k_x}^{(1)}(\omega) = \frac{1}{\omega_{k_x + q,k_x} - \omega - i\frac{\gamma}{2}} + \frac{1}{\omega_{k_x + q,k_x} + \omega + i\frac{\gamma}{2}},
\end{equation}
where $q = \omega/v_s$. The first term is resonant, while the second is counter-resonant, as shown in Fig.~\ref{fig:chi1(omega)}.
As previously discussed, these susceptibility expressions govern the corresponding processes in the high-carrier-density limit. The susceptibility values can be numerically evaluated from these expressions. However, it is useful to derive the analytical forms for the purpose of optimization. In the coming sections, we will do so in two limits: low-mobility and high-mobility.

\subsection{High-Mobility Regime}
\label{sec: High-Mobility Regime}

Here, we seek to derive a general procedure for analytically determining the susceptibilities in the high-mobility (low-decay) regime. We start by considering an electron in its ground state inside the Fermi circle. The probability that this electron interacts with an phonon of wavevector $q = \omega/v_s$ propagating in the $\hat{x}$-direction is governed by the detuning $\Delta \omega$ between the phonon's frequency and the transition frequency corresponding to the electron jumping from $k_x \rightarrow k_x + q$. Consequently, $\Delta \omega$ is a function of the electron's initial $x$-direction wavevector $k_x$ but is invariant in the initial $y$-direction wavevector $k_y$:
\begin{equation} \label{eq: Delta omega}
\Delta \omega = \omega_{k_x + q,k_x} - \omega_0 = \frac{\hbar q (2k_x + q)}{2m} - v_s q = \frac{\hbar q}{m} (k_x - k_{x,0}),
\end{equation}
where $\omega_0 = v_s q$ denotes the phonon's angular frequency, and $k_{x,0} = mv_s/\hbar - q/2$ represents the initial wavevector corresponding to a fully resonant electron-phonon interaction. 

\begin{figure*}[!tb]
    \begin{subfigure}{\columnwidth}
		\centering
	    \includegraphics[width=\linewidth]{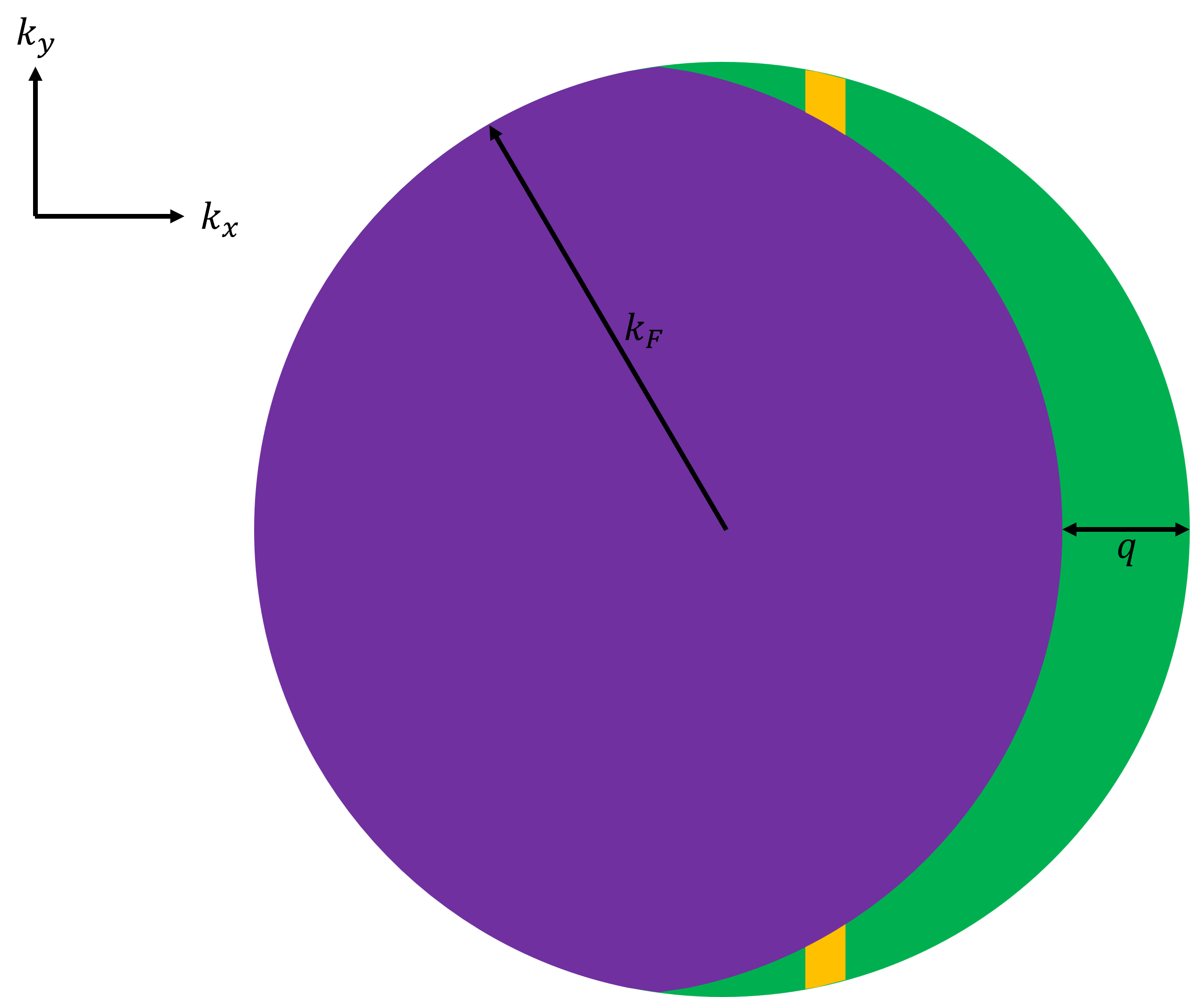}
		\caption{}
		\label{fig:nearresonancebandzoomedout}
	\end{subfigure}
    \begin{subfigure}{\columnwidth}
		\centering
	    \includegraphics[width=0.9\linewidth]{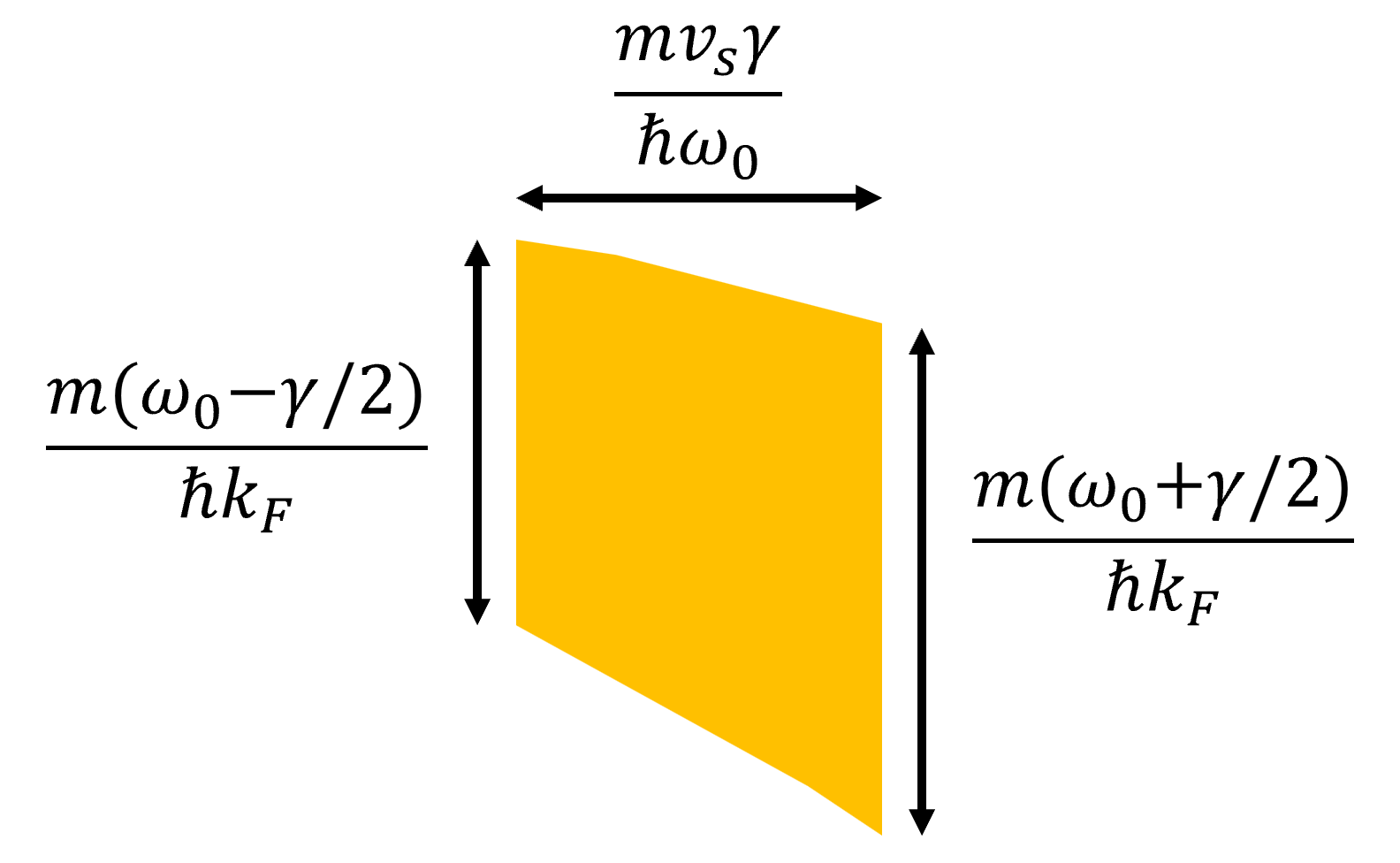}
		\caption{}
		\label{fig:nearresonancebandzoomedin}
	\end{subfigure}
	\caption{Phase-space diagram of near-resonance band spanning the electron-phonon detuning range $\Delta\omega \approx -\gamma/2$ to $\Delta\omega \approx \gamma/2$, zoomed out to show the position of the orange/light-colored band relative to the overall Fermi circle (a), and zoomed in to show the dimensions of each half of the band (b).}
	\label{fig:nearresonanceband}
\end{figure*}

We therefore focus on the near-resonance region, i.e., the band of electrons around the initial wavevector $k_{x,0}$ ranging in detuning from $\Delta \omega \approx -\gamma/2$ to $\Delta \omega \approx \gamma/2$. In the high-carrier-density/high-mobility regime, the $k_x$-span of this band is much less than the diameter $2k_F$ of the Fermi circle. We solve for the $k_x$-span of the near-resonance band from Eq.~\eqref{eq: Delta omega} as follows:
\begin{equation}
k_{x,\textrm{span}} = \frac{m}{\hbar q} (\Delta \omega)_\textrm{span} \approx \frac{2 m \gamma}{\hbar q}.
\end{equation}
Conveniently, the $k_x$-span of the near-resonance band is proportional to the electronic decay rate $\gamma$ (and hence inversely proportional to the mobility $\mu$). Intuitively, this is due to the fact that the electron-phonon detuning varies linearly with $k_x$. Applying the high-mobility requirement $k_{x,\textrm{span}} \ll 2k_F$, we derive the following condition for $\gamma$:
\begin{equation} \label{eq: gamma high-mobility regime}
\gamma \ll \frac{\hbar q k_F}{m},
\end{equation}
which corresponds to the following condition for the mobility $\mu$, upon applying the definition $\mu = q_e/(m \gamma)$:
\begin{equation} \label{eq: mu high-mobility regime}
\mu \gg \frac{q_e}{\hbar q k_F}.
\end{equation}
Since the susceptibility corresponding to each initial electron is invariant in $k_y$, the total susceptibility corresponding to all electrons with a particular initial $k_x$ is proportional to the $k_y$-span of valid initial electronic states with the given $k_x$-value. In the limit of high carrier density (i.e., $k_{x,0} \ll k_F$), the $k_y$-span given in Eq.~\eqref{eq: ky span} for $k_x$ values near $k_{x,0}$ can be approximately linearized in $k_x$ as follows:
\begin{align}
\begin{split}
k_{y,\textrm{span}} &= 2\Big(\sqrt{k_F^2 - k_x^2} - \sqrt{k_F^2 - (k_x + q)^2}\Big) \\
&\approx 2 k_F \bigg(-\frac{k_x^2}{2 k_F^2} + \frac{(k_x + q)^2}{2 k_F^2}\bigg) \\
&= \frac{q (2k_x + q)}{k_F}.
\end{split}
\end{align}
We can express any given $k_x$ near $k_{x,0}$ as $k_x = k_{x,0} + \Delta k_x$, where $\Delta k_x$ is the deviation of the initial wavevector from that corresponding to a resonant electron-phonon interaction. Then, per Eq.~\eqref{eq: Delta omega}, the detuning $\Delta \omega$ varies linearly with the wavevector deviation $\Delta k_x$ as $\Delta \omega = \hbar q (\Delta k_x)/m$. Substituting this relationship as well as the definition of $k_{x,0}$ into the above expression, we find that to first-order, the $k_y$-span varies linearly with the detuning $\Delta \omega$:
\begin{align}
\begin{split}
k_{y,\textrm{span}} &\approx \frac{q (2k_{x,0} + q)}{k_F} + \frac{2 q \Delta k_x}{k_F} \\
&= \frac{2 m \omega_0}{\hbar k_F} + \frac{2 m \Delta \omega}{\hbar k_F},
\end{split}
\end{align}
where we have also substituted the phonon frequency $\omega_0 = v_s q$. Using the fact that $dk_x = m d(\Delta \omega)/(\hbar q) = m v_s d(\Delta \omega)/(\hbar \omega_0)$, a generic susceptibility term $f(\Delta \omega)$ can be integrated over phase space in the following manner:
\begin{align}
\begin{split} \label{eq: chi(Delta omega) phase space integral}
&\int dk_x \int dk_y f(\Delta \omega) \\
&\approx \frac{2 m}{\hbar k_F} \int_{-\infty}^{\infty} \frac{m v_s}{\hbar \omega_0} d(\Delta \omega) (\omega_0 + \Delta \omega) f(\Delta \omega) \\
&= \frac{2 m^2 v_s}{\hbar^2 k_F \omega_0} \int_{-\infty}^{\infty} d(\Delta \omega) (\omega_0 + \Delta \omega) f(\Delta \omega) \\
&= \frac{2 m^2 v_s}{\hbar^2 k_F \omega_0} 
\begin{cases}
\omega_0 \int_{-\infty}^{\infty} d(\Delta \omega) f(\Delta \omega), & f(\Delta \omega) \textrm{ even}, \\
\int_{-\infty}^{\infty} d(\Delta \omega) (\Delta \omega) f(\Delta \omega), & f(\Delta \omega) \textrm{ odd}
\end{cases}.
\end{split}
\end{align}
Note that we only need to consider the resonant terms for $f(\Delta \omega)$, since the counter-resonant terms are negligible relative to the resonant terms in the high-mobility limit. Intuitively, these results are explained by Fig.~\ref{fig:nearresonanceband}.
Two observations are worth making here. First, we note that the $k_y$-span of valid initial states in the near-resonance band varies inversely with the Fermi wavevector $k_F$ (as shown in Fig.~\ref{fig:nearresonanceband}(b)), ensuring that the overall susceptibility also varies inversely with $k_F$ as well. It might seem counterintuitive that a \textit{larger} Fermi circle would yield a \textit{smaller} span of valid initial states. This can be visually explained based on Fig.~\ref{fig:nearresonanceband}(a) by the fact that for the given value of $k_x$, as the Fermi wavevector is increased, the total range of $k_y$ included inside the Fermi circle also increases, but the part of that range inside the forbidden (purple/circle-minus-crescent) zone grows even faster due to a reduced circle curvature. Therefore, the range of $k_y$ included inside the valid (green/crescent) region actually decreases as the Fermi wavevector increases. Since $k_F$ varies as $\sqrt{n}$ (where $n$ is the carrier density), this implies that any susceptibility term varies as $n^{-1/2}$ (with the exceptions of the real parts of $\chi^{(1)}$ and $\chi^{(2)}$, discussed later in this section).

The second observation regards the scaling of the near-resonance band area with the decay rate $\gamma$. For convenience, we define a new decay parameter $\gamma' = \gamma/2$. For each electron in the near-resonance band, the average $N^{\textrm{th}}$-order susceptibility is proportional to $1/\gamma'^N$. If a single-electron susceptibility term is even in the detuning $\Delta \omega$, then the total susceptibility corresponding to that term is simply determined by multiplying the average single-electron susceptibility by the area of the near-resonance band (which is proportional to $\gamma'$, as shown in Fig.~\ref{fig:nearresonanceband}(b)). Therefore, the overall susceptibility for the process represented by an even term is proportional to $1/\gamma'^{N - 1}$. However, if a single-electron susceptibility term is odd in $\Delta \omega$, then the contributions from the electrons featuring positive and negative electron-phonon detuning values largely cancel each other out. In this case, the net contribution comes from the \textit{difference} in the number of electrons between the negative-detuning and positive-detuning cases. As Fig.~\ref{fig:nearresonanceband}(b) shows, this difference is proportional to $\gamma'^2$ rather than $\gamma'$ when integrated over the band. As a result, the overall susceptibility for the process represented by an odd term is proportional to $1/\gamma'^{N - 2}$.

It is worth addressing the special cases of $\textrm{Re}[\chi^{(1)}]$ and $\textrm{Re}[\chi^{(2)}]$. In the case of $\textrm{Re}[\chi^{(1)}]$, the integral does not converge. Therefore, the contributing electron-phonon interactions are concentrated far off-resonance rather than near-resonance. This is evidenced by the fact that the resonant and counter-resonant terms for $\textrm{Re}[\chi^{(1)}]$ are always approximately equal even in the high-mobility limit. Consequently, $\textrm{Re}[\chi^{(1)}]$ is invariant in the mobility. It is also important to note that since the contributions to $\textrm{Re}[\chi^{(1)}]$ are far off-resonance, the initial electronic states are primarily concentrated around $k_x = k_F$. Since $\Delta \omega \propto k_x$ for $k_x \gg q$, and $\textrm{Re}[\chi^{(1)}] \propto 1/(\Delta \omega)$ for far-off-resonance electrons, the susceptibility for each electron varies inversely with $k_F$. On the other hand, the size of the region of valid initial states varies directly with $k_F$ (i.e., the radius of the Fermi circle). As a result, these two variations cancel out, and $\textrm{Re}[\chi^{(1)}]$ is invariant in the Fermi wavevector and thus independent of carrier density. 

The case of $\textrm{Re}[\chi^{(2)}]$ is complicated by the fact that the near-resonance contributions cancel out when integrated. On the other hand, it also converges to zero for high detuning values, unlike $\textrm{Re}[\chi^{(1)}]$. Consequently, in the high-mobility limit, the electronic contributions to $\textrm{Re}[\chi^{(2)}]$ are neither dominantly near-resonance or far-off-resonance, but rather in between the two limits. We will thus tackle the problem of $\textrm{Re}[\chi^{(2)}]$ in the high-mobility limit solely through numerical means.

\subsection{Low-Mobility Regime}
\label{sec: Low-Mobility Regime}

Here, we focus on deriving a procedure for analytically determining the susceptibilities in the low-mobility (high-decay) limit. Conceptually, this corresponds to the regime in which the $k_x$-span of the near-resonance band is much greater than the diameter $2k_F$ of the Fermi circle, such that all electrons are included in the band and the electron-phonon detuning for any electron is negligible compared to the electronic decay rate $\gamma$. Following a derivation analogous to the high-mobility regime, the following conditions on $\gamma$ and the mobility $\mu$ hold for the low-mobility regime:
\begin{align}
\gamma &\gg \frac{\hbar q k_F}{m}, \\
\mu &\ll \frac{q_e}{\hbar q k_F}.
\end{align}
In this regime, the far-off-resonance electrons will dominate the susceptibility terms.

We start by considering how the real-valued susceptibility terms are reduced. In general, the real part of $f^{(N)}$ can be expressed as a power series in the electron-phonon detuning $\Delta \omega$ as follows:
\begin{equation} \label{eq: real susceptibility terms generic}
\textrm{Re}[f^{(N)}(\Delta \omega)] = \sum_{m = 0}^{\frac{1}{2} (N - \frac{1}{2} + (-1)^N \frac{1}{2})} \frac{c_{N,m} \gamma'^{2m} (\Delta \omega)^{N - 2m}}{\prod_{n = 1}^N a_n^2 (\Delta \omega)^2 + \gamma'^2},
\end{equation}
where each $a_n$ is a positive integer and each $c_{N,m}$ represents a coefficient that derives from expanding the numerator. In the low-mobility limit, we can make the approximation $|\Delta \omega| \ll \gamma'$. As such, we can reduce each summation to the leading term in $\gamma'$. This yields a real part of $f^{(N)}$ that varies linearly with the detuning for odd $N$, while being invariant in the detuning for even $N$ (assuming that this result does not cancel out):
\begin{equation} \label{eq: low-mobility real susceptibility generic}
\textrm{Re}[f^{(N)}(\Delta \omega)] \approx 
\begin{cases}
\frac{c_{N,N/2}}{\gamma'^N}, & N \textrm{ even}, \\
\frac{c_{N,(N-1)/2} \Delta \omega}{\gamma'^{N + 1}}, & N \textrm{ odd}
\end{cases}.
\end{equation}
Consequently, in the low-mobility limit, the real part of $\chi^{(N)}$ varies with the mobility $\mu$ as $\mu^N$ ($\mu^{N+1}$) for even (odd) $N$, assuming that the leading term does not cancel out. To determine the overall $\chi^{(N)}$ values, we devise a scheme to integrate the single-electron values over the entire phase space of valid initial states rather than just the near-resonance region. To this end, we approximate the $k_y$-span by applying the approximation $q \ll k_F$ but \textit{not} $k_x \ll k_F$:
\begin{align}
\begin{split} \label{eq: ky span far-off-resonance}
k_{y,\textrm{span}}(k_x) &= 2 \Big(\sqrt{k_F^2 - k_x^2} - \sqrt{k_F^2 - (k_x + q)^2}\Big) \\
&\approx 2 \sqrt{k_F^2 - k_x^2} - 2 \sqrt{k_F^2 - k_x^2} \bigg(1 - \frac{q k_x}{k_F^2 - k_x^2}\bigg) \\
&= \frac{2 q k_x}{\sqrt{k_F^2 - k_x^2}}.
\end{split}
\end{align}
The available phase-space area, which we label $A_\mathrm{phase}$, is solved by simply integrating the $k_y$-span over $k_x$ as follows:
\begin{align}
\begin{split} \label{eq: available phase space area}
A_\textrm{phase} = \int_0^{k_F} dk_x k_{y,\textrm{span}}(k_x) &\approx \int_0^{k_F} dk_x \frac{2 q k_x}{\sqrt{k_F^2 - k_x^2}} \\
&= \frac{2 \omega_0}{v_s} \int_0^{k_F} dk_x \frac{k_x}{\sqrt{k_F^2 - k_x^2}} \\
&= \frac{2 \omega_0 k_F}{v_s} \int_0^{\pi/2} d\theta \cos{\theta} \\
&= \frac{2 \omega_0 k_F}{v_s}.
\end{split}
\end{align}
Since the single-electron processes feeding into $\textrm{Re}[\chi^{(N)}]$ for even $N$ are invariant in the detuning, the overall susceptibility can simply be calculated by multiplying $\textrm{Re}[f^{(N)}]$ by the phase-space area $A_\mathrm{phase}$. To determine the overall $\textrm{Re}[\chi^{(N)}]$ for odd values of $N$, however, we need to integrate the detuning $\Delta \omega$ over this phase-space area. Since the electrons in the overall range of valid initial states are concentrated near $k_x = k_F$, we can make the approximation $k_x \gg q$ and thus $\Delta \omega \approx \omega_{k_x + q,k_x} \approx \hbar q k_x/m$. We thus integrate $\Delta \omega$ over the valid initial phase-space as follows:
\begin{align}
\begin{split} \label{eq: generic integral low-mobility}
\int_{-q/2}^{k_F} dk_x k_{y,\textrm{span}}(k_x) \Delta \omega &\approx \int_0^{k_F} dk_x \frac{2 q k_x}{\sqrt{k_F^2 - k_x^2}} \frac{\hbar q k_x}{m} \\
&= \frac{2 \hbar \omega_0^2}{m v_s^2} \int_0^{k_F} dk_x \frac{k_x^2}{\sqrt{k_F^2 - k_x^2}} \\
&= \frac{2 \hbar \omega_0^2 k_F^2}{m v_s^2} \int_0^{\pi/2} d\theta \cos^2{\theta} \\
&= \frac{\pi \hbar \omega_0^2 k_F^2}{2 m v_s^2}.
\end{split}
\end{align}
Note that the result as $k_F^2$. This is due to two factors. The first is that the available phase space area (corresponding to the number of available electrons) scales as $k_F$. The second is that the probability that an electron interacts with the phonon fields also scales as $k_F$ in the given regime, since $\expect{\Delta \omega} \propto k_F$. As such, $\textrm{Re}[\chi^{(N)}]$ scales linearly with the carrier density $n$ for odd $N$. On the other hand, for even $N$, $\textrm{Re}[\chi^{(N)}]$ scales as $n^{1/2}$, since the electron-phonon interaction probability in this case is invariant in the detuning $\Delta \omega$.

It is also worth examining the integral of $(\Delta \omega)^n$ more generally over the valid initial phase space. This is especially useful for the cases where the leading term cancels out, leaving behind secondary terms that vary as $(\Delta \omega)^n$ where $n > 1$:
\begin{align}
\begin{split} \label{eq: generic higher-order integral low-mobility}
&\int_{-q/2}^{k_F} dk_x k_{y,\textrm{span}}(k_x) (\Delta \omega)^n \\
&\approx \int_0^{k_F} dk_x \frac{2 q k_x}{\sqrt{k_F^2 - k_x^2}} \bigg(\frac{\hbar q k_x}{m}\bigg)^n \\
&= \frac{2 \hbar^n \omega_0^{n+1}}{m^n v_s^{n+1}} \int_0^{k_F} dk_x \frac{k_x^{n+1}}{\sqrt{k_F^2 - k_x^2}} \\
&= \frac{2 \hbar^n \omega_0^{n+1} k_F^{n+1}}{m^n v_s^{n+1}} \int_0^{\pi/2} d\theta \cos^{n+1}{\theta} \\
&= \frac{\sqrt{\pi} \hbar^n \omega_0^{n+1} k_F^{n+1}}{m^n v_s^{n+1}} \frac{\Gamma \Big(\frac{n + 2}{2}\Big)}{\Gamma \Big(\frac{n + 3}{2}\Big)}.
\end{split}
\end{align}
This result varies as $k_F^{n + 1}$, since the number of available electrons scales linearly with $k_F$, while the interaction probability for each electron scales as $k_F^n$.

Finally, we comment on the imaginary parts of $\chi^{(N)}$. Separating these into individual terms corresponding to specific processes, the imaginary parts of the individual susceptibility terms $f^{(N)}$ take the following form:
\begin{align}
\begin{split}
&\textrm{Im}[f^{(N)}(\Delta \omega)] = \\
&\quad \sum_{m = 1}^{\frac{1}{2} (N + \frac{1}{2} - (-1)^N \frac{1}{2})} \frac{c_{N,m} \gamma'^{2m - 1} (\Delta \omega)^{N - 2m + 1}}{\prod_{n = 1}^N a_n^2 (\Delta \omega)^2 + \gamma'^2}.
\end{split}
\end{align}
Using the same argument as with the real parts, the leading term reduces to the following in the low-mobility limit:
\begin{equation}
\textrm{Im}[f^{(N)}(\Delta \omega)] \approx 
\begin{cases}
\frac{c_{N,N/2} \Delta \omega}{\gamma'^{N + 1}}, & N \textrm{ even}, \\
\frac{c_{N,(N+1)/2}}{\gamma'^N}, & N \textrm{ odd}
\end{cases}.
\end{equation}
The leading term for $\textrm{Im}[\chi^{(N)}]$ therefore scales with mobility as $\mu^{N + 1}$ ($\mu^N$) for even (odd) $N$. However, it is critical to note that in the low-mobility regime, the resonant and counter-resonant terms largely cancel out, corresponding to the fact that stimulated emission largely balances out absorption. As a result, the overall $\textrm{Im}[\chi^{(N)}]$ for a generic $N$ must be calculated by making a higher-order expansion of the terms to account for the slight difference between the absorption and emission probabilities.

\subsection{First-Order Susceptibility}
\label{sec: First-Order Susceptibility}

We now apply our findings for the low-mobility and high-mobility limits to derive the first-order susceptibility in these limits. In parallel, we numerically determine the susceptibility as a function of mobility. We start with the real part of $\chi^{(1)}$, which governs the dielectric constant for the 2DEG. As discussed in the previous sections, $\textrm{Re}[\chi^{(1)}]$ follows a special rule in the high-mobility limit, while following the generic behavior in the low-mobility limit.

We thus focus first on the low-mobility limit. Substituting Eq.~\eqref{eq: chi(1) denominator terms} into Eq.~\eqref{eq: real susceptibility terms generic} and simplifying per Eq.~\eqref{eq: low-mobility real susceptibility generic}, we find that $\textrm{Re}[f_{k_x}^{(1)}]$ approximately reduces to the following:
\begin{equation}
\textrm{Re}[f^{(1)}(\Delta \omega)] \approx \frac{2 \Delta \omega}{\gamma'^2}.
\end{equation}
Note that the real part of $\chi^{(1)}$ thus scales with mobility as $\mu^2$ in the low-mobility limit. Integrating $\Delta \omega$ over the phase-space area of the initial electronic states as in Eq.~\eqref{eq: generic integral low-mobility} and multiplying by the constants shown in Eq.~\eqref{eq: chi(1) integral}, we find the following analytical expression for the real part of $\chi^{(1)}$ in the low-mobility limit:
\begin{align}
\begin{split}
\textrm{Re}[\chi^{(1)}(\omega_0)] &\approx \bigg(\frac{q_e^2 v_s^2}{2 \pi^2 \hbar \epsilon_0 t_\mathrm{2DEG} \omega_0^2}\bigg) \frac{2}{\gamma'^2} \bigg(\frac{\pi \hbar \omega_0^2 k_F^2}{2 m v_s^2}\bigg) \\
&= \frac{q_e^2 k_F^2}{2 \pi \epsilon_0 t_\mathrm{2DEG} m \gamma'^2} \\
&= \frac{2 m k_F^2 \mu^2}{\pi \epsilon_0 t_\mathrm{2DEG}},
\end{split}
\end{align}
where we substituted the relationship $\gamma' = q_e/(2 m \mu)$ in the last line. Note that this susceptibility is independent of the phonon frequency $\omega_0$ in the limit $q \ll k_F$. This is because the total number of electrons (encapsulated by the phase space area) and the real part of the inverse detuning per electron each varies linearly with $\omega_0$, thus cancelling out the inverse dependence of the dipole interaction probability with $\omega_0^2$.

Next, we examine the real part of $\chi^{(1)}$ in the high-mobility limit. As discussed in Sec.~\ref{sec: High-Mobility Regime}, $\textrm{Re}[\chi^{(1)}]$ is dominated by the far-off-resonance interactions, unlike the general case for the high-mobility limit. Using the approximation $|\Delta \omega| \gg \gamma'$, we find that $\textrm{Re}[f_{k_x}^{(1)}]$ approximately reduces to:
\begin{equation} \label{eq: high-mobility chi(1) generic}
\textrm{Re}[f^{(1)}(\Delta \omega)] \approx \frac{2}{\Delta \omega}.
\end{equation}
As with the generic low-mobility case, we use the fact that the valid initial states are concentrated near $k_x = k_F$, leading to the approximation $k_x \gg q$ and thus $\Delta \omega \approx \omega_{k_x + q,k_x} \approx \hbar q k_x/m$. Integrating $(\Delta \omega)^{-1}$ over the valid initial phase-space area yields the following result:
\begin{align}
\begin{split}
\int_{-q/2}^{k_F} dk_x k_{y,\textrm{span}}(k_x) (\Delta \omega)^{-1} &\approx \int_0^{k_F} dk_x \frac{2 q k_x}{\sqrt{k_F^2 - k_x^2}} \frac{m}{\hbar q k_x} \\
&= \frac{2 m}{\hbar} \int_0^{k_F} \frac{dk_x}{\sqrt{k_F^2 - k_x^2}} \\
&= \frac{2 m}{\hbar} \int_0^{\pi/2} d\theta \\
&= \frac{\pi m}{\hbar}.
\end{split}
\end{align}
Multiplying this by 2 (per Eq.~\eqref{eq: high-mobility chi(1) generic}) and by constants shown in Eq.~\eqref{eq: chi(1) integral}, we find the following analytical expression for the real part of $\chi^{(1)}$ in the high-mobility limit:
\begin{align}
\begin{split}
\textrm{Re}[\chi^{(1)}(\omega_0)] &\approx \bigg(\frac{q_e^2 v_s^2}{2 \pi^2 \hbar \epsilon_0 t_\mathrm{2DEG} \omega_0^2}\bigg) 2 \bigg(\frac{\pi m}{\hbar}\bigg) \\
&= \frac{q_e^2 v_s^2 m}{\pi \hbar^2 \epsilon_0 t_\mathrm{2DEG} \omega_0^2}.
\end{split}
\end{align}
Note that this is independent of both mobility and carrier density. The invariance in mobility can be explained by the fact that the dielectric screening is dominated by far-off-resonance electron-phonon interactions, causing the detuning to dominate over the decay rate, as explained earlier. The invariance in carrier density $n$ is explained by the fact that the number of electrons available to interact with phonons varies as $\sqrt{n}$ (since the corresponding phase-space area varies as $k_F$), while the probability that any given electron actually interacts with a phonon varies as $1/\sqrt{n}$ (since in the far-off-resonance limit, this probability scales inversely with the average detuning, which in turn varies as $k_F$ and thus as $\sqrt{n}$).

We now turn to the imaginary part of $\chi^{(1)}$, which governs the net phonon absorption rate and thus the spectral broadening for the phonon energy levels. As discussed in Sec.~\ref{sec: Low-Mobility Regime}, the imaginary parts follow a special rule in the low-mobility limit due to cancellation between absorptive and emissive processes. We thus start by analyzing the high-mobility limit. Here, we can approximately ignore the second (counter-resonant) term in $f_{k_x}^{(1)}$ in Eq.~\eqref{eq: chi(1) denominator terms}, yielding:
\begin{align}
f^{(1)}(\Delta \omega) &\approx \frac{1}{\Delta \omega - i\gamma'}, \\
\textrm{Im}[f^{(1)}(\Delta \omega)] &\approx \frac{\gamma'}{(\Delta \omega)^2 + \gamma'^2}.
\end{align}
Using the near-resonance approximation and the corresponding procedure in Eq.~\eqref{eq: chi(Delta omega) phase space integral}, we integrate this over the near-resonance band (noting that the integrand is even in $\Delta \omega$) and multiply by the coefficients in Eq.~\eqref{eq: chi(1) integral}, leading to the following result for the imaginary part of $\chi^{(1)}$ in the high-mobility limit:
\begin{align}
\begin{split}
&\textrm{Im}[\chi^{(1)}(\omega_0)] \\
&\approx \bigg(\frac{q_e^2 v_s^2}{2 \pi^2 \hbar \epsilon_0 t_\mathrm{2DEG} \omega_0^2}\bigg) \frac{2 m^2 v_s}{\hbar^2 k_F} \int_{-\infty}^{\infty} d(\Delta \omega) \frac{\gamma'}{(\Delta \omega)^2 + \gamma'^2} \\
&= \bigg(\frac{q_e^2 v_s^2}{2 \pi^2 \hbar \epsilon_0 t_\mathrm{2DEG} \omega_0^2}\bigg) \frac{2 m^2 v_s}{\hbar^2 k_F} \pi \\
&= \frac{q_e^2 v_s^3 m^2}{\pi \hbar^3 \epsilon_0 t_\mathrm{2DEG} \omega_0^2 k_F}.
\end{split}
\end{align}
As such, the imaginary part of $\chi^{(1)}$ is invariant in $\gamma'$ and hence independent of mobility in the high-mobility limit. This is because the number of electrons in the near-resonance region scales linearly with $\gamma'$, while the interaction probability for each electron scales inversely with $\gamma'$, causing the overall interaction rate to be constant in $\gamma'$.

Finally, we solve for the imaginary part of $\chi^{(1)}$ in the low-mobility limit. If we take the resonant and counter-resonant terms (corresponding to phonon absorption and emission, respectively) in Eq.~\eqref{eq: chi(1) denominator terms} individually, then the magnitude of the imaginary part of each of those terms should approximately reduce to $1/\gamma'$, since $\gamma' \gg |\Delta \omega| \gg \omega_0$ in the low-mobility limit. However, this dynamic is fundamentally altered by the fact that the two terms cancel each other out to first-order. This requires us to make a higher-order approximation, altering the mobility-dependence of the net susceptibility:
\begin{align}
\begin{split}
&\textrm{Im}[f^{(1)}_{k_x}](\omega_0) \\
&= \frac{\gamma'}{(\omega_{k_x + q,k_x} - \omega_0)^2 + \gamma'^2} - \frac{\gamma'}{(\omega_{k_x + q,k_x} + \omega_0)^2 + \gamma'^2} \\
&\approx \frac{4 \omega_0 \omega_{k_x + q,k_x}}{\gamma'^3} \\
&\approx \frac{4 \omega_0 \Delta \omega}{\gamma'^3}.
\end{split}
\end{align}
Intuitively, this result can be conceptualized as follows: the probability of the 2DEG electrons absorbing a phonon from the field (represented by the resonant term) is almost (though not fully) cancelled out by the probability of 2DEG electrons emitting a phonon into the field. The former (deamplifying) process narrowly edges out the latter (amplifying) process, yielding a positive imaginary part of the electron-phonon interaction probability. As with the real part of $f_{k_x}^{(1)}$, the imaginary part is approximately linear in the electron-phonon detuning $\Delta \omega$. We therefore integrate over the phase-space area of initial electronic states as in Eq.~\eqref{eq: generic integral low-mobility} and mutiply by the coefficients in Eq.~\eqref{eq: chi(1) integral}:
\begin{align}
\begin{split}
\textrm{Im}[\chi^{(1)}(\omega_0)] &\approx \bigg(\frac{q_e^2 v_s^2}{2 \pi^2 \hbar \epsilon_0 t_\mathrm{2DEG} \omega_0^2}\bigg) \frac{4 \omega_0}{\gamma'^3} \bigg(\frac{\pi \hbar \omega_0^2 k_F^2}{2 m v_s^2}\bigg) \\
&= \frac{q_e^2 \omega_0 k_F^2}{\pi \epsilon_0 m t_\mathrm{2DEG} \gamma'^3} \\
&= \frac{8 m^2 \omega_0 k_F^2 \mu^3}{\pi \epsilon_0 q_e t_\mathrm{2DEG}}.
\end{split}
\end{align}
Although the overall susceptibilities corresponding to the absorptive and emissive processes individually scale linearly with the mobility $\mu$, the net process scales as $\mu^3$. As with the real part of $\chi^{(1)}$ in the low-mobility regime, this result also varies linearly with carrier density $n \propto k_F^2$, due to a combination of the overall number of available electrons varying and the single-electron interaction probability each varying as $n^{1/2}$.

\begin{figure*}[!tb]
    \begin{subfigure}{\columnwidth}
		\centering
	    \includegraphics[width=\linewidth]{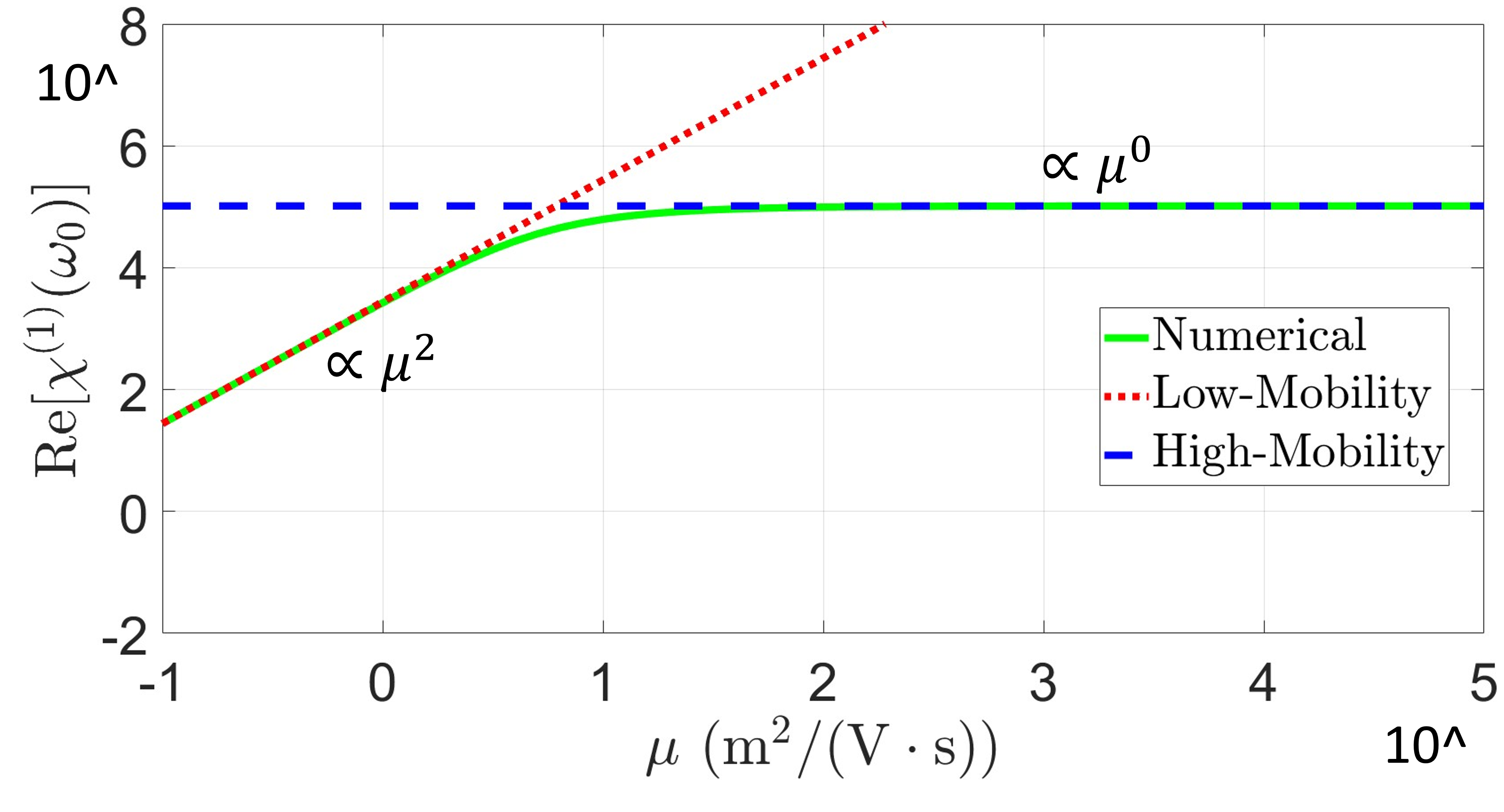}
		\caption{}
		\label{fig:chi1realresults}
	\end{subfigure}
    \begin{subfigure}{\columnwidth}
		\centering
	    \includegraphics[width=\linewidth]{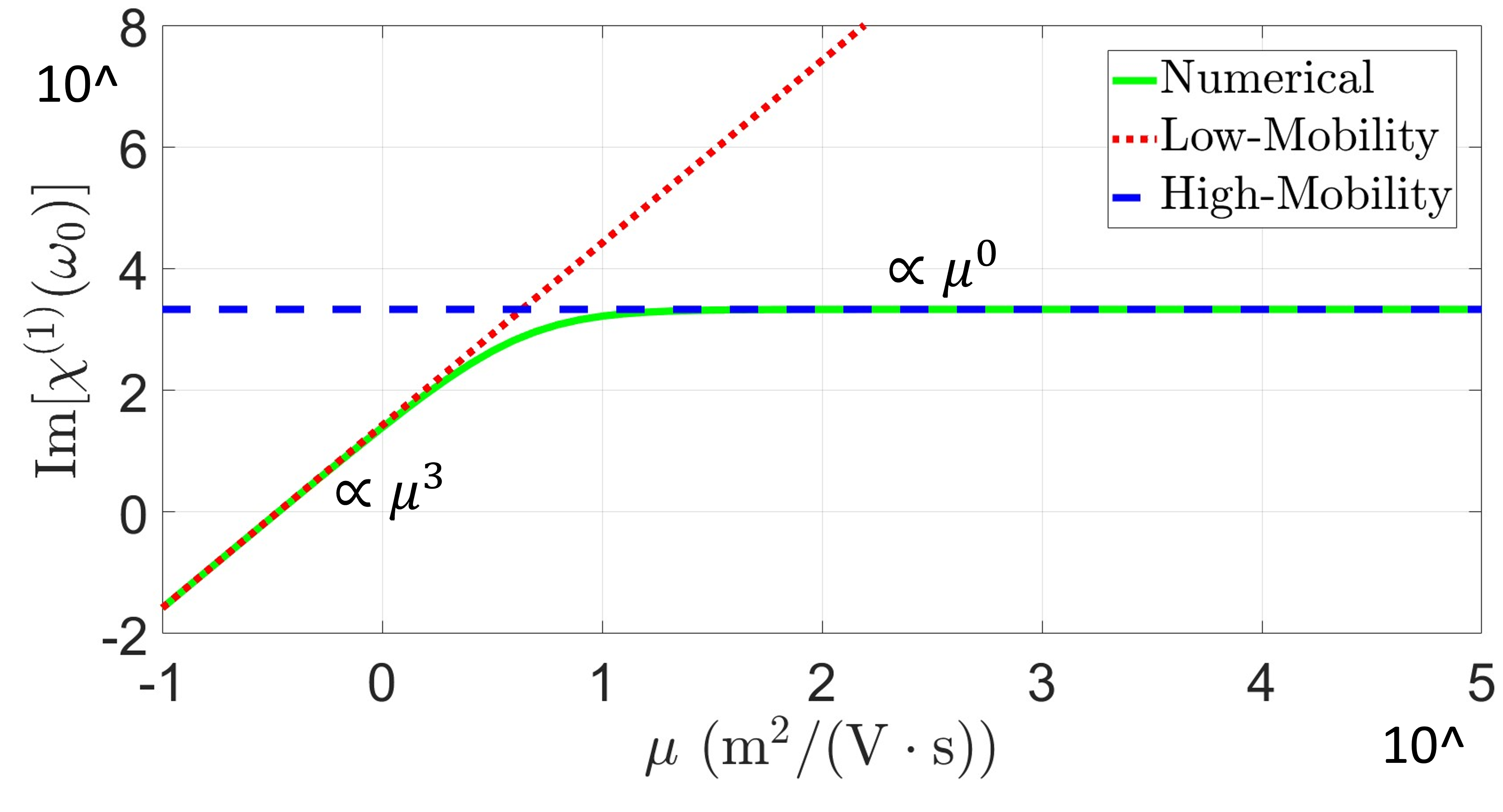}
		\caption{}
		\label{fig:chi1imagresults}
	\end{subfigure}
	\caption{Results for the real (a) and imaginary (b) parts of $\chi^{(1)}(\omega_0)$ as functions of the carrier mobility $\mu$, including numerical (solid, green), low-mobility analytical (dotted, red), and high-mobility analytical (dashed, blue) calculations. We use a phonon angular frequency of $\omega_0 = 2\pi \times 10^9 \textrm{ s}^{-1}$, speed of sound $v_s = 4 \times 10^3$ m/s, a carrier density $n = 2 \times 10^{15} \textrm{ m}^{-2}$, a carrier effective mass $m = 0.067 m_0$, and a 2DEG thickness $t_\mathrm{2DEG} = 2 \times 10^{-8} \textrm{ m}$.}
	\label{fig:chi1results}
\end{figure*}

Figure~\ref{fig:chi1results} depicts the numerically-calculated results for the real and imaginary parts of $\chi^{(1)}$ (given a phonon angular frequency of $\omega_0 = 2\pi \times 10^9 \textrm{ s}^{-1}$, speed of sound $v_s = 4 \times 10^3$ m/s, a carrier density $n = 2 \times 10^{15} \textrm{ m}^{-2}$, a carrier effective mass $m = 0.067 m_0$ relative to the bare electron mass $m_0$, and a 2DEG thickness $t_\mathrm{2DEG} = 2 \times 10^{-8} \textrm{ m}$), along with the analytical results in the high-mobility and low-mobility limits.
As desired, the numerical and analytical results match in the low-mobility and high-mobility regimes, with the real (imaginary) part varying as $\mu^2$ ($\mu^3$) in the low-mobility regime, and with both parts plateauing in the high-mobility regime. Note that for all mobility values, $\textrm{Re}[\chi^{(1)}]$ vastly exceeds $\textrm{Im}[\chi^{(1)}]$. The dielectric constant can thus be approximated as fully real rather than complex, which features important implications when calculating the wave-mixing dynamics.

\subsection{Second-Order Susceptibility}
\label{sec: Second-Order Susceptibility}

We now derive the second-order susceptibility, using numerical means to calculate the susceptibility for all mobility values and analytical means to specifically derive the low-mobility limit. We start with sum-frequency generation, governed by $\chi^{(2)}(\omega_1,\omega_2)$. In the low-mobility limit, we can reduce $\textrm{Re}[f_{k_x}^{(2)}]$ in Eq.~\eqref{eq: chi(2) denominator terms} to the following using the approximation $|\Delta \omega| \gg \omega_{k_x + q_1 + q_2,k_x},\omega_1,\omega_2$:
\begin{equation}
\textrm{Re}[f^{(2)}_\mathrm{SF}] \approx -\frac{4}{\gamma'^2}.
\end{equation}
The sum-frequency-generation susceptibility thus varies with mobility as $\mu^2$ in this regime, following the general low-mobility rule laid out in Eq.~\eqref{eq: low-mobility real susceptibility generic}. We can analytically solve for the overall susceptibility in the low-mobility regime by multiplying by the available phase-space area in Eq.~\eqref{eq: available phase space area} and the coefficients in Eq.~\eqref{eq: chi(2) integral}. It is important to consider, however, that for two of the terms in Eq.~\eqref{eq: chi(2) denominator terms}, the width of the phase-space area is $q_1 = \omega_1/v_s$ (see Figs.~\ref{fig:chi2(omega1,omega2)}(a) and (d)), while for the other two terms, the width is $q_2 = \omega_2/v_s$ (see Figs.~\ref{fig:chi2(omega1,omega2)}(b) and (c)). Consequently, for the available phase-space area expression in Eq.~\eqref{eq: available phase space area}, we apply the replacement $\omega_0 \rightarrow (\omega_1 + \omega_2)/2$, yielding the following overall second-order susceptibility in the low-mobility limit:
\begin{align}
\begin{split}
\textrm{Re}[\chi^{(2)}(\omega_1,\omega_2)] &\approx \bigg(\frac{q_e^3 v_s^3}{2 \pi^2 \hbar^2 \epsilon_0 t_\mathrm{2DEG} \omega_1 \omega_2 (\omega_1 + \omega_2)}\bigg) \\
&\quad \times \bigg(-\frac{4}{\gamma'^2}\bigg) \bigg(\frac{2 (\omega_1 + \omega_2) k_F}{2 v_s}\bigg) \\
&= -\frac{2 q_e^3 v_s^2 k_F}{\pi^2 \hbar^2 \epsilon_0 t_\mathrm{2DEG} \omega_1 \omega_2 \gamma'^2} \\
&= -\frac{8 q_e v_s^2 m^2 k_F \mu^2}{\pi^2 \hbar^2 \epsilon_0 t_\mathrm{2DEG} \omega_1 \omega_2}.
\end{split}
\end{align}
Note the linear variation in $k_F$, implying that the susceptibility scales with the carrier density as $n^{1/2}$. This is because the number of available electrons (corresponding to the available phase-space area) varies linearly with $k_F$, while the interaction probability per electron is independent of $k_F$.

\begin{figure}[!tb]
	\centering
	\includegraphics[width=\columnwidth]{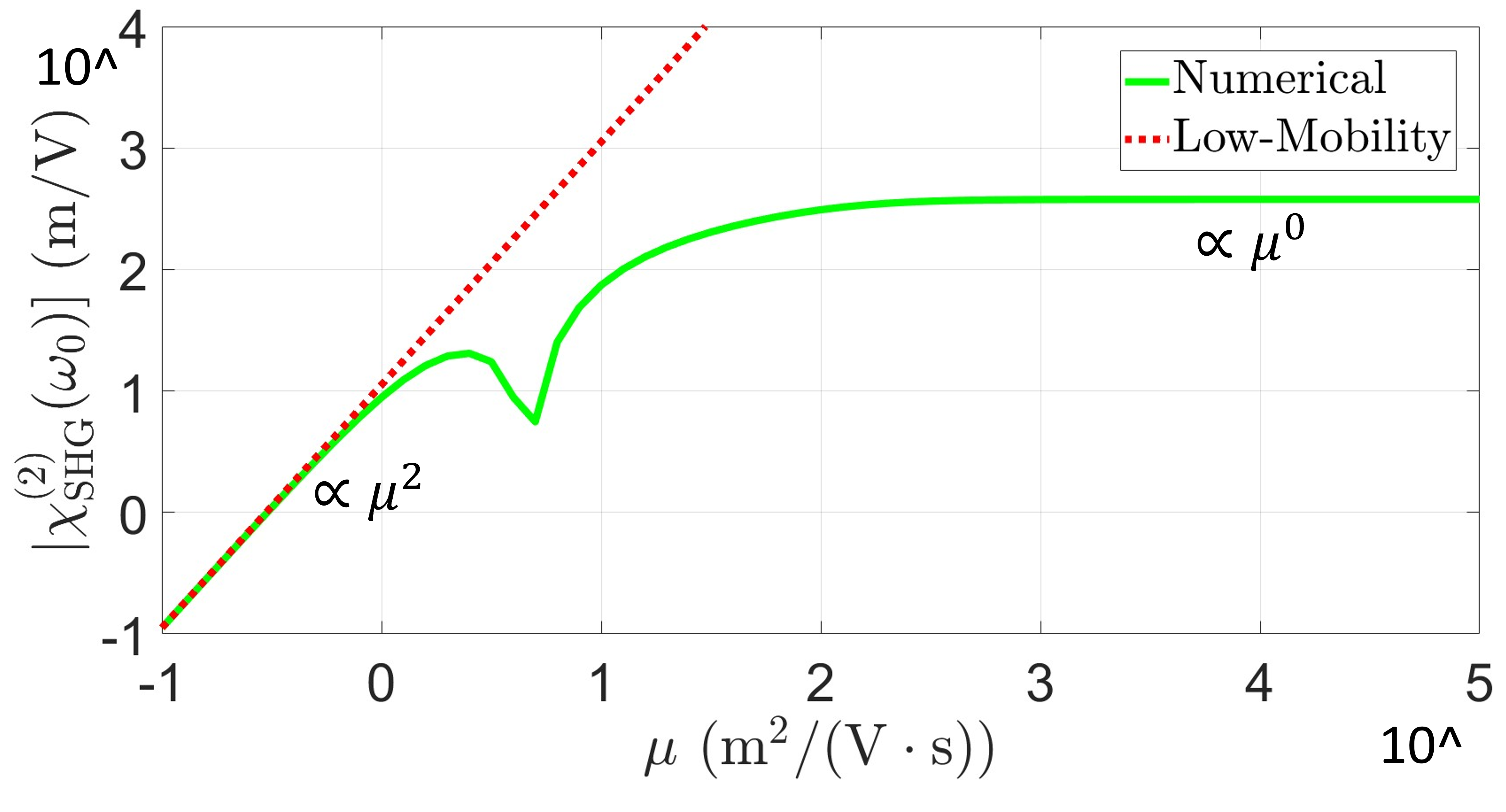}
	\caption{Results for the amplitude of the real part of $\chi^{(2)}(\omega_0,\omega_0)$ as a function of the carrier mobility $\mu$, including numerical (solid, green) and low-mobility analytical (dotted, red) calculations. We use a phonon angular frequency of $\omega_0 = 2\pi \times 10^9 \textrm{ s}^{-1}$, speed of sound $v_s = 4 \times 10^3$ m/s, a carrier density $n = 2 \times 10^{15} \textrm{ m}^{-2}$, a carrier effective mass $m = 0.067 m_0$, and a 2DEG thickness $t_\mathrm{2DEG} = 2 \times 10^{-8} \textrm{ m}$.}
	\label{fig:chi2results}
\end{figure}

Figure~\ref{fig:chi2results} depicts the numerically-calculated results for the amplitude of the real part of $\chi^{(2)}(\omega_1,\omega_2)$ in the degenerate limit $\omega_1 = \omega_0 = \omega_2$ representing the second-harmonic-generation process (we label this quantity $\chi^{(2)}_\mathrm{SHG}(\omega_0)$), along with the analytical result in the low-mobility limit, given the same parameters as in the $\chi^{(1)}$ calculation.
As desired, the numerical and analytical results match in the low-mobility regime, with the susceptibility varying as $\mu^2$. It is interesting to note that the susceptibility plateaus in the high-mobility regime. As discussed in Sec.~\ref{sec: High-Mobility Regime}, the near-resonance electronic contributions to the real part of $\chi^{(2)}$ cancel out in the high-mobility regime, leaving behind electronic contributions that are somewhat off-resonance (but not far-off-resonance, since $\textrm{Re}[\chi^{(2)}]$ converges to zero for large detuning values). Along with the very low value of $\gamma$, this causes the effect of the decay rate $\gamma$ on the electron interaction probabilities to be negligible, leading $\textrm{Re}[\chi^{(2)}]$ to be constant in mobility in the high-mobility limit.

Next, we turn to parametric amplification, focusing specifically on the degenerate case ($\omega_1 = \omega_0 = \omega_2$). Examining $\textrm{Re}[f_{k_x}^{(2)}]$ from Eq.~\eqref{eq: chi(2) parametric denominator terms} in the low-mobility limit, we note that the amplitude of each of the terms can be approximated as $1/\gamma'^2$ to first order, as with second-harmonic generation. However, the key difference here is that the leading terms cancel out for both the resonant case and the counter-resonant case, due to destructive interference of probability waves in the degenerate traveling-wave parametric amplifier. Consequently, we shift to the secondary term, which equals $2(\Delta \omega)^2/\gamma'^4$ (where $\Delta \omega = \omega_{k_x + q_0,k_x} - \omega_0$). As a result, the parametric amplification susceptibility varies with mobility as $\mu^4$ instead of $\mu^2$:
\begin{equation}
\textrm{Re}[f^{(2)}_\mathrm{PA}(\Delta \omega)] \approx \frac{8 (\Delta \omega)^2}{\gamma'^4}.
\end{equation}
To analytically solve for the low-mobility susceptibility, we need to integrate $(\Delta \omega)^2$ over the available phase-space area. Since the electron-phonon transitions in the low-mobility limit are concentrated far-off-resonance, we approximate $\Delta \omega \approx \hbar q k_x/m$, enabling us to apply the generic higher-order integral shown in Eq.~\eqref{eq: generic higher-order integral low-mobility} as follows:
\begin{equation}
\int_{-q/2}^{k_F} dk_x k_{y,\mathrm{span}}(k_x) (\Delta \omega)^2 \approx \frac{4 \hbar^2 \omega_0^3 k_F^3}{3 m^2 v_s^3}.
\end{equation}
Multiplying this $f^{(2)}_\mathrm{PA}$ and the coefficients in Eq.~\eqref{eq: chi(2) integral}, we find the following degenerate parametric amplification susceptibility in the low-mobility limit:
\begin{align}
\begin{split}
\textrm{Re}[\chi^{(2)}(2\omega_0,-\omega_0)] &\approx \bigg(\frac{q_e^3 v_s^3}{2 \pi^2 \hbar^2 \epsilon_0 t_\mathrm{2DEG} 2\omega_0^3}\bigg) \\
&\quad \times \bigg(\frac{8}{\gamma'^4}\bigg) \bigg(\frac{4 \hbar^2 \omega_0^3 k_F^3}{3 m^2 v_s^3}\bigg) \\
&= \frac{8 q_e^3 k_F^3}{3 \pi^2 \epsilon_0 t_\mathrm{2DEG} m^2 \gamma'^4} \\
&= \frac{128 m^2 k_F^3 \mu^4}{3 \pi^2 \epsilon_0 q_e t_\mathrm{2DEG}}.
\end{split}
\end{align}
Note that this is independent of the frequency $\omega_0$. This is because the number of available electrons varies linearly with $\omega_0$, while the interaction probability for each electron varies inversely with $\omega_0$, and the two variations cancel out. Furthermore, since the interaction probability per electron now varies as $k_F^2$, the susceptibility scales as $k_F^3$, implying a variation with carrier density as $n^{3/2}$.

\begin{figure}[!tb]
	\centering
	\includegraphics[width=\columnwidth]{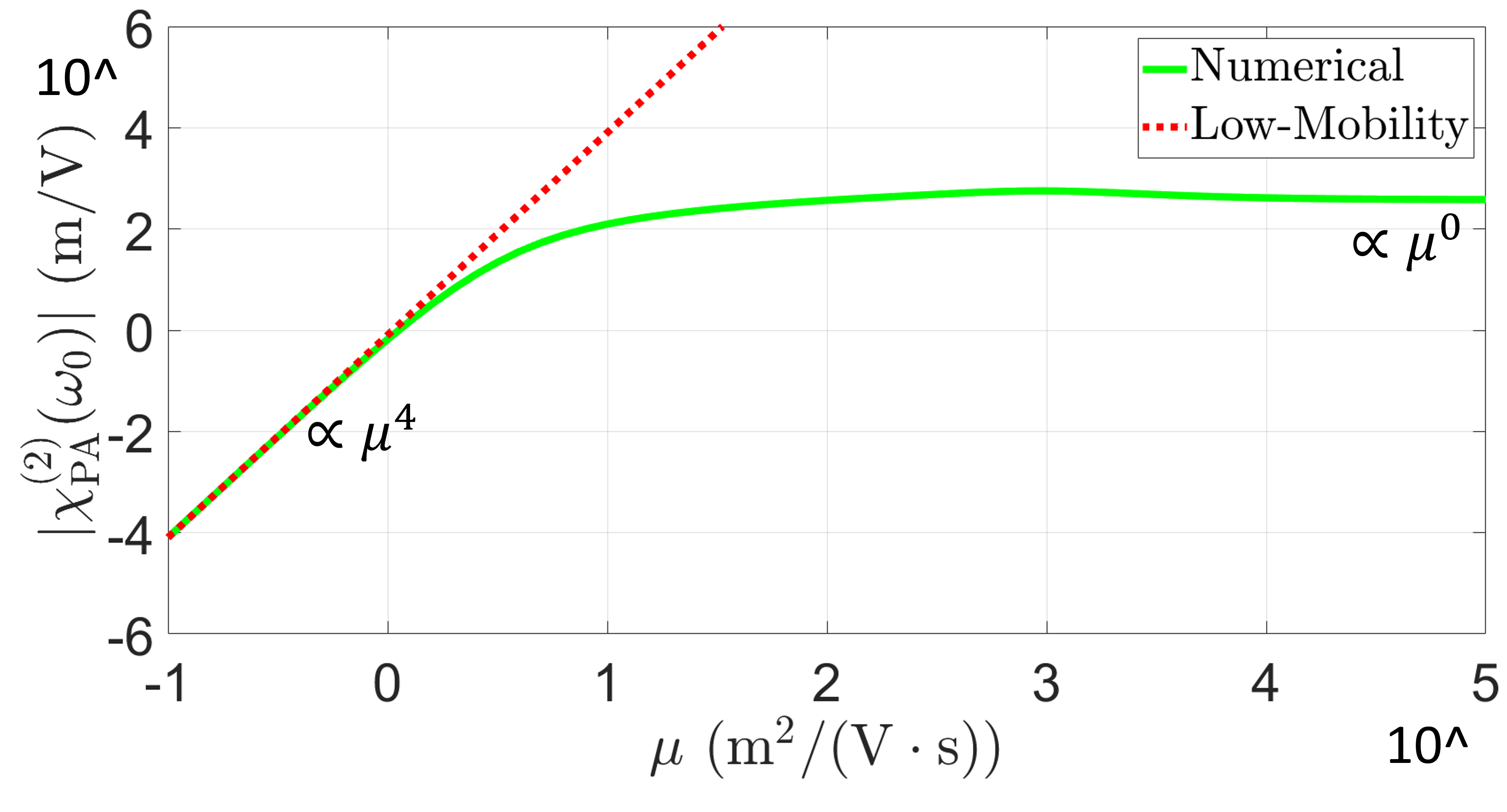}
	\caption{Results for the amplitude of the real part of $\chi^{(2)}(2\omega_0,-\omega_0)$ as a function of the carrier mobility $\mu$, including numerical (solid, green) and low-mobility analytical (dotted, red) calculations. We use a phonon angular frequency of $\omega_0 = 2\pi \times 10^9 \textrm{ s}^{-1}$, speed of sound $v_s = 4 \times 10^3$ m/s, a carrier density $n = 2 \times 10^{15} \textrm{ m}^{-2}$, a carrier effective mass $m = 0.067 m_0$, and a 2DEG thickness $t_\mathrm{2DEG} = 2 \times 10^{-8} \textrm{ m}$.}
	\label{fig:chi2parametricresults}
\end{figure}

Figure~\ref{fig:chi2parametricresults} depicts the numerically-calculated results for the amplitude of the real part of $\chi^{(2)}(2\omega_0,-\omega_0)$ (we label this quantity $\chi^{(2)}_\mathrm{PA}(\omega_0)$). 
As desired, the numerical and analytical results match in the low-mobility regime, with the susceptibility varying as $\mu^4$. The susceptibility levels out in the high-mobility regime for the same reason as in the second-harmonic-generation case.

\subsection{Third-Order Kerr Susceptibility}
\label{sec: Third-Order Kerr Susceptibility}

We conclude our susceptibility calculations by deriving the third-order Kerr susceptibility $\chi^{(3)}(\omega_0,\omega_0,-\omega_0)$ governing the Kerr shift and the degenerate four-wave-mixing process. In the low-mobility limit, we simplify $\textrm{Re}[f_{k_x}^{(3)}]$ in Eq.~\eqref{eq: chi Kerr denominator terms} to the following, since $|\Delta \omega| \gg \omega_{k_x + 2q,k_x},\omega$:
\begin{equation}
\textrm{Re}[f^{(3)}(\Delta \omega)] \approx (-4 + 2 + 2 - 4) \frac{\gamma'^2 \Delta \omega}{\gamma'^6} = -\frac{4 \Delta \omega}{\gamma'^4},
\end{equation}
where $\Delta \omega = \omega_{k_x + q,k_x} - \omega_0$. Following the general rule for the low-mobility regime as outlined in Eq.~\eqref{eq: low-mobility real susceptibility generic}, the Kerr susceptibility varies with mobility as $\mu^4$ in this regime. We analytically solve for the overall susceptibility in the low-mobility limit by integrating over the phase-space area of initial electronic states using the procedure in Eq.~\eqref{eq: generic integral low-mobility} and multiplying by the coefficients in Eq.~\eqref{eq: chi(3) integral}:
\begin{align}
\begin{split}
&\textrm{Re}[\chi^{(3)}(\omega_0,\omega_0,-\omega_0)] \\
&\approx \bigg(\frac{q_e^4 v_s^4}{\pi^2 \hbar^3 \epsilon_0 t_\mathrm{2DEG} \omega_0^4}\bigg) \bigg(-\frac{4}{\gamma'^4}\bigg) \bigg(\frac{\pi \hbar \omega_0^2 k_F^2}{2 m v_s^2}\bigg) \\
&= -\frac{2 q_e^4 v_s^2 k_F^2}{\pi \hbar^2 \epsilon_0 m t_\mathrm{2DEG} \omega_0^2 \gamma'^4} \\
&= -\frac{32 v_s^2 m^3 k_F^2 \mu^4}{\pi \hbar^2 \epsilon_0 m t_\mathrm{2DEG} \omega_0^2}.
\end{split}
\end{align}
The quadratic variation in $k_F$ implies a linear scaling with the carrier density $n$. This is because the number of available electrons and the interaction probability per electron each scales as $n^{1/2}$.

We now turn to the high-mobility limit. Here, we have to account for the fact that the transition frequency $\omega_{k_x + 2q,k_x + q}$ for the electron upon absorbing a second phonon of wavevector $q$ exceeds that upon absorbing the first phonon $\omega_{k_x + q,k_x}$ by a constant ($k_x$-independent) offset frequency, which we label $\omega'$:
\begin{align}
\begin{split}
\omega' &= \omega_{k_x + 2q,k_x + q} - \omega_{k_x + q,k_x} \\
&= \frac{\hbar q (2(k_x + q) + q)}{2m} - \frac{\hbar q (2k_x + q)}{2m} \\
&= \frac{\hbar q^2}{m}.
\end{split}
\end{align}
In terms of this offset, $\textrm{Re}[f_{k_x}^{(3)}]$ in Eq.~\eqref{eq: chi Kerr denominator terms} takes the following form, considering only the resonant (first and second) terms:
\begin{align}
\begin{split}
&\textrm{Re}[f^{(3)}(\Delta \omega)] \\
&\approx \sum_{\pm} \frac{1}{(\Delta \omega \mp i\gamma') (2 \Delta \omega + \omega' - i\gamma') (\Delta \omega - i\gamma')} \\
&= \frac{(2 \Delta \omega + \omega')(\Delta \omega)^2 - \gamma'^2(\Delta \omega \pm 3 \Delta \omega \pm \omega')}{((\Delta \omega)^2 + \gamma'^2)^2 ((2 \Delta \omega + \omega')^2 + \gamma'^2)},
\end{split}
\end{align}
where the top and bottom indices in the $\pm$ and $\mp$ operations represent the first and second terms of Eq.~\eqref{eq: chi Kerr denominator terms}, respectively. We integrate this over the near-resonance band (using the procedure in Eq.~\eqref{eq: chi(Delta omega) phase space integral}) and multiply by the coefficients in Eq.~\eqref{eq: chi(3) integral}, leading to the following result for the overall Kerr susceptibility in the high-mobility limit:
\begin{widetext}
\begin{align}
\begin{split}
\textrm{Re}[\chi^{(3)}(\omega_0,\omega_0,-\omega_0)] &\approx \bigg(\frac{q_e^4 v_s^4}{\pi^2 \hbar^3 \epsilon_0 t_\mathrm{2DEG} \omega_0^4}\bigg) \frac{2 m^2 v_s}{\hbar^2 k_F} \int_{-\infty}^{\infty} d(\Delta \omega) (\omega_0 + \Delta \omega) \frac{(2 \Delta \omega + \omega')(\Delta \omega)^2 - \gamma'^2(\Delta \omega \pm 3 \Delta \omega \pm \omega')}{((\Delta \omega)^2 + \gamma'^2)^2 ((2 \Delta \omega + \omega')^2 + \gamma'^2)} \\
&= \beta \frac{\pi (3 \gamma'^2 + \omega' \omega_0)}{\gamma' (9 \gamma'^2 + \omega'^2)} \\
&= \frac{2 \pi \beta m \mu}{q_e} \bigg(\frac{3 q_e^2 + 4 \omega' \omega_0 m^2 \mu^2}{9 q_e^2 + 4 \omega'^2 m^2 \mu^2}\bigg),
\end{split}
\end{align}
\end{widetext}
where $\beta$ is defined as follows:
\begin{equation}
\beta = \frac{2 q_e^4 v_s^5 m^2}{\pi^2 \hbar^5 \epsilon_0 \omega_0^5 k_F t_\mathrm{2DEG}}.
\end{equation}
Note that the Kerr susceptibility varies linearly with $\mu$ in the high-mobility limit. However, the slope of this variation is 3 times greater in the ultra-high-mobility limit than in the medium-high-mobility limit. This is because the linear variation originates from different sources for the two cases. In the medium-high-mobility case, the electronic spectrum effectively becomes harmonic, thus making $\textrm{Re}[f^{(3)}(\Delta \omega)]$ odd in $\Delta \omega$ and yielding $\expect{\textrm{Re}[f^{(3)}(\Delta \omega)} \propto 1/\gamma'^3$. As such, integrating the product of this function and $\Delta \omega$ over the near-resonance band results in $\textrm{Re}[\chi^{(3)}(\omega_0,\omega_0,-\omega_0)] \propto 1/\gamma' \propto \mu$. On the other hand, in the ultra-high-mobility limit, the size of the electronic anharmonicity in phase space exceeds the width of the near-resonance band, making $\textrm{Re}[f^{(3)}(\Delta \omega)]$ even in $\Delta \omega$ and yielding $\expect{\textrm{Re}[f^{(3)}(\Delta \omega)} \propto 1/\gamma'^2$. Integrating this over the near-resonance band also results in $\textrm{Re}[\chi^{(3)}(\omega_0,\omega_0,-\omega_0)] \propto 1/\gamma' \propto \mu$, but with a different slope than the medium-high-mobility case.

\begin{figure}[!tb]
	\centering
	\includegraphics[width=\columnwidth]{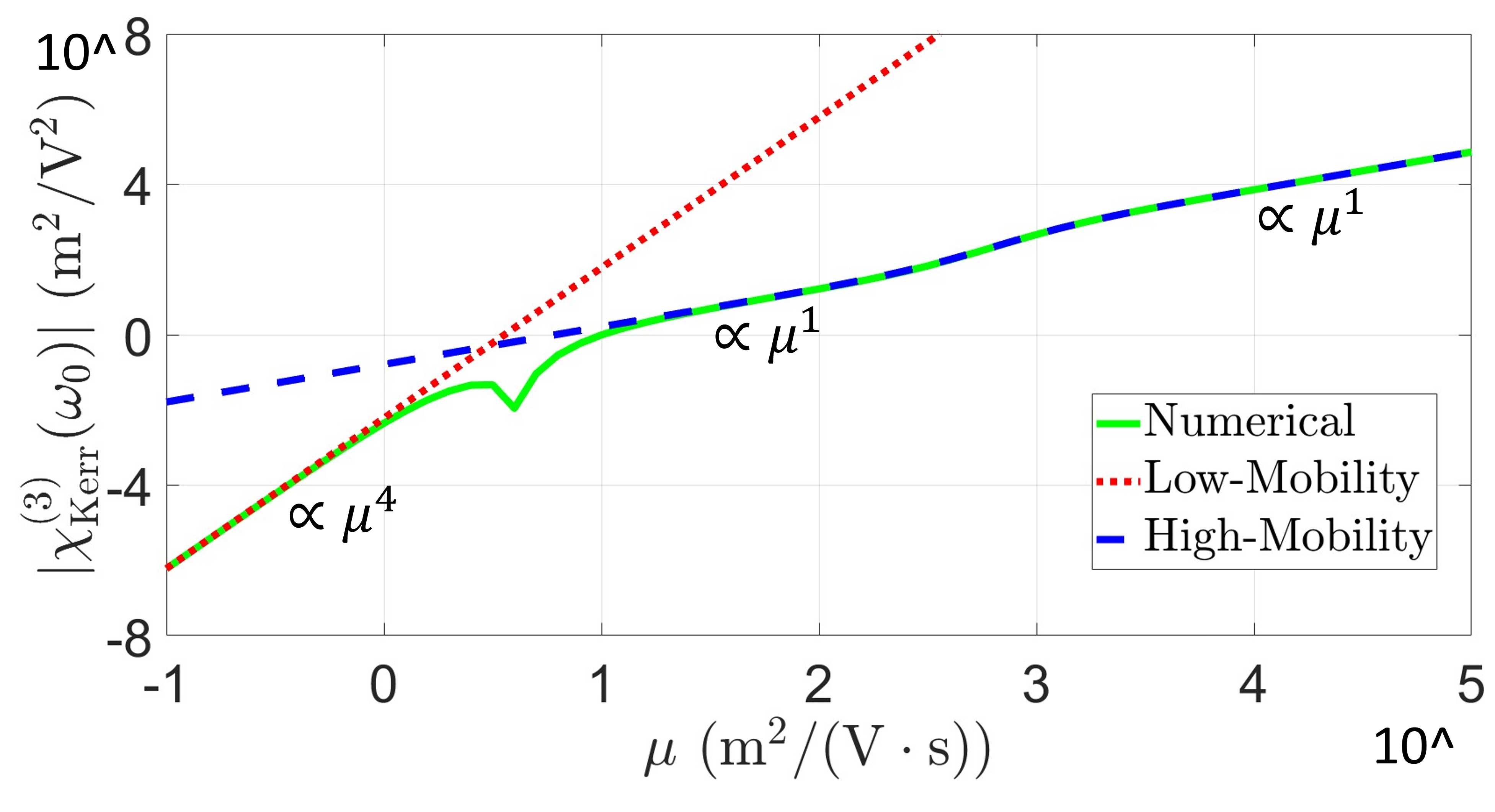}
	\caption{Results for the amplitude of the real part of $\chi^{(3)}(\omega_0,\omega_0,-\omega_0)$ as a function of the carrier mobility $\mu$, including numerical (solid, green), low-mobility analytical (dotted, red), and high-mobility analytical (dashed, blue) calculations. We use a phonon angular frequency of $\omega_0 = 2\pi \times 10^9 \textrm{ s}^{-1}$, speed of sound $v_s = 4 \times 10^3$ m/s, a carrier density $n = 2 \times 10^{15} \textrm{ m}^{-2}$, a carrier effective mass $m = 0.067 m_0$, and a 2DEG thickness $t_\mathrm{2DEG} = 2 \times 10^{-8} \textrm{ m}$.}
	\label{fig:chi3results}
\end{figure}

Figure~\ref{fig:chi3results} depicts the numerically-calculated results for the amplitude of the real part of $\chi^{(3)}(\omega_0,\omega_0,-\omega_0)$, which we label $\chi^{(3)}_\mathrm{Kerr}(\omega_0)$.
We use the same parameters as in the $\chi^{(1)}$ and $\chi^{(2)}$ calculations. As desired, the numerical and analytical results match in the low-mobility and high-mobility regimes, varying as $\mu^4$ in the low-mobility regime and as $\mu^1$ in the high-mobility regime. The aforementioned change in the proportionality coefficient with $\mu$ between the medium-high-mobility and ultra-high-mobility regimes is apparent in the intercept shift in the log-log plot around $10^3 \textrm{ m}^2/(\textrm{V} \cdot \textrm{s})$.

\section{Calculation of Mixing Dynamics}
\label{sec: Calculation of Mixing Dynamics}

The $N^\textrm{th}$-order susceptibility manifests itself in the $(N+1)$-wave-mixing process. For a quantitative understanding, we can substitute the expression for $\expect{d_\mathrm{eff}(q_p)}$ in Eq.~\eqref{eq: dipole moment expectation value} into the interaction Hamiltonian in Eq.~\eqref{eq: interaction Hamiltonian multiple modes}, yielding an expansion in product of field amplitudes:
\begin{align}
\begin{split}
\label{eq: interaction Hamiltonian with chi(N)}
H_\mathrm{int} &= -\epsilon_0 V_\mathrm{2DEG} \sum_n \sum_p \sum_{\substack{p_1,...,p_N \\ \omega_{p_1} + ... + \omega_{p_N} = \omega}} \\
&\quad\quad \chi^{(N)}(\omega_{p_1},...,\omega_{p_n}) E^*(\omega) E(\omega_{p_1})...E(\omega_{p_N}).
\end{split}
\end{align}
Intuitively, this corresponds to the fact that the different field modes are coupled to one another via interactions with the 2DEG electrons. At the elemental level, the field modes $p_1$ through $p_N$ simultaneously polarize the electrons, which in turn interact with the field mode $p$ through field-dipole coupling. It is apparent that the real part of the susceptibility corresponds to the conversion efficiency between modes (or to an energy shift in the acoustic states, for the case of mode self-coupling), while the imaginary part yields a broadening of the acoustic states caused by loss due to phonon absorption by the electrons.

The Hamiltonian corresponding to the interaction of a phonon from each of the modes $p_1,...,p_N$ and a phonon of frequency $\omega = \omega_{p_1} + ... + \omega_{p_N}$ can be expressed in operator form as follows:
\begin{equation}
H = \hbar \Big(g a^{\dag} a_1...a_N + g^* a_1^{\dag}...a_N^{\dag} a\Big),
\end{equation}
where, as evidenced by Eq.~\eqref{eq: interaction Hamiltonian with chi(N)}, the $(N+1)$-phonon interaction rate $g$ is given by the following:
\begin{align}
\begin{split} \label{eq: (N+1)-phonon interaction rate}
g &= -\frac{\epsilon_0 V_\mathrm{2DEG}}{\hbar} \chi^{(N)}(\omega_{p_1},...,\omega_{p_n}) \\
&\quad \times E_\mathrm{zpf}^*(\omega) E_\mathrm{zpf}(\omega_{p_1})...E_\mathrm{zpf}(\omega_{p_N}),
\end{split}
\end{align}
where $E_\mathrm{zpf}$ represents the zero-point electric field experienced by the 2DEG electrons. Intuitively, since $\chi^{(N)}$ represents the total induced electron polarization per unit amplitude of the $N$ polarizing fields $p_1,...,p_N$, multiplying it by $V_\mathrm{2DEG} E_\mathrm{zpf}(\omega_{p_1})...E_\mathrm{zpf}(\omega_{p_N})$ yields the product of the probability that an electron becomes polarized by absorbing $N$ phonons and the resulting dipole moment of the polarized electron. This polarized electron then interacts with the $(N+1)$-st field (represented here by the frequency $\omega$) through the dipole interaction. The net process is electron-mediated coupling of $N+1$ phonons from the respective fields, with the interaction rate given by Eq.~\eqref{eq: (N+1)-phonon interaction rate}.

In general, we can determine the time-evolution of the output electric field $\omega$ in terms of the amplitudes of the input electric fields using the Heisenberg equation of motion as follows:
\begin{align}
\begin{split}
\dot{E}(\omega) &= E_\mathrm{zpf}(\omega) \dot{a} \\
&= -\frac{i}{\hbar} E_\mathrm{zpf}(\omega) [a,H] \\
&= -i E_\mathrm{zpf}(\omega) g a_1...a_N \\
&= i \frac{\epsilon_0 V_\mathrm{2DEG}}{\hbar} \chi^{(N)}(\omega_{p_1},...,\omega_{p_n}) \Big|E_\mathrm{zpf}(\omega)\Big|^2 \\
&\quad \times E(\omega_1)...E(\omega_N).
\end{split}
\end{align}
The zero-point electric field $E_\mathrm{zpf}$ is determined by substituting $A_\mathrm{ph} = \sqrt{v_s/(2V_\omega)}$ (corresponding to a vacuum fluctuation spanning the mode volume) into Eq.~\eqref{eq: electric field diagonal dielectric exponential envelope}, yielding:
\begin{equation} \label{eq: E_zpf}
E_\mathrm{zpf}(\omega) = -\frac{C}{\epsilon(\omega)} \sqrt{\frac{\hbar \omega}{2 V_\omega}},
\end{equation}
where $V_\omega$ is the overall mode volume for the acoustic wave. From this, we can determine the spatial evolution of the output field amplitude $A_\mathrm{ph}(\omega)$ in terms of the input amplitudes $A_\mathrm{ph}(\omega_1),...A_\mathrm{ph}(\omega_N)$:
\begin{align}
\begin{split} \label{eq: spatial evolution of output field}
&\partial_x A_\mathrm{ph}(\omega) = \\
&\quad i \frac{\epsilon_0}{\hbar} \frac{\chi^{(N)}(\omega_{p_1},...,\omega_{p_n}) \sqrt{\omega \omega_1...\omega_N}}{\epsilon^*(\omega) \epsilon(\omega_1)...\epsilon(\omega_N)} \bigg(-C \sqrt{\frac{\hbar}{v_s}}\bigg)^{N+1} \\
&\quad\quad \times \frac{V_\mathrm{2DEG}}{2 V_\omega} A_\mathrm{ph}(\omega_1)...A_\mathrm{ph}(\omega_N).
\end{split}
\end{align}
Note that the conversion efficiency is proportional to the ratio between the 2DEG volume $V_\mathrm{2DEG}$ and the overall mode volume $V_\omega$ (which simply reduces to the ratio between the 2DEG thickness $t_\mathrm{2DEG}$ and the mode depth $L_{z,\omega}$ if the 2DEG covers the piezoelectric material's surface). This is because the fraction of the total mode energy stored in the 2DEG field is proportional to the ratio of the two volumes. 

\begin{figure}[!tb]
	\centering
	\includegraphics[width=\columnwidth]{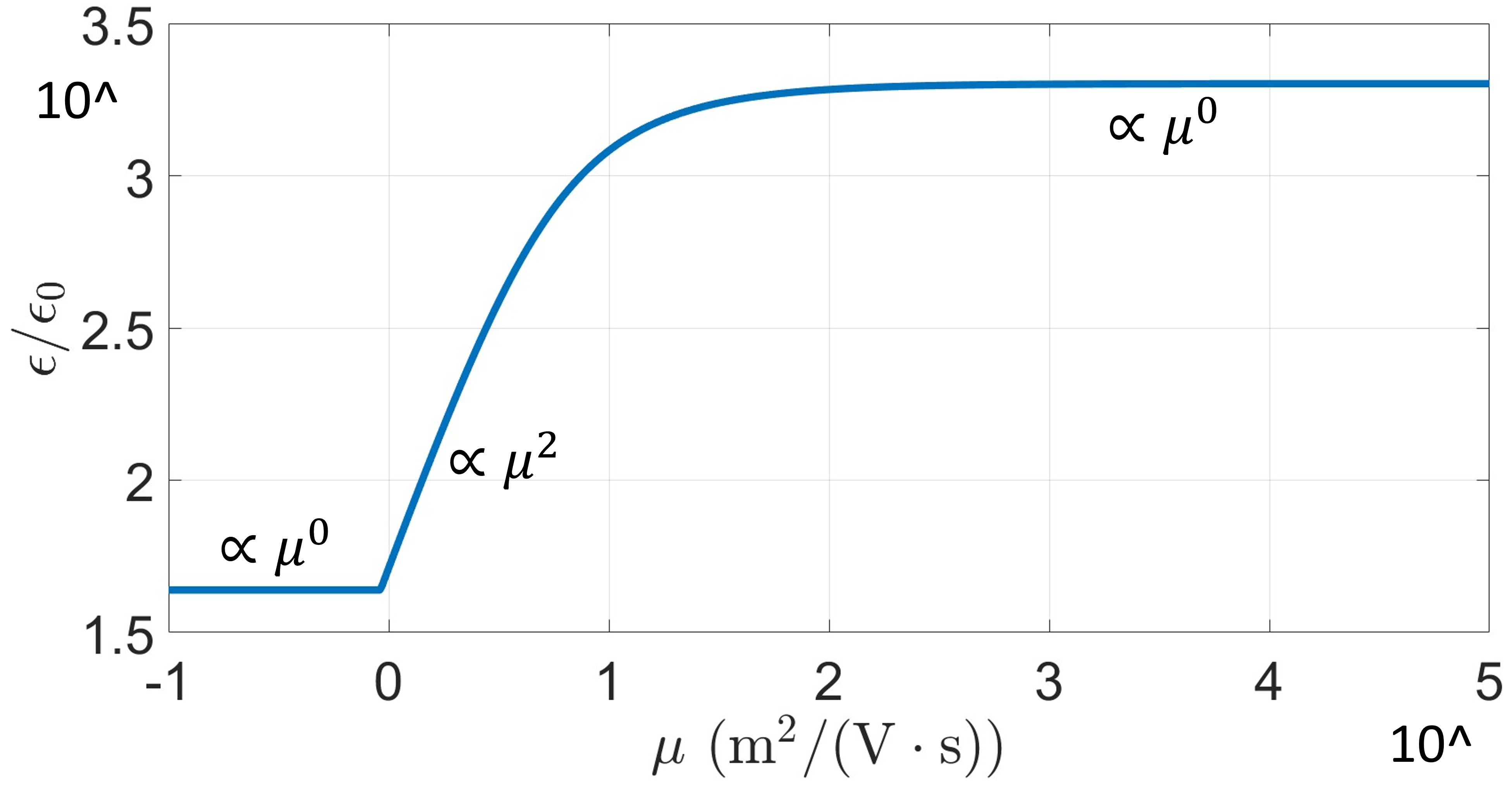}
	\caption{Dielectric constant in units of $\epsilon_0$ as a function of the carrier mobility $\mu$, given a piezoelectric material dielectric constant $\epsilon_p = 43.6 \epsilon_0$, a phonon angular frequency of $\omega_0 = 2\pi \times 10^9 \textrm{ s}^{-1}$, speed of sound $v_s = 4 \times 10^3$ m/s, a carrier density $n = 2 \times 10^{15} \textrm{ m}^{-2}$, a carrier effective mass $m = 0.067 m_0$, and a 2DEG thickness $t_\mathrm{2DEG} = 2 \times 10^{-8} \textrm{ m}$.}
	\label{fig:dielectricconstant}
\end{figure}

It is also worth noting that the conversion efficiency varies inversely with $\epsilon^{N+1}$, where $\epsilon$ is the effective 2DEG dielectric constant, which is in turn proportional to $\textrm{Re}[\chi^{(1)}]$. Naively, we would assume that $\epsilon = \epsilon_0 \textrm{Re}[\chi^{(1)}]$. However, there are 2 critical caveats. The first is that due to the thin-film nature of the 2DEG (i.e., the fact that the 2DEG thickness is much smaller than the field wavelength), the dielectric screening is strongly suppressed. As explained in Appendix~\ref{sec: Relating 2DEG Polarization to Induced Electric Field}, this attenuates the dielectric constant by a factor $a$, which is a function of the ratio between the 2DEG thickness $t_\mathrm{2DEG}$ and the field's half-wavelength $l$:
\begin{equation}
a = \frac{t_\mathrm{2DEG}}{2 \pi l} \bigg(2 \log{\bigg(\frac{l}{t_\mathrm{2DEG}}\bigg) + 3}\bigg).
\end{equation}
The second caveat is that when $\textrm{Re}[\chi^{(1)}]$ is low enough that the corresponding dielectric constant is below the dielectric constant of the piezoelectric material $\epsilon_p$, the latter provides the effective dielectric constant for the 2DEG, since the fact that the 2DEG is much thinner than the piezoelectric material ensures that the fringe induced field from the piezoelectric material fully penetrates the 2DEG. As such, the dielectric constant can be approximated as follows:
\begin{equation}
\epsilon(\omega_0) \approx
\begin{cases}
a \epsilon_0 \textrm{Re}[\chi^{(1)}(\omega_0)], & a \epsilon_0 \textrm{Re}[\chi^{(1)}] > \epsilon_p, \\
\epsilon_p, & a \epsilon_0 \textrm{Re}[\chi^{(1)}(\omega_0)] < \epsilon_p
\end{cases}.
\end{equation}
The 2DEG dielectric constant is plotted in Fig.~\ref{fig:dielectricconstant} given $\epsilon_p = 43.6 \epsilon_0$, with all the other parameters the same as in the susceptibility plots.
Given the dielectric constant and the susceptibilities, we solve for the rates of 3 processes: linear absorption (corresponding to $\textrm{Im}[\chi^{(1)}]$), as well as second-harmonic generation and parametric amplification (corresponding to $\textrm{Re}[\chi^{(2)}]$). We also derive the figure-of-merit for the Kerr nonlinearity by calculating the ratio between the phonon Kerr shift (corresponding to $\textrm{Re}[\chi^{(3)}]$) and the phonon spectral broadening (corresponding to $\textrm{Im}[\chi^{(1)}]$, given a high-Q acoustic cavity, which has recently been achieved \cite{shao2019phononicband}).

\subsection{Linear Absorption}
\label{sec: Linear Absorption}

\begin{figure}[!tb]
	\centering
	\includegraphics[width=\columnwidth]{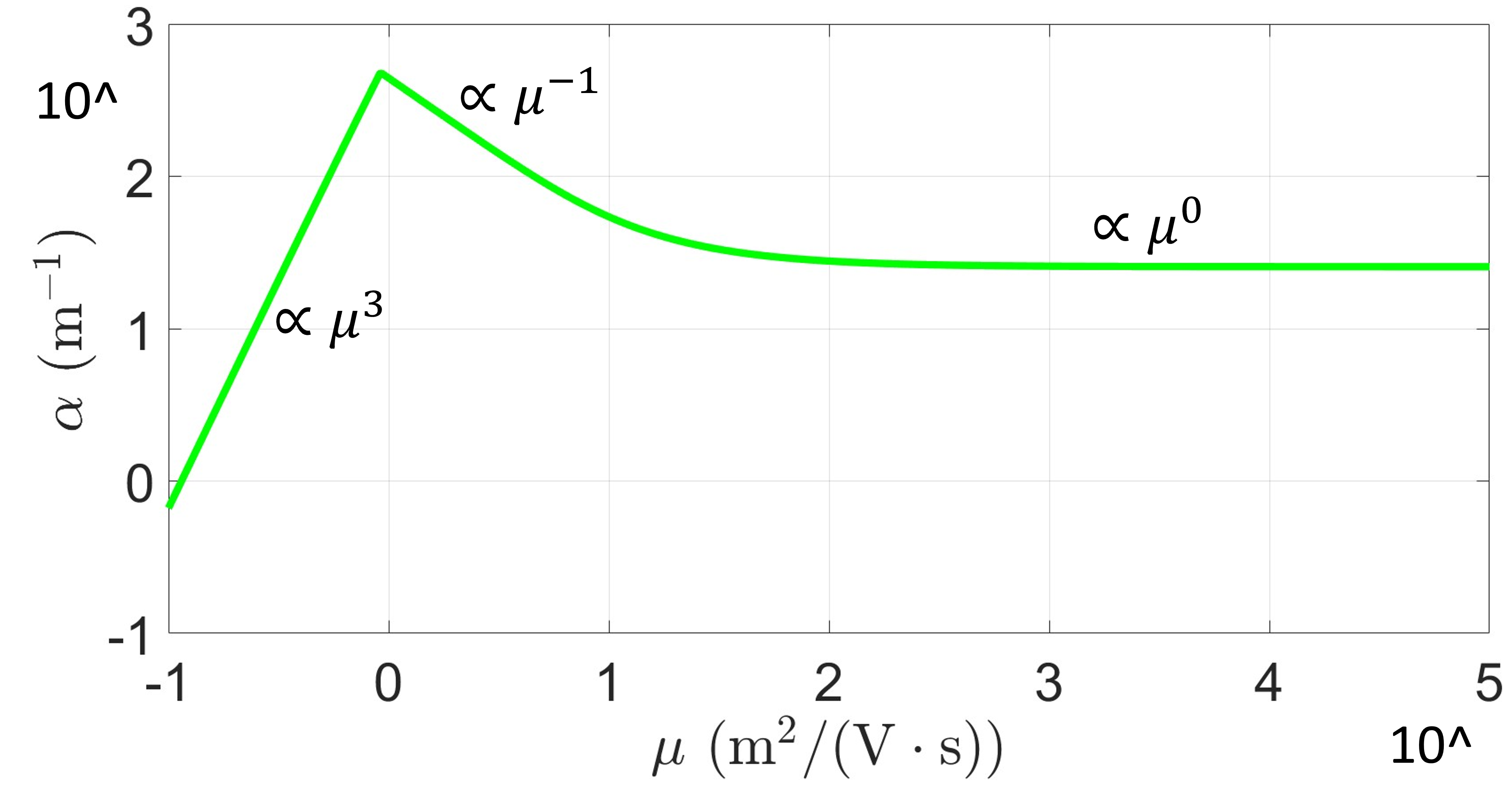}
	\caption{Linear absorption coefficient per unit length $\alpha$ as a function of the carrier mobility $\mu$, given a piezoelectric material dielectric constant $\epsilon_p = 43.6 \epsilon_0$, $C^2 = 6.1 \epsilon_0$, a phonon angular frequency of $\omega_0 = 2\pi \times 10^9 \textrm{ s}^{-1}$, mode depth $L_z = 2 \times 10^{-6}$ m, speed of sound $v_s = 4 \times 10^3$ m/s, a carrier density $n = 2 \times 10^{15} \textrm{ m}^{-2}$, a carrier effective mass $m = 0.067 m_0$, and a 2DEG thickness $t_\mathrm{2DEG} = 2 \times 10^{-8} \textrm{ m}$.}
	\label{fig:linearabsorption}
\end{figure}

We start with the case of linear absorption. Based on Eq.~\eqref{eq: spatial evolution of output field}, the spatial evolution of a single field propagating through the heterostructure and experiencing linear absorption takes the following form:
\begin{equation}
\partial_x A_\mathrm{ph}(\omega_0) = -\alpha A_\mathrm{ph}(\omega_0),
\end{equation}
where $\alpha$ is the amplitude attenuation rate, given as:
\begin{align}
\begin{split}
\alpha &= -i \frac{\epsilon_0}{\hbar} \frac{(i \textrm{Im}[\chi^{(1)}(\omega)]) \sqrt{\omega_0^2}}{\epsilon^*(\omega_0) \epsilon(\omega_0)} \bigg(-C \sqrt{\frac{\hbar}{v_s}}\bigg)^2 \frac{V_\mathrm{2DEG}}{2 V} \\
&= \frac{\epsilon_0 C^2 V_\mathrm{2DEG} \omega_0}{2 v_s V} \frac{\textrm{Im}[\chi^{(1)}(\omega_0)]}{|\epsilon(\omega_0)|^2}.
\end{split}
\end{align}
Figure~\ref{fig:linearabsorption} depicts the absorption rate given $C^2 = 6.1 \epsilon_0$ and effective mode depth $L_z = 2 \times 10^{-6} \textrm{ m}$ (i.e., the half-wavelength of the $\omega_0$ field) along with the same parameters used in the susceptibility calculations, assuming that the 2DEG covers the piezoelectric material such that $V_\mathrm{2DEG}/V = t_\mathrm{2DEG}/L_z$.
At very low mobility, the imaginary part of $\chi^{(1)}$ varies as $\mu^3$, while the dielectric constant is flat in mobility, causing the absorption rate to vary as $\mu^3$. For moderately low mobility, though, the dielectric-constant-squared varies as $\mu^4$, causing the absorption rate to decrease with mobility as $\mu^{-1}$. The linear absorption coefficient reaches a constant value of about $25 \textrm{ m}^{-1}$ at the high-mobility limit (since both the dielectric constant and the imaginary part of $\chi^{(1)}$ are constant at high mobility), corresponding to an effective phonon coherence length of 40 mm.

\subsection{Second-Harmonic Generation}
\label{sec: Second-Harmonic Generation}

Next, we solve for the second-harmonic-generation coupling strength. From Eq.~\eqref{eq: spatial evolution of output field}, the output field spatially evolves as follows:
\begin{equation}
\partial_x A_\mathrm{ph}(2\omega_0) = i G A_\mathrm{ph}(\omega_1) A_\mathrm{ph}(\omega_2),
\end{equation}
where $2\omega_0 = \omega_1 + \omega_2$, and $G$ is the coupling strength, given as:
\begin{equation}
G = \frac{\epsilon_0}{\hbar} \frac{\textrm{Re}[\chi^{(2)}(\omega_1,\omega_2)] \sqrt{2\omega_0 \omega_1 \omega_2}}{\epsilon^*(2\omega_0) \epsilon(\omega_1) \epsilon(\omega_2)} \bigg(-C \sqrt{\frac{\hbar}{v_s}}\bigg)^3 \frac{V_\mathrm{2DEG}}{2 V_{2\omega_0}},
\end{equation}
where $V_{2\omega_0}$ is the mode volume corresponding to the $2\omega_0$ field. Here, the half-wavelength yields an effective mode depth of $L_{z,2\omega_0} = 1 \times 10^{-6} \textrm{ m}$. For degenerate second-harmonic generation (i.e, $\omega_1 = \omega_0 = \omega_2$), the coupling strength becomes the following:
\begin{equation}
G = -\frac{\epsilon_0 C^3 V_\mathrm{2DEG}}{2 V_{2\omega_0}} \sqrt{\frac{2 \hbar \omega_0^3}{v_s^3}} \frac{\textrm{Re}[\chi^{(2)}(\omega_0,\omega_0)]}{\epsilon^*(2\omega_0) |\epsilon(\omega_0)|^2}.
\end{equation}
Figure~\ref{fig:secondharmonic} depicts the second-harmonic coupling strength using the same parameters as in the linear absorption case (except for the mode depth, as previously mentioned):
\begin{figure}[!tb]
	\centering
	\includegraphics[width=\columnwidth]{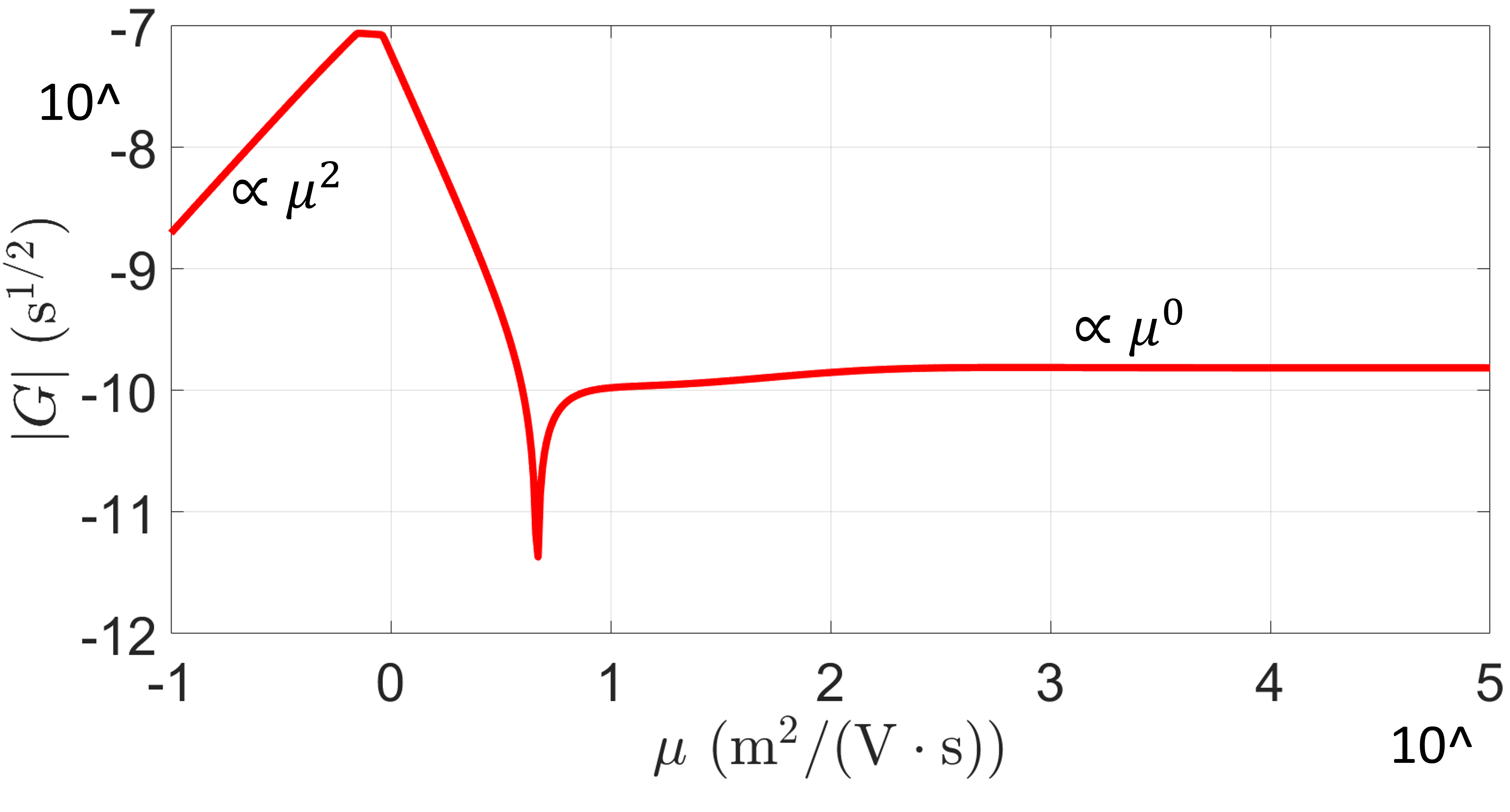}
	\caption{Amplitude of coupling strength $G$ for second-harmonic generation as a function of the carrier mobility $\mu$, given a piezoelectric material dielectric constant $\epsilon_p = 43.6 \epsilon_0$, $C^2 = 6.1 \epsilon_0$, a phonon angular frequency of $\omega_0 = 2\pi \times 10^9 \textrm{ s}^{-1}$, $2\omega_0$-field mode depth $L_{z,2\omega_0} = 1 \times 10^{-6}$ m, speed of sound $v_s = 4 \times 10^3$ m/s, a carrier density $n = 2 \times 10^{15} \textrm{ m}^{-2}$, a carrier effective mass $m = 0.067 m_0$, and a 2DEG thickness $t_\mathrm{2DEG} = 2 \times 10^{-8} \textrm{ m}$.}
	\label{fig:secondharmonic}
\end{figure}
At very low mobility, $\chi^{(2)}$ varies as $\mu^2$, while the dielectric constant is flat in mobility, causing $|G|$ to vary as $\mu^2$. Once the dielectric constant starts rising with mobility, however, the strong inverse dependence of $|G|$ with the dielectric constant (specifically, the $\epsilon^{-3}$ variation) causes $|G|$ to sharply drop with increasing mobility despite the fact that $\chi^{(2)}$ increases with mobility in this regime. However, at the high-mobility limit, both $\chi^{(2)}$ and the dielectric constant reach constant values, causing $|G|$ to level out at $7.6 \times 10^{-11} \textrm{ s}^{1/2}$.

\subsection{Parametric Amplification}
\label{sec: Parametric Amplification}

\begin{figure*}[t]
	\centering
	\includegraphics[width=\textwidth]{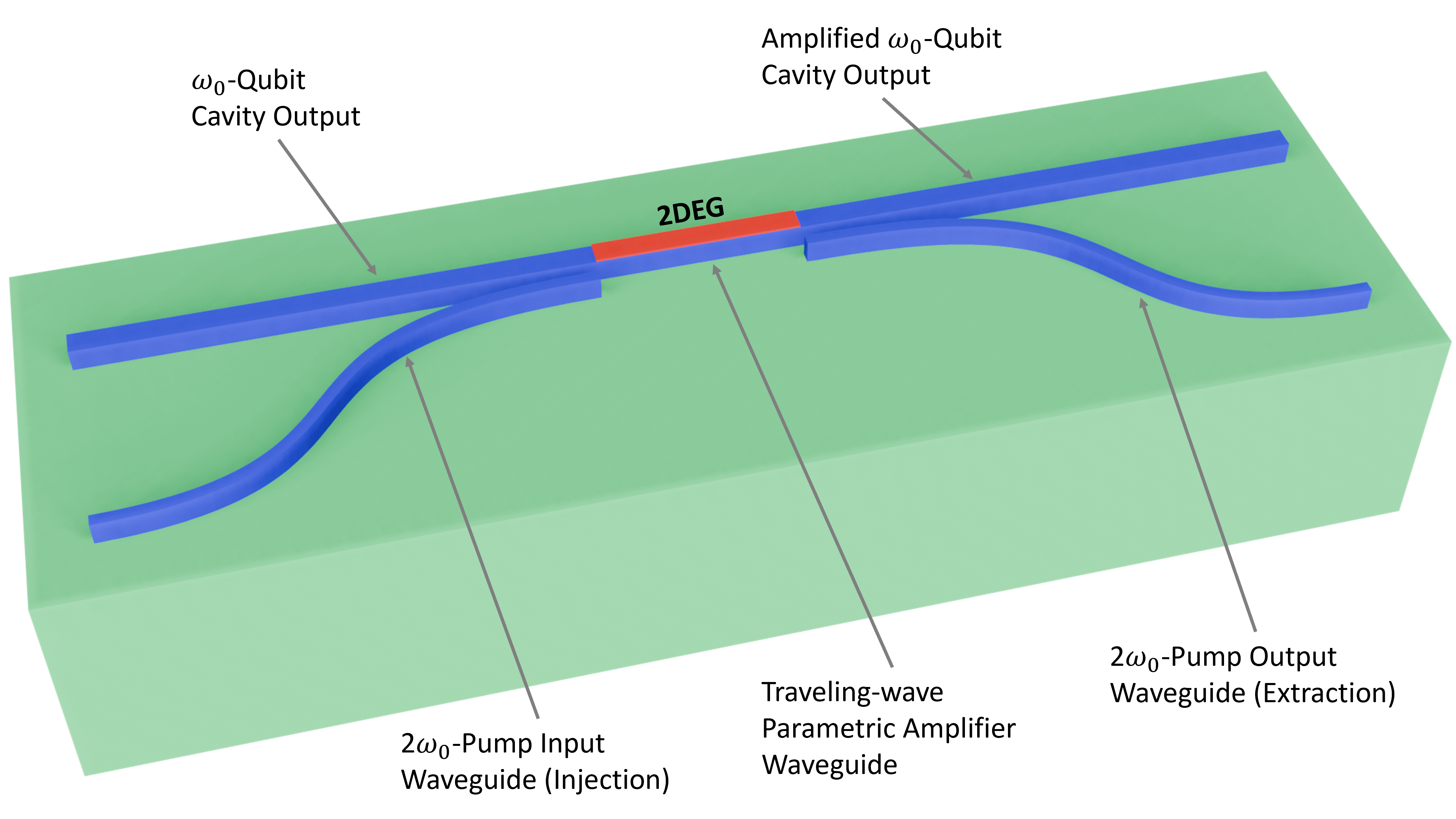}
	\caption{Diagram of the degenerate traveling-wave parametric amplifier system. Note that the S-curve on the left carries the pump input, while the curve on the right drains the pump phonons remaining after the 3-wave-mixing process. The region in between represents the core amplifier region, with the 2DEG (red) on top of the piezoelectric material (blue).}
	\label{fig:parametricamplifier}
\end{figure*}

Here, we solve for the parametric amplification gain per unit length. We specifically design a phase-sensitive, degenerate, traveling-wave parametric amplifier due to its ability to achieve the minimum possible additive quantum noise for an amplification process \cite{caves1982quantumlimits, yurke1989observationparametric, levandovsky1999amplitudesqueezing, castellanosbeltran2008amplificationsqueezing, karlsson2016transmissionsystems}. It is worth noting that resonant, degenerate parametric amplifiers could alternatively be designed to minimize the pump power required for a given gain, though a phase-preserving amplifier can never achieve the same additive quantum noise as a phase-sensitive (traveling-wave) amplifier \cite{caves1982quantumlimits, clerk2010introductionquantum}.

A diagram of a traveling-wave parametric amplifier is shown in Fig.~\ref{fig:parametricamplifier}. Given an input signal field with frequency $\omega_1$ and a pump field with frequency $2\omega_0 = \omega_1 + \omega_2$, the output field $\omega_2$ evolves in a manner analogous to the output field in Sec.~\ref{sec: Second-Harmonic Generation}:
\begin{equation} \label{eq: parametric amplifier omega2 evolution}
\partial_x A_\mathrm{ph}(\omega_2) = i G_{p,1} A_\mathrm{ph}(2\omega_0) A_\mathrm{ph}^*(\omega_1),
\end{equation}
where $G_{p,1}$ is the parametric conversion efficiency per unit length per pump phonon current density amplitude (defined as the square-root of the number of pump phonons per unit time per unit cross-sectional area), defined as:
\begin{align}
\begin{split}
&G_{p,1} = \\
&\quad \frac{\epsilon_0}{\hbar} \frac{\textrm{Re}[\chi^{(2)}(2\omega_0,-\omega_1)] \sqrt{2\omega_0 \omega_1 \omega_2}}{\epsilon(2\omega_0) \epsilon^*(\omega_1) \epsilon^*(\omega_2)} \bigg(-C \sqrt{\frac{\hbar}{v_s}}\bigg)^3 \frac{V_\mathrm{2DEG}}{2 V_{\omega_2}}.
\end{split}
\end{align}
Conversely, if we consider $\omega_2$ as the input signal field, then the time-evolution of $\omega_1$ is modeled by switching $\omega_1$ and $\omega_2$ in Eq.~\eqref{eq: parametric amplifier omega2 evolution}, yielding:
\begin{equation} \label{eq: parametric amplifier omega1 evolution}
\partial_x A_\mathrm{ph}(\omega_1) = i G_{p,1} A_\mathrm{ph}(2\omega_0) A_\mathrm{ph}^*(\omega_2).
\end{equation}
We now specifically consider the case of a degenerate traveling-wave parametric amplifier (i.e., $\omega_1 = \omega_0 = \omega_2$). If we set the phase of the pump field $A_\mathrm{ph}(2\omega_0)$ such that it is in phase with $-i A_\mathrm{ph}(\omega_0)/A_\mathrm{ph}^*(\omega_0)$, then the time-evolution of the signal/idler field amplitude is solved by summing Eqs.~\eqref{eq: parametric amplifier omega2 evolution} and~\eqref{eq: parametric amplifier omega1 evolution}:
\begin{equation}
\partial_x |A_\mathrm{ph}(\omega_0)| = 2 |G_{p,1}| \Big|A_\mathrm{ph}(2\omega_0)\Big| |A_\mathrm{ph}(\omega_0)|.
\end{equation}
As such, the amplitude $|A_\mathrm{ph}(\omega_0)|$ grows exponentially at a rate $2 |G_{p,1}| \Big|A_\mathrm{ph}(2\omega_0)\Big|$. The signal/idler field power $P(\omega_0)$ is proportional to the field \textit{intensity} $|A_\mathrm{ph}(\omega_0)|^2$ and thus evolves at twice the rate of the amplitude:
\begin{equation}
P(\omega_0,x) = P(\omega_0,0) e^{4 |G_{p,1}| |A_\mathrm{ph}(2\omega_0)| x}.
\end{equation}
The gain per unit length per pump phonon current density amplitude is thus expressed in dB form as:
\begin{equation}
G_{p,2} = \bigg(\frac{10}{\ln{(10)}}\bigg) 4 |G_{p,1}| = \frac{40 |G_{p,1}|}{\ln{(10)}}. 
\end{equation}
An alternative metric for parametric amplification is the gain per unit length per pump power amplitude (defined as the square-root of the pump power). This is solved from $G_{p,2}$ in the following manner:
\begin{align}
\begin{split}
G_p &= \frac{G_{p,2}}{\sqrt{\hbar 2 \omega_0 L_y L_{z,2\omega_0}}} \\
&= \frac{20 \epsilon_0 C^3 V_\mathrm{2DEG}}{\ln{(10)} V_{\omega_0}} \sqrt{\frac{\omega_0^2}{v_s^3 L_y  L_{z,2\omega_0}}} \frac{|\textrm{Re}[\chi^{(2)}(2\omega_0,-\omega_0)]|}{|\epsilon(2\omega_0)| |\epsilon(\omega_0)|^2}.
\end{split}
\end{align}
Note that $G_p$ features an inverse dependence on the pump mode's cross-sectional area $L_y L_{z,2\omega_0}$ in addition to the standard $V_\mathrm{2DEG}/V_{\omega_0}$ variation for $G_{p,1}$. This is because for a given pump power, the pump field amplitude increases with decreasing cross-sectional area. Since the gain is proportional to the pump field amplitude rather than the power, this causes the gain to increase for lower cross-sectional area values. 

\begin{figure}[!tb]
	\centering
	\includegraphics[width=\columnwidth]{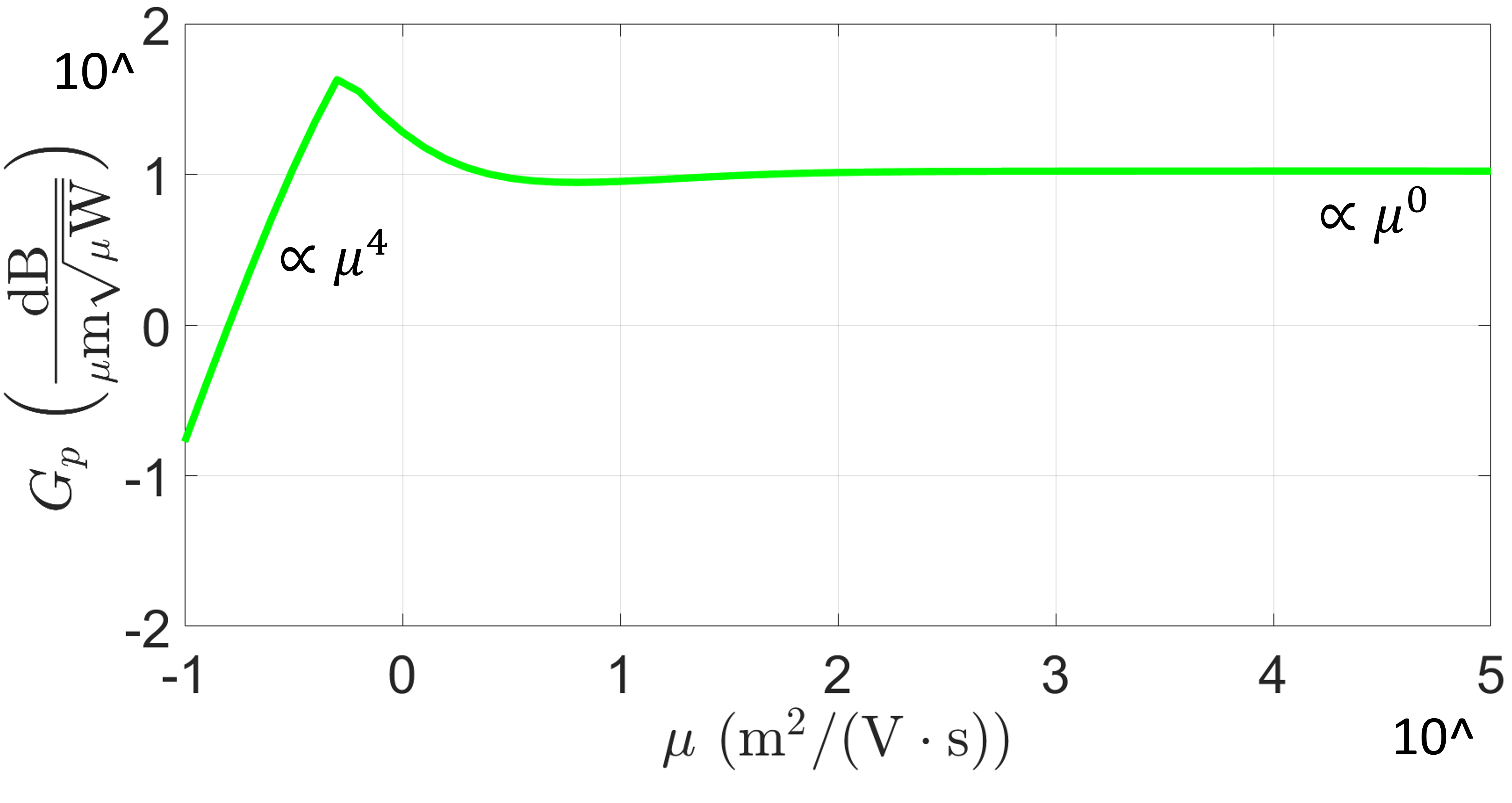}
	\caption{Degenerate parametric amplification gain per unit length per unit power amplitude $G_p$ as a function of the carrier mobility $\mu$, given a piezoelectric material dielectric constant $\epsilon_p = 43.6 \epsilon_0$, $C^2 = 6.1 \epsilon_0$, a signal/idler angular frequency of $\omega_0 = \pi \times 10^{10} \textrm{ s}^{-1}$, signal-field mode depth $L_{z,\omega} = 4 \times 10^{-7}$ m, pump-field mode depth $L_{z,2\omega} = 2 \times 10^{-7}$ m, width $L_y = 1.2 \times 10^{-5}$ m, speed of sound $v_s = 4 \times 10^3$ m/s, a carrier density $n = 2 \times 10^{15} \textrm{ m}^{-2}$, a carrier effective mass $m = 0.067 m_0$, and a 2DEG thickness $t_\mathrm{2DEG} = 2 \times 10^{-8} \textrm{ m}$.}
	\label{fig:parametricgain}
\end{figure}

Figure~\ref{fig:parametricgain} depicts the parametric amplification gain per unit length per pump power amplitude, given the same parameters as in the second-harmonic-generation case, except that we use a signal/idler frequency of 5 GHz (corresponding to $\omega_0 = \pi \times 10^{10} \textrm{ s}^{-1}$) and a mode width of 15 wavelengths (i.e., $L_y = 12 \textrm{ } \mu\textrm{m}$). The mode depths corresponding to the signal and pump fields, respectively, are the half-wavelengths $L_{z,\omega_0} = 400 \textrm{ nm}$ and $L_{z,2\omega_0} = 200 \textrm{ nm}$.
At very low mobility, $\chi^{(2)}(2\omega_0,-\omega_0)$ varies as $\mu^4$, while the dielectric constant is invariant in mobility, causing $G_p$ to vary as $\mu^4$, peaking at $43 \textrm{ dB}/(\mu\textrm{m} \sqrt{\mu\textrm{W}})$, corresponding to a 20-dB gain in 4.7 $\mu$m (about 2.4\% of the Rayleigh length) given a 10-nW pump power input. As the mobility is increased from this point, though, the strong inverse dependence of the gain with the dielectric constant causes $G_p$ to decline with increasing mobility for the intermediate-mobility case, as with the second-harmonic case. Eventually, at the high-mobility limit, both $\chi^{(2)}(2\omega_0,-\omega_0)$ and the dielectric constant become constant, causing $G_p$ to level out at $11 \textrm{ dB}/(\mu\textrm{m} \sqrt{\mu\textrm{W}})$, corresponding to a 20-dB gain in 19 $\mu$m (about 9.5\% of the Rayleigh length) given a 10-nW pump power input.

\subsection{Kerr Nonlinearity}
\label{sec: Kerr Nonlinearity}

\begin{figure*}[t]
	\centering
	\includegraphics[width=\textwidth]{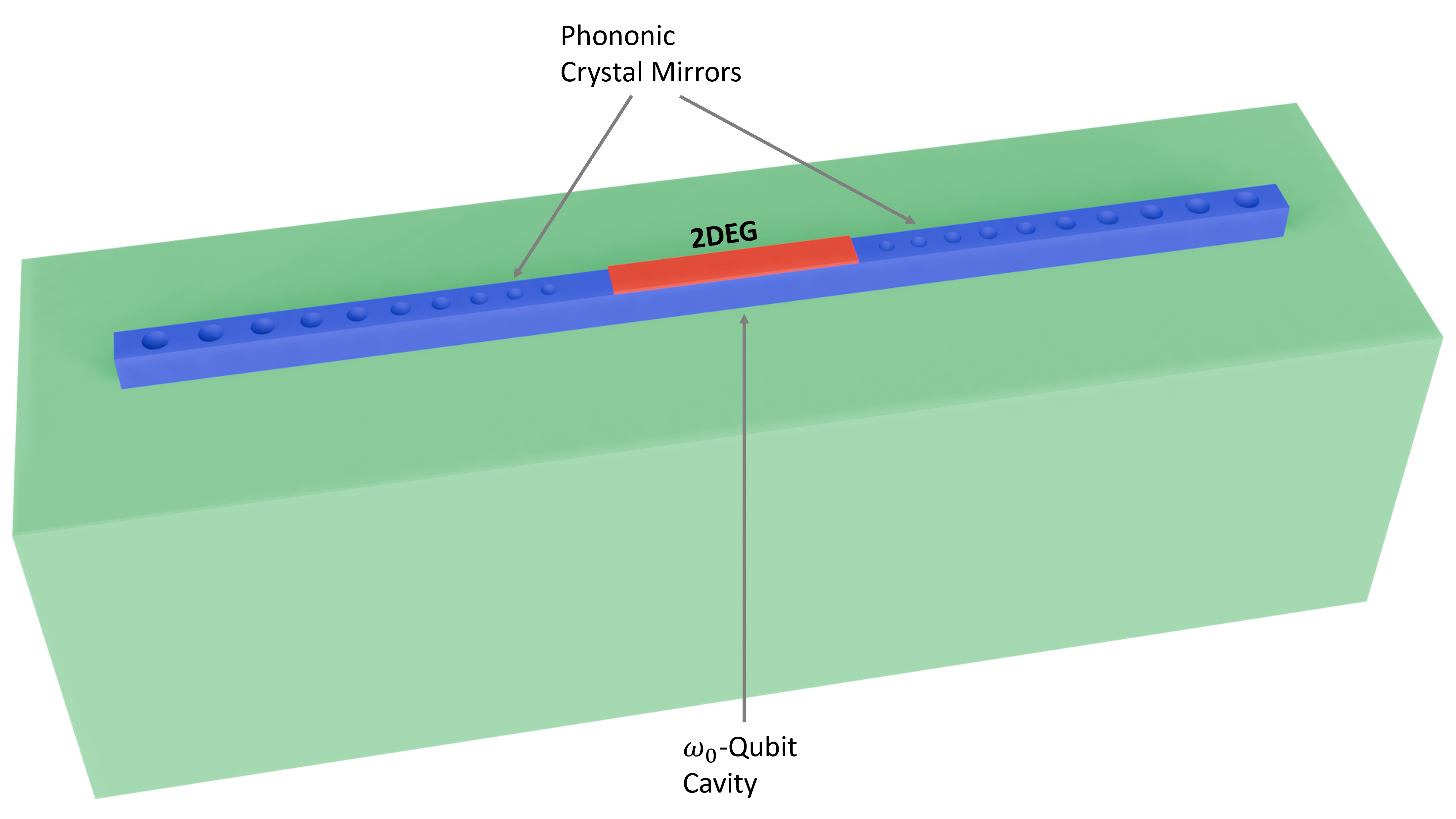}
	\caption{Diagram of the acoustic cavity, with the 2DEG (red) stacked on top of the piezoelectric material (blue) in the central region. The coupling between the two gives rise to a Kerr nonlinearity through 4-wave-mixing.}
	\label{fig:kerrcavity}
\end{figure*}

Finally, we solve for the Kerr nonlinearity figure-of-merit (F.O.M.), which we define as the ratio between the Kerr-induced anharmonicity and the spectral broadening. This will determine whether the electron-phonon interaction induces enough anharmonicity in the phonon spectrum such that the system can be used as an artificial atom. A diagram of the acoustic cavity giving rise to the Kerr nonlinearity is shown in Fig.~\ref{fig:kerrcavity}. From Eq.~\eqref{eq: (N+1)-phonon interaction rate}, the anharmonicity induced by the Kerr shift in a phonon ladder of fundamental angular frequency $\omega_0$ is calculated by doubling the Kerr coupling coefficient, which we calculate from the third-order Kerr susceptibility $\chi^{(3)}_\mathrm{Kerr}$ as follows:
\begin{align}
\begin{split}
2 |g_\mathrm{Kerr}| &= 2 \frac{\epsilon_0 V_\mathrm{2DEG}}{\hbar} \chi^{(3)}_\mathrm{Kerr}(\omega_0) \Big|E_\mathrm{zpf}(\omega_0)\Big|^4.
\end{split}
\end{align}
On the other hand, given a high-Q acoustic cavity (which has recently been achieved \cite{shao2019phononicband}), the spectral broadening is given by the linear absorption per unit time, which corresponds to $\textrm{Im}[\chi^{(1)}]$. We convert the result for the linear amplitude attenuation coefficient per unit propagation distance $\alpha$ from Sec.~\ref{sec: Linear Absorption} into per-unit-time absorption by doubling it (to convert from amplitude attenuation to energy absorption) and multiplying by the speed of sound $v_s$, yielding:
\begin{equation}
\gamma_{ph} = 2 \frac{\epsilon_0 V_\mathrm{2DEG}}{\hbar} \textrm{Im}[\chi^{(1)}(\omega_0)] \Big|E_\mathrm{zpf}(\omega_0)\Big|^2,
\end{equation}
where $\gamma_{ph}$ is the phononic spectral broadening. The Kerr-nonlinearity figure-of-merit (F.O.M.) is therefore solved as follows:
\begin{align}
\begin{split}
\textrm{F.O.M.} &= \frac{2 |g_\mathrm{Kerr}|}{\gamma_{ph}} \\
&= \frac{\chi^{(3)}_\mathrm{Kerr}(\omega_0) |E_\mathrm{zpf}(\omega_0)|^2}{\textrm{Im}[\chi^{(1)}(\omega_0)]} \\
&= \frac{C^2 \hbar \omega_0 \chi^{(3)}_\mathrm{Kerr}(\omega_0)}{2 V \epsilon^2(\omega) \textrm{Im}[\chi^{(1)}(\omega_0)]}.
\end{split}
\end{align}
As this expression shows, the Kerr nonlinearity is maximized by minimizing the acoustic field's mode volume. This is because the Kerr shift varies with the electric field intensity per phonon, which varies inversely with mode volume. 

\begin{figure}[!tb]
	\centering
	\includegraphics[width=\columnwidth]{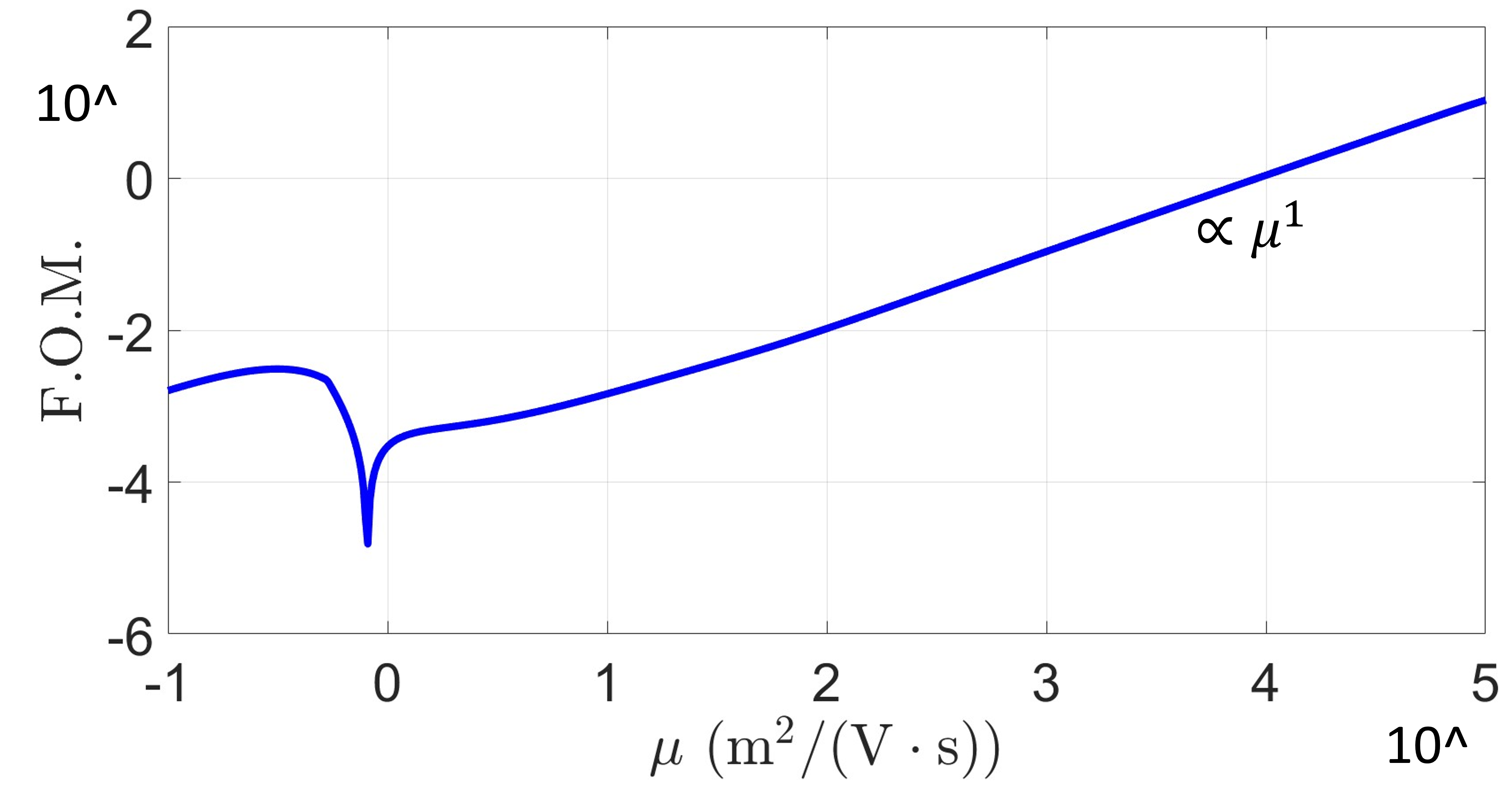}
	\caption{Kerr nonlinearity figure-of-merit (F.O.M.) as a function of the carrier mobility $\mu$, given a piezoelectric material dielectric constant $\epsilon_p = 43.6 \epsilon_0$, $C^2 = 6.1 \epsilon_0$, a phonon angular frequency of $\omega_0 = \pi \times 10^{10} \textrm{ s}^{-1}$, mode length of $4 \times 10^{-7}$ m along each dimension, speed of sound $v_s = 4 \times 10^3$ m/s, a carrier density $n = 2 \times 10^{15} \textrm{ m}^{-2}$, a carrier effective mass $m = 0.067 m_0$, and a 2DEG thickness $t_\mathrm{2DEG} = 2 \times 10^{-8} \textrm{ m}$.}
	\label{fig:kerrfigureofmerit}
\end{figure}

Figure~\ref{fig:kerrfigureofmerit} depicts the Kerr nonlinearity figure-of-merit using the same parameters as in the linear absorption and second-harmonic generation cases, except that the angular frequency $\omega_0$ is changed to $\pi \times 10^{10} \textrm{ s}^{-1}$ to align with the 5-GHz transmon frequency.
We minimize the mode volume by setting the mode length equal to the half-wavelength of 400 nm along each dimension. At high mobilities, the Kerr figure-of-merit scales linearly with $\mu$. This is because $\chi^{(3)}_\mathrm{Kerr}$ also scales linearly with $\mu$ in this regime, while both the real and imaginary parts of $\chi^{(1)}$ are constant in mobility. At very high mobilies (above $10^4 \textrm{ m}^2/(\textrm{V} \cdot \textrm{s})$), the figure-of-merit exceeds unity, indicating that the phononic system can be used as an artificial atom.

\section{Comparison of Quantum and Classical Results in Fluid Limit}
\label{sec: Comparison of Quantum and Classical Results in Fluid Limit}

Although the previous sections treated the carrier states as coherent fermionic states, it is interesting to analyze the system in the incoherent (fluid) carrier limit for the purpose of comparing the quantum results for the linear absorption (corresponding to $\textrm{Im}[\chi^{(1)}]$) and the sum-frequency generation (corresponding to $\textrm{Re}[\chi^{(2)}]$) processes to classical results from previous studies. Here, we operate in the low-mobility limit and effectively treat the electrons as bosonic. Due to the latter, all electrons are capable of interacting with phonons, and due to the former, the probability of such an interaction is approximately uniform for all of the electrons. Furthermore, we only consider the resonant interaction processes, in line with previous theoretical studies of various types of interactions \cite{wherrett1983comparisontheories, lehmberg1976negativenonlinear, yajima1978ultrafast, khoo1980timedelayed, panock1977twolaser}.

\subsection{Linear Absorption in Fluid Limit}
\label{sec: Linear Absorption in Fluid Limit}

We start by considering the linear absorption. In the low-mobility regime, the resonant part of $\textrm{Im}[f_{k_x}^{(1)}]$ from Eq.~\eqref{eq: chi(1) denominator terms} reduces to:
\begin{equation}
\textrm{Im}[f^{(1)}_\mathrm{res}] = \frac{\gamma'}{(\omega_{k_x + q,k_x} - \omega)^2 + \gamma'^2} \approx \frac{1}{\gamma'},
\end{equation}
which is uniform for all electrons, as expected. Then, if we make the assumption that all electrons in the Fermi circle are capable of absorbing phonons, we solve for the imaginary part of $\chi^{(1)}$ simply by multiplying this expression by the Fermi circle area and the coefficients in Eq.~\eqref{eq: chi(1) integral}, yielding:
\begin{align}
\begin{split}
\textrm{Im}[\chi^{(1)}(\omega_0)] &\approx \bigg(\frac{q_e^2 v_s^2}{2 \pi^2 \hbar \epsilon_0 t_\mathrm{2DEG} \omega_0^2}\bigg) \frac{1}{\gamma'} \Big(\pi k_F^2\Big) \\
&= \frac{q_e^2 v_s^2 n}{\hbar \epsilon_0 t_\mathrm{2DEG} \omega_0^2 \gamma'} \\
&= \frac{2 q_e v_s^2 m n \mu}{\hbar \epsilon_0 t_\mathrm{2DEG} \omega_0^2},
\end{split}
\end{align}
where we substituted $k_F^2 = 2\pi n$ in the second line (where $n$ is the area carrier density) and $\gamma' = q_e/(2m\mu)$ in the third line. 

Now, we apply another consequence of the fluid treatment of electrons: that all ground-state electrons feature zero momentum ($\bm{k} = 0$). Then, a momentum kick from absorbing a phonon of wavevector $q$ would elevate a ground-state electron's energy to $\hbar q^2/(2m)$. If we assume that the electron-phonon interaction is resonant, then $\hbar q^2/(2m)$ must equal $v_s q$ (i.e., $\hbar \omega_0 = 2 m v_s^2$). Substituting this into the above expression for the imaginary part of $\chi^{(1)}$ yields the following result:
\begin{equation}
\textrm{Im}[\chi^{(1)}](\omega_0) \approx \frac{q_e n_\mathrm{bulk} \mu}{\epsilon_0 \omega_0} = \frac{\omega_c \epsilon}{\epsilon_0 \omega_0},
\end{equation}
where we defined a bulk volumetric electron density $n_\mathrm{bulk} = n/t_\mathrm{2DEG}$, along with the frequency parameter $\omega_c = n_\mathrm{bulk} q_e \mu/\epsilon$. It is worth noting that the imaginary part of $\chi^{(1)}$ varies linearly with both carrier density and mobility via $\omega_c$. The former is due to the fact that a higher carrier density leads to a larger number of electrons available for interacting with phonons, while the latter is because for a given electron, the electron-phonon interaction probability scales inversely with the spectral broadening in the limit in which the detuning is much smaller than the broadening.

Next, we determine the phonon absorption per unit length $\alpha$ from the imaginary part of $\chi^{(1)}$ by multiplying by $\epsilon_0 V_\mathrm{2DEG} |E(\omega_0)|^2/\hbar$ to determine the absorption per unit time and dividing by the propagation speed $v_s$. Recall that the zero-point electric field intensity is a function of the effective dielectric constant $\epsilon_\mathrm{eff}$, the elasticity $\kappa$, and the frequency $\omega_0$:
\begin{equation}
|E_\mathrm{zpf}(\omega_0)|^2 = \frac{e^2}{|\epsilon_\mathrm{eff}|^2} |S_\mathrm{zpf}|^2 = \frac{K^2 \epsilon \hbar \omega_0}{2 |\epsilon_\mathrm{eff}|^2 V_\mathrm{piezo}},
\end{equation}
where $\epsilon$ is the dielectric screening (i.e., the real part of $\chi^{(1)}$) and $K^2 = e^2/(\epsilon \kappa)$ represents the square of the electromechanical coupling. This accords with Eq.~\eqref{eq: E_zpf}, except for the fact that $\epsilon_\mathrm{eff}$ incorporates both the real and imaginary parts of $\chi^{(1)}$ for the sake of generality:
\begin{equation}
\epsilon_\mathrm{eff} = \epsilon_0 \Big(\textrm{Re}[\chi^{(1)}(\omega_0)] + i \textrm{Im}[\chi^{(1)}(\omega_0)]\Big) = \epsilon \bigg(1 + i \frac{\omega_c}{\omega_0}\bigg).
\end{equation}
The absorption per unit length $\alpha$ thus becomes the following:
\begin{align}
\begin{split}
\alpha &= \frac{\epsilon_0 V_\mathrm{2DEG}}{\hbar v_s} \textrm{Im}[\chi^{(1)}](\omega_0) \frac{K^2 \epsilon \hbar \omega_0}{2 |\epsilon_\mathrm{eff}|^2 V_\mathrm{piezo}} \\
&= \frac{\epsilon_0 V_\mathrm{2DEG}}{\hbar v_s} \frac{\omega_c \epsilon}{\epsilon_0 \omega_0} \frac{K^2 \epsilon \hbar \omega_0}{2 \epsilon^2 (1 + \omega_c^2/\omega_0^2) V_\mathrm{2DEG}} \\
&= \frac{K^2}{2} \frac{\omega_c}{v_s} \bigg(1 + \frac{\omega_c^2}{\omega_0^2}\bigg)^{-1}.
\end{split}
\end{align}
where in the first line, we inserted $v_s$ in the denominator to convert from per-unit-time to per-unit-length absorption, while in the second line, we used the fact that $V_\mathrm{piezo} = V_\mathrm{2DEG}$ for a bulk piezoelectric semiconductor. Note that this result matches that derived previously in the classical limit \cite{white1962amplificationultrasonic}.

\subsection{Sum-Frequency Generation in Fluid Limit}
\label{sec: Sum-Frequency Generation in Fluid Limit}

We now turn to the question of resolving the real part of $\chi^{(2)}$ in for the purpose of deriving the dynamics of sum-frequency generation. In the low-mobility regime, the resonant part of $\textrm{Re}[f_{k_x}^{(2)}]$ from Eq.~\eqref{eq: chi(2) denominator terms} reduces to the following:
\begin{widetext}
\begin{align}
\begin{split}
&\textrm{Re}[f^{(2)}_\mathrm{res}] \\
&= 
\bigg(\frac{1}{(\omega_{k_x + q_1 + q_2,k_x} - (\omega_1 + \omega_2) - i\frac{\gamma}{2}) (\omega_{k_x + q_1,k_x} - \omega_1 - i\frac{\gamma}{2})} + \frac{1}{(\omega_{k_x + q_1 + q_2,k_x} - (\omega_1 + \omega_2) - i\frac{\gamma}{2}) (\omega_{k_x + q_2,k_x} - \omega_2 - i\frac{\gamma}{2})}\bigg) \\
&\approx -\frac{2}{\gamma'^2},
\end{split}
\end{align}
\end{widetext}
which is also uniform for all electrons. Then, assuming that all electrons are capable of undergoing second-order interaction with phonons, we solve for the real part of $\chi^{(2)}$ by multiplying this expression by the Fermi circle area and the coefficients in Eq.~\eqref{eq: chi(2) integral}, yielding the following result:
\begin{align}
\begin{split}
\textrm{Re}[\chi^{(2)}(\omega_1,\omega_2)] &\approx \bigg(\frac{q_e^3 v_s^3}{2 \pi^2 \hbar^2 \epsilon_0 t_\mathrm{2DEG} \omega_1 \omega_2 (\omega_1 + \omega_2)}\bigg) \\
&\quad \times \bigg(-\frac{2}{\gamma'^2}\bigg) \Big(\pi k_F^2\Big) \\
&= -\frac{2 q_e^3 v_s^3 n}{\hbar^2 \epsilon_0 t_\mathrm{2DEG} \omega_1 \omega_2 (\omega_1 + \omega_2) \gamma'^2} \\
&= -\frac{8 q_e v_s^3 m^2 n \mu^2}{\hbar^2 \epsilon_0 t_\mathrm{2DEG} \omega_1 \omega_2 \omega_3},
\end{split}
\end{align}
where $\omega_3 = \omega_1 + \omega_2$, and we substituted $k_F^2 = 2\pi n$ and $\gamma'^2 = q_e^2/(4 m^2 \mu^2)$ in the second and third lines, respectively. 

Applying the ground-state approximation described in the linear absorption derivation, along with the single-phonon resonance condition that $\hbar q^2/(2m)$ must equal $v_s q$ (i.e., $\hbar^2 \omega^2 = 4m^2v_s^4$, where $\omega_1 \approx \omega \approx \omega_2$), we find that the real part of $\chi^{(2)}$ reduces to the following:
\begin{equation}
\textrm{Re}[\chi^{(2)}(\omega_1,\omega_2)] \approx -\frac{2 q_e n_\mathrm{bulk} \mu^2}{\epsilon_0 v_s \omega_3} = -\frac{2\epsilon \omega_c\mu}{\epsilon_0 v_s \omega_3},
\end{equation}
where $n_\mathrm{bulk}$ and $\omega_c$ are defined as in the linear absorption calculation, and $\omega_3 = \omega_1 + \omega_2$.

Next, we use the real part of $\chi^{(2)}$ to determine how the $\omega_3$ field evolves in time given input fields $\omega_1$ and $\omega_2$. To that end, the rate of change of the field $E_3$ due to the sum-frequency-generation process can be calculated by promoting the fields to operators, such that $E_n = E_{\mathrm{zpf},n} a_n$, where $a_n$ is the lowering operator for the mode $n$ and $E_{\mathrm{zpf},n}$ is the zero-point electric field amplitude. Recall that the electric field amplitude is directly proportional to the strain field amplitude:
\begin{equation}
E_n = -\frac{e}{\epsilon_n} S_n = -\frac{e}{\epsilon \Gamma_n} S_n,
\end{equation}
where $\Gamma_n$ is defined as follows for the given mode $n$:
\begin{equation}
\Gamma_n = 1 + i\frac{\omega_c}{\omega_n}.
\end{equation}
Also, recall that the zero-point electric field intensity varies linearly with the phonon frequency and inversely with the mode volume as follows:
\begin{equation}
|E_{\mathrm{zpf},n}|^2 = \frac{e^2}{|\epsilon_n|^2} |S_{\mathrm{zpf},n}|^2 = \frac{K^2 \epsilon \hbar \omega_n}{2 |\epsilon \Gamma_n|^2 V} = \frac{K^2 \hbar \omega_n}{2 \epsilon |\Gamma_n|^2 V}.
\end{equation}
The time-evolution of $E_3$ due to sum-frequency generation is thus solved by applying the 3-wave-mixing Hamiltonian:
\begin{widetext}
\begin{align}
\begin{split}
\dot{E_3}^{(2)} &= -i\frac{E_{\mathrm{zpf},3}}{\hbar} [a_3,H_\mathrm{int}^{(2)}] \\
&= i \frac{\epsilon_0 V}{\hbar} \textrm{Re}[\chi^{(2)}](\omega_1,\omega_2) \Big|E_{\mathrm{zpf},3}\Big|^3 E_2 E_1 [a_3,a_3^{\dag}] \\
&\approx i \frac{\epsilon_0 V}{\hbar} \bigg(-\frac{2 \epsilon \omega_c \mu}{\epsilon_0 v_s \omega_3}\bigg) \bigg(\frac{K^2 \hbar \omega_3}{2 \epsilon |\Gamma_3|^2 V}\bigg) \bigg(-\frac{e}{\epsilon \Gamma_2} S_2\bigg) \bigg(-\frac{e}{\epsilon \Gamma_1} S_1\bigg) \\
&= -i K^2 \frac{e^2 \mu \omega_c}{\epsilon^2 v_s} \frac{1}{\Gamma_1 \Gamma_2 |\Gamma_3|^2} S_2 S_1.
\end{split}
\end{align}
\end{widetext}
Finally, we solve for $E_3^{(2)}$ by dividing both sides by $-i \omega_3$ (since $\dot{E_3}^{(2)} = -i \omega_3 E_3^{(2)}$):
\begin{equation}
E_3^{(2)} = K^2 \frac{e^2 \mu}{\epsilon^2 v_s} \frac{\omega_c}{\omega_3} \frac{1}{\Gamma_1 \Gamma_2 |\Gamma_3|^2} S_2 S_1.
\end{equation}
Note that this result matches that calculated in the classical limit \cite{conwellganguly1971mixing} if $K^2 \rightarrow 1$ and $|\Gamma_3|^2 \rightarrow 1$.

\section{Conclusion}

Using time-dependent perturbation theory, we have theoretically derived the nonlinear acoustic susceptibilities up to an arbitrary order for a technologically feasible heterostructure consisting of a 2DEG stacked on top of a piezoelectric material. We have also used this susceptibility to derive the equations of motion for various nonlinear quantum phononic processes. We presented the first, second, and third order susceptibilites in detail, both analytically and numerically, and their variations with important parameters such as carrier concentration and mobility. Our results demonstrate that both the second-order and third-order susceptibilities are maximized at high mobility, demonstrating the advantage that a 2DEG creates by virtue of its ultra-high mobility. We note that in the case of classical acoustoelectric mediation of phonon-mixing nonlinearities \cite{hackett2023mixing}, the application of bias fields causes profound modification of these nonlinear processes. These modifications range from amplification or deamplification in space of the waves involved in the mixing processes, to relaxation of phase-matching conditions, to direct modification of the phononic susceptibility. Though the treatment of the interaction of these quantum phononic mixing processes with bias fields is beyond the scope of this paper, we expect that in this regime, the effect of bias fields will be similarly profound, and we plan to analyze it in future work.

It is particularly noteworthy that the Kerr nonlinearity continues to increase with mobility, even at high mobility values. If we are able to achieve a sufficiently high-mobility in a 2DEG and fabricate a device such that a heterostructure island mostly fills the inside of an acoustic cavity, then the phenomena such as phononic cavity blockades (analogous to photon blockades \cite{birnbaum2005photonblockade, rabl2011photonblockade}) and other types of single-phonon level quantum interactions or even logic gates are achievable. The relevant metric to enter this regime is the Kerr shift per phonon inside the cavity relative to the cavity linewidth, given a high-Q acoustic cavity (which has recently been achieved \cite{shao2019phononicband}). If we can produce such a system with a high enough mobility and small enough cavity linewidth such that only the transition from the ground state to a single phonon excitation is effectively resonant with the cavity, then this system could provide a new type of artificial atom, analogous to a superconducting circuit qubit and at the same microwave frequencies but with a length scale of microns instead of centimeters. Furthermore, the large second-order nonlinearity of our heterostructure system at high mobilities could be used  to create a parametric phonon amplifier. This combination of artificial atom cavities and traveling-wave parametric amplifiers, together with bus waveguides, make a complete analog to the primary hardware components of superconducting circuit quantum computers. Together with a method for microwave-to-phonon transduction (well-established interdigital electrode transducers) and for tuning the coupling rate between cavities \cite{shao2022electricalcontrol, taylor2022reconfigurablequantum}, one has the pieces necessary to perform state preparation, single and multi-qubit gates via tunable cavity-cavity coupling, and room temperature readout after parametric amplification. Thus, it could be possible to make microwave frequency phononic quantum processors based on these heterostructures that are highly analogous to superconducting circuit quantum computers and operating at the same frequency and temperature but with many orders of magnitude higher qubit density. We note these qubits would require no shielding from the environment, as there are no phononic modes in the vacuum to exclude and their size compared to a microwave wavelength gives them a vanishingly small electromagnetic dipole moment. Finally, their reduced size and wholly different mechanism for generating nonlinearity may lead to reduced decoherence from sources such as charge fluctuations and two-level systems.

\begin{acknowledgements}
This work is partially supported by DARPA contract DARPA-PA-23-03-01. The views, opinions and/or findings expressed are those of the authors and
should not be interpreted as representing the official views or policies of
the Department of Defense or the U.S. Government. 

Distribution Statement A: Approved for Public Release, Distribution Unlimited.

\end{acknowledgements}

\appendix

\section{Calculation of Higher-Order Susceptibilities from Time-Dependent Perturbation Theory}
\label{sec: Calculation of Higher-Order Susceptibilities from Time-Dependent Perturbation Theory}

We now seek to derive this using the overall Hamiltonian $H = H_0 + H_\mathrm{int}(t)$ and applying time-dependent perturbation theory. From the 2DEG self-energy in Eq.~\eqref{eq: 2DEG self-energy}, we deduce that $\ket{\Psi^{(n)}(t)}$ can be expressed as a generic superposition of 2DEG eigenstates $\ket{l}$ in the following manner:
\begin{equation}
\ket{\Psi^{(n)}(t)} = \delta_{n,0} \ket{g} + \sum_{l > g} a_l^{(n)}(t) e^{-i \omega_{lg} t} e^{-\frac{\gamma_l}{2} t} \ket{l}.
\end{equation}
The goal is to solve for the time-varying coefficients $a_l^{(n)}(t)$. For the zeroth-order wavefunction $\ket{\Psi^{(0)}(t)}$, the 2DEG is fully in the ground state $\ket{g}$, yielding $a_l^{(0)}(t) = \delta_{l,g}$. To solve for the higher-order coefficients, we start with the Schrodinger-equation relationship linking $\ket{\Psi^{(n)}(t)}$ with $\ket{\Psi^{(n-1)}(t)}$:
\begin{equation}
i\hbar \frac{\partial \ket{\Psi^{(n)}(t)}}{\partial t} = H_0 \ket{\Psi^{(n)}(t)} + H_\mathrm{int}(t) \ket{\Psi^{(n-1)}(t)},
\end{equation}
which takes the following form when each side is expanded in the 2DEG basis:
\begin{align}
\begin{split}
&\sum_l \dot{a_l}^{(n)}(t) e^{-i \omega_{lg} t} e^{-\frac{\gamma_l}{2} t} \ket{l} = \\
&\quad -\frac{i}{\hbar} \sum_l a_l^{(n)}(t) e^{-i \omega_{lg} t} e^{-\frac{\gamma_l}{2} t} H_\mathrm{int}(t) \ket{l}.
\end{split}
\end{align}
Applying $\bra{m} e^{i \omega_{mg} t} e^{\frac{\gamma_m}{2} t}$ to both sides and integrating over time, we find the following recursive expression for the coefficients:
\begin{align}
\begin{split}
&a_m^{(n)}(t) \\
&= \frac{i}{\hbar} \sum_l \int_{-\infty}^t dt' a_l^{(n-1)}(t') \braket{m|H_\mathrm{int}(t')|l} e^{i \Big(\omega_{ml} - i \frac{\gamma_{ml}}{2}\Big) t'} \\
&= \frac{i}{\hbar} \sum_r E(\omega_r) \sum_l d_{\mathrm{eff},ml}(q_r) \\
&\quad\quad \times \int_{-\infty}^t dt' a_l^{(n-1)}(t') e^{i \Big(\omega_{ml} - \omega_r - i \frac{\gamma_{ml}}{2}\Big) t'},
\end{split}
\end{align}
where $\gamma_{ml} = \gamma_m - \gamma_l$, and we substituted the interaction Hamiltonian from Eq.~\eqref{eq: interaction Hamiltonian multiple modes} in the second line. Starting with $a_l^{(0)}(t) = \delta_{l,g}$, we find that the time-dependence in the integral becomes entirely contained in an easily-integrable exponential expression. Using the recursive process from the above expression yields the following analytical result for the coefficients:
\begin{widetext}
\begin{equation}
a_{l_n}^{(n)}(t) = \frac{1}{\hbar^n} \sum_{l_1,...,l_{n-1}} \sum_{r_1,...,r_n} \frac{d_{\mathrm{eff},l_n l_{n-1}}(q_{r_n}) ... d_{\mathrm{eff},l_1 g}(q_{r_1}) E(\omega_{r_1}) ... E(\omega_{r_n}) e^{i \Big(\omega_{l_n g} - (\omega_{r_1} + ... + \omega_{r_n}) - i \frac{\gamma_{l_n g}}{2}\Big) t}}{(\omega_{l_n g} - (\omega_{r_1} + ... + \omega_{r_n}) - i\frac{\gamma_{l_n g}}{2}) ... (\omega_{l_1 g} - \omega_{r_1} - i\frac{\gamma_{l_1 g}}{2})}.
\end{equation}
\end{widetext}
This intuitively corresponds to an $n$-step process from $g$ to $l_n$ through the intermediate states $l_1,...,l_{n-1}$, with the $m^{\textrm{th}}$ step mediated through an absorption of a phonon with frequency $\omega_{r_m}$ and wavevector $q_{r_m} = \omega_{r_m}/v_s$. It is also important to derive the complex conjugate of these coefficients:
\begin{widetext}
\begin{align}
\begin{split}
a_{l'_{n'}}^{(n')*}(t) 
&= \frac{1}{\hbar^{n'}} \sum_{l'_1,...,l'_{n'-1}} \sum_{r'_1,...,r'_{n'}} \frac{d_{\mathrm{eff},g l'_1}(-q_{r'_1}) ... d_{\mathrm{eff},l'_{n'-1} l'_{n'}}(-q_{r'_{n'}}) E(-\omega_{r'_1}) ... E(-\omega_{r'_{n'}}) e^{i \Big(\omega_{g l'_{n'}} + (\omega_{r'_1} + ... + \omega_{r'_{n'}}) - i \frac{\gamma_{l'_{n'} g}}{2}\Big) t}}
{(\omega_{l'_{n'} g} - (\omega_{r'_1} + ... + \omega_{r'_{n'}}) + i\frac{\gamma_{l'_{n'} g}}{2}) ... (\omega_{l'_1 g} - \omega_{r'_1} + i\frac{\gamma_{l'_1 g}}{2})} \\
&= \frac{1}{\hbar^n} \sum_{l'_1,...,l'_{n'-1}} \sum_{r'_1,...,r'_{n'}} \frac{d_{\mathrm{eff},g l'_1}(q_{r'_1}) ... d_{\mathrm{eff},l'_{n'-1} l'_{n'}}(q_{r'_{n'}}) E(\omega_{r'_1}) ... E(\omega_{r'_{n'}}) e^{i \Big(\omega_{g l'_{n'}} - (\omega_{r'_1} + ... + \omega_{r'_{n'}}) - i \frac{\gamma_{l'_{n'} g}}{2}\Big) t}}
{(\omega_{l'_{n'} g} + (\omega_{r'_1} + ... + \omega_{r'_{n'}}) + i\frac{\gamma_{l'_{n'} g}}{2}) ... (\omega_{l'_1 g} + \omega_{r'_1} + i\frac{\gamma_{l'_1 g}}{2})},
\end{split}
\end{align}
\end{widetext}
where in the first line we used the facts that $E^*(\omega) = E(-\omega)$ for any field frequency $\omega$ and $d^*_{\mathrm{eff},fi}(q) = d_{\mathrm{eff},if}(-q)$ for the corresponding field wavevector $q = \omega/v_s$, and in the second line we reversed the signs of the field frequencies $\omega_{r'_1},...,\omega_{r'_{n'}}$ in order to ensure that the time-dependent phase corresponding to the fields would align with the non-conjugated coefficients. 

Having derived the coefficients for the wavefunction, we now solve for the $N^\textrm{th}$-order expectation value of the effective dipole moment corresponding to the wavevector $q_p = -q_{p_1} - ... - q_{p_N}$:
\begin{widetext}
\begin{align}
\begin{split}
\expect{d_\mathrm{eff}^{(N)}(q_p)}_t &= \sum_{j = 0}^N \braket{\Psi^{(j)}(t)|d_\mathrm{eff}(q_p)|\Psi^{(N - j)}(t)} \\
&= \frac{1}{\hbar^N} \sum_{j = 0}^N \sum_{l'_j, l_{N - j}} a_{l'_j}^{(j)*} a_{l_{N - j}}^{(N - j)} e^{i \Big(\omega_{l'_j g} + i \frac{\gamma_{l'_j g}}{2}\Big) t} e^{-i \Big(\omega_{l_{N - j} g} - i \frac{\gamma_{l_{N - j} g}}{2}\Big) t} \braket{l'_j|d_\mathrm{eff}(q_p)|l_{N - j}} \\
&= \frac{1}{\hbar^N} \sum_{j = 0}^N \sum_{l_1,...,l_N} \sum_{r_1,...,r_N} e^{-i (\omega_{r_1} + ... + \omega_{r_N}) t} \\
&\quad \times \frac{d_{\mathrm{eff},g l_N}(q_{r_N}) ... d_{\mathrm{eff},l_{N - j + 2} l_{N - j + 1}}(q_{r_{N - j + 1}})}
{(\omega_{l_N g} + \omega_{r_N} + i\frac{\gamma_{l_N g}}{2}) ... (\omega_{l_{N - j + 1} g} + (\omega_{r_{N - j + 1}} + ... + \omega_{r_N}) + i\frac{\gamma_{l_{N - j + 1} g}}{2})} \\
&\quad \times d_{\mathrm{eff},l_{N - j + 1} l_{N - j}}(q_p) \\
&\quad \times \frac{d_{\mathrm{eff},l_{N - j} l_{N - j - 1}}(q_{r_{N - j}}) ... d_{\mathrm{eff},l_1 g}}{(\omega_{l_{N - n} g} - (\omega_{r_1} + ... + \omega_{r_{N - n}}) - i\frac{\gamma_{l_{N - n} g}}{2}) ... (\omega_{l_1 g} - \omega_{r_1} - i\frac{\gamma_{l_1 g}}{2})} E(\omega_{r_1}) ... E(\omega_{r_N}).
\end{split}
\end{align}
\end{widetext}
Calculating the $N^\textrm{th}$-order susceptibility $\chi^{(N)}$ using Eq.~\eqref{eq: chi(N)}, we find that the field frequencies $\omega_{r_1},...,\omega_{r_N}$ are constrained such that $\omega_{r_1} + ... + \omega_{r_N} = \omega_{p_1} + ... + \omega_{p_N}$, as desired:
\begin{widetext}
\begin{align}
\begin{split} \label{eq: chi(N) general}
\chi^{(N)}(\omega_{p_1},...,\omega_{p_N})
&= \frac{1}{\hbar^N \epsilon_0 V_\mathrm{2DEG}} \sum_{j = 0}^N \sum_{l_1,...,l_N} \sum_{r_1,...,r_N} \int_{-\infty}^{\infty} dt e^{i ((\omega_{p_1} + ... + \omega_{p_N}) - (\omega_{r_1} + ... + \omega_{r_N})) t} \\
&\quad \times \frac{d_{\mathrm{eff},g l_N}(q_{r_N}) ... d_{\mathrm{eff},l_{N - j + 2} l_{N - j + 1}}(q_{r_{N - j + 1}})}
{(\omega_{l_N g} + \omega_{r_N} + i\frac{\gamma_{l_N g}}{2}) ... (\omega_{l_{N - j + 1} g} + (\omega_{r_{N - j + 1}} + ... + \omega_{r_N}) + i\frac{\gamma_{l_{N - j + 1} g}}{2})} \\
&\quad \times d_{\mathrm{eff},l_{N - j + 1} l_{N - j}}(-q_{p_1} - ... - q_{p_N}) \\
&\quad \times \frac{d_{\mathrm{eff},l_{N - j} l_{N - j - 1}}(q_{r_{N - j}}) ... d_{\mathrm{eff},l_1 g}}{(\omega_{l_{N - n} g} - (\omega_{r_1} + ... + \omega_{r_{N - n}}) - i\frac{\gamma_{l_{N - n} g}}{2}) ... (\omega_{l_1 g} - \omega_{r_1} - i\frac{\gamma_{l_1 g}}{2})} \\
&= \frac{1}{\hbar^N \epsilon_0 V_\mathrm{2DEG}} \sum_{j = 0}^N \sum_{l_1,...,l_N} \mathcal{P} \sum_{p_1,...,p_N} \\
&\quad \times \frac{d_{\mathrm{eff},g l_N}(q_{p_N}) ... d_{\mathrm{eff},l_{N - j + 2} l_{N - j + 1}}(q_{p_{N - j + 1}})}
{(\omega_{l_N g} + \omega_{p_N} + i\frac{\gamma_{l_N g}}{2}) ... (\omega_{l_{N - j + 1} g} + (\omega_{p_{N - j + 1}} + ... + \omega_{p_N}) + i\frac{\gamma_{l_{N - j + 1} g}}{2})} \\
&\quad \times d_{\mathrm{eff},l_{N - j + 1} l_{N - j}}(-q_{p_1} - ... - q_{p_N}) \\
&\quad \times \frac{d_{\mathrm{eff},l_{N - j} l_{N - j - 1}}(q_{p_{N - j}}) ... d_{\mathrm{eff},l_1 g}}{(\omega_{l_{N - n} g} - (\omega_{p_1} + ... + \omega_{p_{N - n}}) - i\frac{\gamma_{l_{N - n} g}}{2}) ... (\omega_{l_1 g} - \omega_{p_1} - i\frac{\gamma_{l_1 g}}{2})},
\end{split}
\end{align}
\end{widetext}
where in the second line, we use the fact that $\int_{-\infty}^{\infty} dt e^{i \omega t} = \delta_{\omega,0}$. It is also worth noting the symbol $\mathcal{P}$ in front of the summation over the modes $p_1,...,p_N$, which denotes a permutation over the mode indices. This is due to the fact that there is no specific requirement that each generic mode frequency $\omega_{r_m}$ equal the corresponding input frequency $\omega_{p_m}$, but rather that the sum of the generic mode frequencies (i.e., $\omega_{r_1} + ... + \omega_{r_N}$) equal the sum of the input frequencies (i.e., $\omega_{p_1} + ... + \omega_{p_n}$). As a result, the set of transitions encapsulated by the real part of $\chi^{(N)}$ conserves the energy of the acoustic field, which is required in order for an electron starting in the ground state $\ket{g}$ to return to that state after transitioning through the intermediate states $\ket{l_1},...,\ket{l_N}$. Since each mode features a well-defined momentum as well, the net momentum of the field is conserved too.

\section{Calculating Susceptibility Through Phase-Space Integrals}
\label{sec: Calculating Susceptibility Through Phase-Space Integrals}

\begin{figure*}[!tb]
    \centering
    \begin{subfigure}{\columnwidth}
        \centering
        \includegraphics[width=\linewidth]{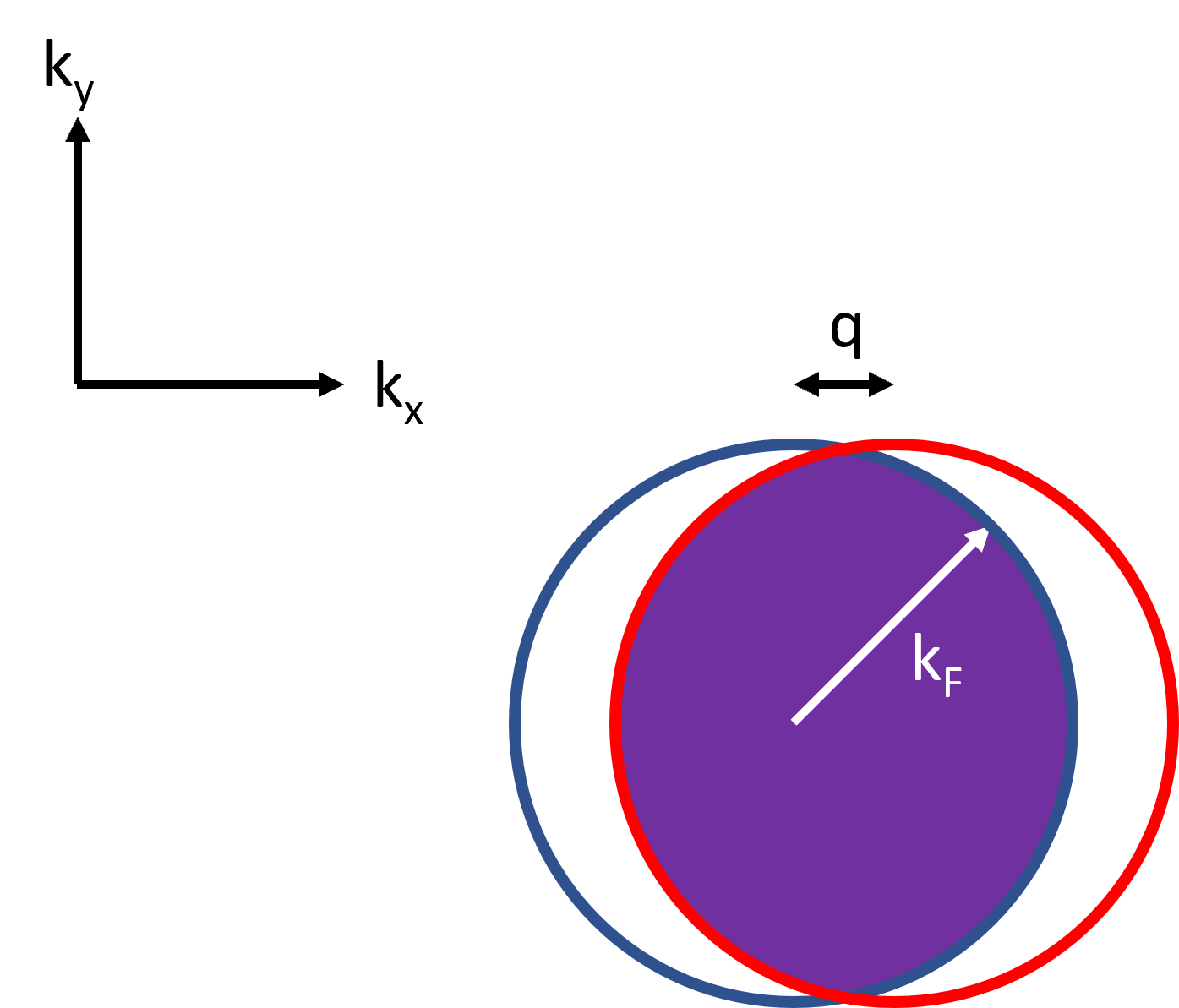}
        \caption{}
        \label{fig:phononabsorption1}
    \end{subfigure}
    \begin{subfigure}{\columnwidth}
        \centering
        \includegraphics[width=\linewidth]{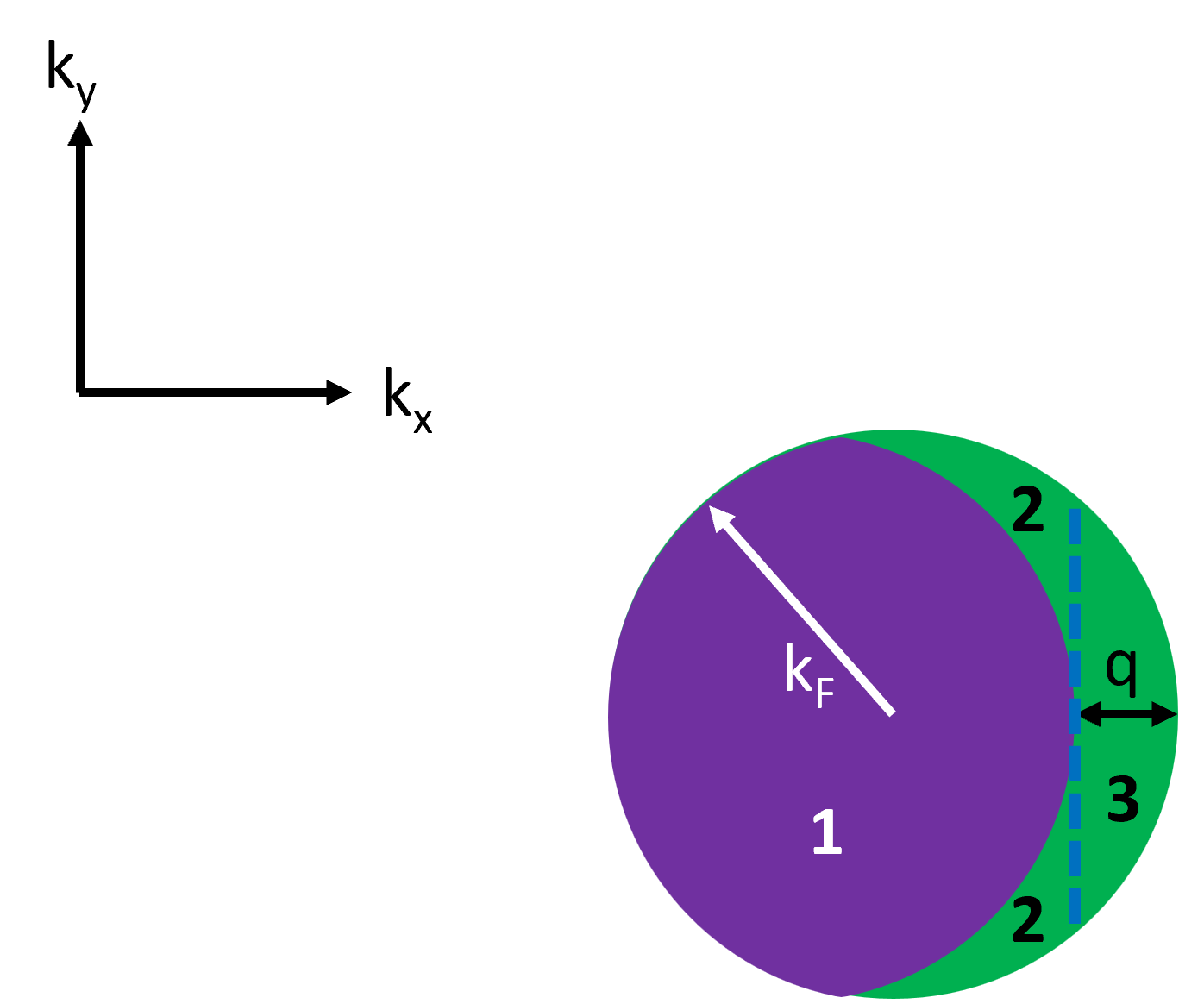}
        \caption{}
        \label{fig:phononabsorption2}
    \end{subfigure}
    \caption{(a) Phase-space diagram of valid states for absorption of a phonon with wavevector $\bm{\hat x} q$ by an electron. The blue (left) circle is the Fermi circle, which in the ground state is completely filled. The red (right) circle is the Fermi circle shifted in the $+\hat{k_x}$-direction by $q$, with the enclosed area representing potential destinations for the excited electrons. However, the shaded (overlapping) area represents states that are banned from serving as final states, since they are already occupied. Consequently, the range of possible destination states are represented by the area within the red (right) circle, less the shaded (overlapping) area; (b) Phase-space diagram of the range of possible initial states. The purple region (Zone 1) represents the forbidden initial states, while the green region (Zones 2 and 3) represents the allowed initial states.}
    \label{fig:phononabsorption}
\end{figure*}

Here, we convert the summation over states in the susceptibility expressions to phase-space integrals. Given a 2DEG area $A_\mathrm{2DEG}$, the following replacement can be used for any summation over generic initial wavevectors $\bm{k}$:
\begin{equation} \label{eq: generic sum of wavevectors}
\sum_{\bm{k}} = \frac{2A_\mathrm{2DEG}}{(2\pi)^2} \int dk_x \int dk_y,
\end{equation}
where the factor of 2 in the numerator is inserted to account for 2 spin states per spatial state. The main challenge is to determine the boundaries of integration for $k_x$ and $k_y$. The key requirement is an occupied initial state paired with an unoccupied final state. To this end, Fig.~\ref{fig:phononabsorption}(a) depicts the initial and final states in phase space.
As explained in the caption, the possible final states for a single excitation from the ground state is represented by the area enclosed by the red (right) circle, minus the shaded (overlapping) area. To represent the zone of forbidden initial states, the shaded zone representing the forbidden final states can be shifted leftward by $q$ to cover the corresponding part of the blue (left) circle. The resulting range of valid initial states is depicted in Fig.~\ref{fig:phononabsorption}(b). The states in the purple region (Zone 1) are forbidden from serving as initial states, while the rest of the states within the circle (green region) are allowed to act as initial states. We separate the range of allowed states into Zone 2 (left of dotted line) and Zone 3 (right of dotted line) in order to streamline the process of establishing the integral boundaries. In Zone 2, $k_x$ ranges from $-q/2$ to $k_F - q$. For each given $k_x$ in this range, the span of $k_y$ is determined as follows:
\begin{equation} \label{eq: ky span}
2\bigg(\sqrt{k_F^2 - k_x^2} - \sqrt{k_F^2 - (q + k_x)^2}\bigg).
\end{equation}
In Zone 3, $k_x$ ranges from $k_F - q$ to $k_F$. The corresponding span of $k_y$ for each value of $k_x$ is the following:
\begin{equation}
2 \sqrt{k_F^2 - k_x^2}.
\end{equation}
Consequently, the conversion of the summation over $\bm{k}$ into integral form, laid out generically in Eq.~\eqref{eq: generic sum of wavevectors}, takes the following concrete form:
\begin{widetext}
\begin{align}
\begin{split} \label{eq: k sum to integral}
\sum_{\bm{k}} &= \frac{2A_\mathrm{2DEG}}{(2\pi)^2} \Bigg(\int_{-q/2}^{k_F - q} dk_x 2\bigg(\sqrt{k_F^2 - k_x^2} - \sqrt{k_F^2 - (q + k_x)^2}\bigg) + \int_{k_F - q}^{k_F} dk_x 2 \sqrt{k_F^2 - k_x^2}\Bigg) \\
&= \frac{A_\mathrm{2DEG}}{2 \pi^2} \Bigg(\int_{-q/2}^{k_F} dk_x 2 \sqrt{k_F^2 - k_x^2} - \int_{-q/2}^{k_F - q} 2 \sqrt{k_F^2 - (q + k_x)^2}\Bigg).
\end{split}
\end{align}
\end{widetext}
We will evaluate this integral both numerically and analytically (in the high-mobility and low-mobility limits) to determine the susceptibility results.

We start by examining the third-order Kerr suscepbility $\chi^{(3)}_\mathrm{Kerr}$ by applying Eq.~\eqref{eq: chi(N) reduced} to the specific case of degenerate four wave mixing. The $\chi^{(3)}$ susceptibility for this process features the input frequency triplet $(\omega,\omega,-\omega_1)$ (where $\omega$ and $\omega_1$ are positive frequencies), yielding the following expression:
\begin{widetext}
\begin{align}
\begin{split} \label{eq: chi(3)(omega,omega,-omega1) summation}
&\chi^{(3)}(\omega,\omega,-\omega_1) \\
&= 2 \frac{q_e^4}{\hbar^3 \epsilon_0 V_\mathrm{2DEG}} \frac{v_s^4}{\omega^2 \omega_1 (2\omega - \omega_1)} \sum_{k_x,k_y} \bigg(\frac{1}{(\omega_{k_x + 2q - q_1,k_x} - (2\omega - \omega_1) - i\frac{\gamma}{2}) (\omega_{k_x + 2q,k_x} - 2\omega - i\frac{\gamma}{2}) (\omega_{k_x + q,k_x} - \omega - i\frac{\gamma}{2})} \\
&\quad\quad + \frac{1}{(\omega_{k_x + q_1,k_x} - \omega_1 + i\frac{\gamma}{2}) (\omega_{k_x + 2q,k_x} - 2\omega - i\frac{\gamma}{2}) (\omega_{k_x + q,k_x} - \omega - i\frac{\gamma}{2})} \\
&\quad\quad + \frac{1}{(\omega_{k_x - q,k_x} + \omega + i\frac{\gamma}{2}) (\omega_{k_x - 2q,k_x} + 2\omega + i\frac{\gamma}{2}) (\omega_{k_x - q_1,k_x} + \omega_1 - i\frac{\gamma}{2})} \\
&\quad\quad + \frac{1}{(\omega_{k_x - q,k_x} + \omega + i\frac{\gamma}{2}) (\omega_{k_x - 2q,k_x} + 2\omega + i\frac{\gamma}{2}) (\omega_{k_x - 2q + q_1,k_x} + (2\omega - \omega_1) + i\frac{\gamma}{2})}\bigg),
\end{split}
\end{align}
\end{widetext}
where we have defined $q = \omega/v_s$ and $q_1 = \omega_1/v_s$, and the factor of 2 has been added in front to account for the permutations of $\omega$. We convert the summation over $k_x$ and $k_y$ into integrals using the procedure from Eq.~\eqref{eq: k sum to integral}. Note that the integral boundaries for all terms can be aligned if we flip the signs of all wavevector arguments for the frequency shift terms $\omega_{k_f,k_i}$, which is valid since the energy spectrum and occupation probabilities are symmetric about the $k_x$-axis. Then, $\chi^{(3)}$ reduces to the following form:
\begin{align}
\begin{split}
&\chi^{(3)}(\omega,\omega,-\omega_1) = \\
&\quad \frac{q_e^4}{\pi^2 \hbar^3 \epsilon_0 t_\mathrm{2DEG}} \frac{v_s^4}{\omega^2 \omega_1 (2\omega - \omega_1)} \\
&\quad \times \Bigg(\int_{-q/2}^{k_F} dk_x 2 \sqrt{k_F^2 - k_x^2} f_{k_x}^{(3)}(\omega,\omega,-\omega_1) \\
&\quad - \int_{-q/2}^{k_F - q} dk_x 2 \sqrt{k_F^2 - (q + k_x)^2} f_{k_x}^{(3)}(\omega,\omega,-\omega_1)\Bigg),
\end{split}
\end{align}
where $f_{k_x}^{(3)}(\omega,\omega,-\omega_1)$ is approximately the following:
\begin{widetext}
\begin{align}
\begin{split}
f_{k_x}^{(3)}(\omega,\omega,-\omega_1) 
&\approx \frac{1}{(\omega_{k_x + q_2,k_x} - \omega_2 - i\frac{\gamma}{2}) (\omega_{k_x + 2q,k_x} - 2\omega - i\frac{\gamma}{2}) (\omega_{k_x + q,k_x} - \omega - i\frac{\gamma}{2})} \\
&\quad + \frac{1}{(\omega_{k_x + q_1,k_x} - \omega_1 + i\frac{\gamma}{2}) (\omega_{k_x + 2q,k_x} - 2\omega - i\frac{\gamma}{2}) (\omega_{k_x + q,k_x} - \omega - i\frac{\gamma}{2})} \\
&\quad + \frac{1}{(\omega_{k_x + q,k_x} + \omega + i\frac{\gamma}{2}) (\omega_{k_x + 2q,k_x} + 2\omega + i\frac{\gamma}{2}) (\omega_{k_x + q_1,k_x} + \omega_1 - i\frac{\gamma}{2})} \\
&\quad + \frac{1}{(\omega_{k_x + q,k_x} + \omega + i\frac{\gamma}{2}) (\omega_{k_x + 2q,k_x} + 2\omega + i\frac{\gamma}{2}) (\omega_{k_x + q_2,k_x} + \omega_2 + i\frac{\gamma}{2})}.
\end{split}
\end{align}
\end{widetext}
The third-order Kerr susceptibility equals the degenerate four-wave-mixing susceptibility in the limit $\omega_1 \rightarrow \omega$.

Next, we examine the second-order susceptibility, focusing first on the case of sum-frequency generation encapsulated by $\chi^{(2)}(\omega_1,\omega_2)$, where two phonons of frequency $\omega_1$ and $\omega_2$ are absorbed and a phonon of frequency $2\omega = \omega_1 + \omega_2$ is emitted. Applying Eq.~\eqref{eq: chi(N) reduced} again, we obtain the following expression for the second-order susceptibility:
\begin{widetext}
\begin{align}
\begin{split} \label{eq: chi(2)(omega_1,omega_2) summation}
&\chi^{(2)}(\omega_1,\omega_2) \\
&= \frac{q_e^3}{\hbar^2 \epsilon_0 V_\mathrm{2DEG}} \frac{v_s^3}{\omega_1 \omega_2 (\omega_1 + \omega_2)} \sum_{k_x,k_y} \bigg(\frac{1}{(\omega_{k_x + q_1 + q_2,k_x} - (\omega_1 + \omega_2) - i\frac{\gamma}{2}) (\omega_{k_x + q_1,k_x} - \omega_1 - i\frac{\gamma}{2})} \\
&\quad\quad + \frac{1}{(\omega_{k_x + q_1 + q_2,k_x} - (\omega_1 + \omega_2) - i\frac{\gamma}{2}) (\omega_{k_x + q_2,k_x} - \omega_2 - i\frac{\gamma}{2})} \\
&\quad\quad + \frac{1}{(\omega_{k_x - q_1 - q_2,k_x} + (\omega_1 + \omega_2) + i\frac{\gamma}{2}) (\omega_{k_x - q_2,k_x} + \omega_2 + i\frac{\gamma}{2})} \\
&\quad\quad + \frac{1}{(\omega_{k_x - q_1 - q_2,k_x} + (\omega_1 + \omega_2) + i\frac{\gamma}{2}) (\omega_{k_x - q_1,k_x} + \omega_1 + i\frac{\gamma}{2})}\bigg),
\end{split}
\end{align}
\end{widetext}
where $q_1 = \omega_1/v_s$ and $q_2 = \omega_2/v_s$. Converting the summation over wavevectors to integrals using the same procedure and symmetry arguments as in the $\chi^{(3)}$ calculation, we find that $\chi^{(2)}$ reduces to the following form:
\begin{align}
\begin{split}
&\chi^{(2)}(\omega_1,\omega_2) = \\
&\quad \frac{q_e^3}{2 \pi^2 \hbar^2 \epsilon_0 t_\mathrm{2DEG}} \frac{v_s^3}{\omega_1 \omega_2 (\omega_1 + \omega_2)} \\
&\quad \times \Bigg(\int_{-q/2}^{k_F} dk_x 2 \sqrt{k_F^2 - k_x^2} f_{k_x}^{(2)}(\omega_1,\omega_2) \\
&\quad - \int_{-q/2}^{k_F - q} dk_x 2 \sqrt{k_F^2 - (q + k_x)^2} f_{k_x}^{(2)}(\omega_1,\omega_2)\Bigg),
\end{split}
\end{align}
where $f_{k_x}^{(2)}(\omega_1,\omega_2)$ is approximately the following:
\begin{widetext}
\begin{align}
\begin{split}
f_{k_x}^{(2)}(\omega_1,\omega_2) &\approx 
\bigg(\frac{1}{(\omega_{k_x + 2q,k_x} - 2\omega - i\frac{\gamma}{2}) (\omega_{k_x + q_1,k_x} - \omega_1 - i\frac{\gamma}{2})} + \frac{1}{(\omega_{k_x + 2q,k_x} - 2\omega - i\frac{\gamma}{2}) (\omega_{k_x + q_2,k_x} - \omega_2 - i\frac{\gamma}{2})} \\
&\quad\quad + \frac{1}{(\omega_{k_x + 2q,k_x} + 2\omega + i\frac{\gamma}{2}) (\omega_{k_x + q_2,k_x} + \omega_2 + i\frac{\gamma}{2})} + \frac{1}{(\omega_{k_x + 2q,k_x} + 2\omega + i\frac{\gamma}{2}) (\omega_{k_x + q_1,k_x} + \omega_1 + i\frac{\gamma}{2})}\bigg).
\end{split}
\end{align}
\end{widetext}
The susceptibility governing the second-harmonic generation process equals the above susceptibility for the general sum-frequency generation in the limit $\omega_1 = \omega_2$.

Finally, we use Eq.~\eqref{eq: chi(N) reduced} once more to briefly derive the linear susceptibility $\chi^{(1)}(\omega)$ corresponding to the absorption and emission of a phonon of frequency $\omega$:
\begin{align}
\begin{split} \label{eq: chi(1)(omega) summation}
\chi^{(1)}(\omega) &= \frac{q_e^2}{\hbar \epsilon_0 V_\mathrm{2DEG}} \frac{v_s^2}{\omega^2} \sum_{k_x,k_y} \bigg(\frac{1}{\omega_{k_x + q,k_x} - \omega - i\frac{\gamma}{2}} \\
&\quad\quad + \frac{1}{\omega_{k_x - q,k_x} + \omega + i\frac{\gamma}{2}}\bigg),
\end{split}
\end{align}
where $q = \omega/v_s$. Converting to an integral form over phase space in a manner analogous to the $\chi^{(3)}$ and $\chi^{(2)}$ calculations, we find the following $\chi^{(1)}$:
\begin{align}
\begin{split}
&\chi^{(1)}(\omega) = \\
&\quad \frac{q_e^2}{2 \pi^2 \hbar \epsilon_0 t_\mathrm{2DEG}} \frac{v_s^2}{\omega^2} \Bigg(\int_{-q/2}^{k_F} dk_x 2 \sqrt{k_F^2 - k_x^2} f_{k_x}^{(1)}(\omega) \\
&\quad - \int_{-q/2}^{k_F - q} dk_x 2 \sqrt{k_F^2 - (q + k_x)^2} f_{k_x}^{(1)}(\omega)\Bigg),
\end{split}
\end{align}
where $f_{k_x}^{(1)}(\omega)$ is defined as follows:
\begin{equation}
f_{k_x}^{(1)}(\omega) = \frac{1}{\omega_{k_x + q,k_x} - \omega - i\frac{\gamma}{2}} + \frac{1}{\omega_{k_x + q,k_x} + \omega + i\frac{\gamma}{2}}.
\end{equation}

\section{Relating 2DEG Polarization to Induced Electric Field}
\label{sec: Relating 2DEG Polarization to Induced Electric Field}

Here, we seek to derive the relationship between the internal electric field induced by the polarization in the 2DEG versus the polarization itself. The key difference between a bulk material and a thin-film 2DEG is that the charges contributing to the polarization in the latter must be modeled as finite rather than infinite distributions, as we will show.

\begin{figure}[!tb]
	\centering
	\includegraphics[width=\linewidth]{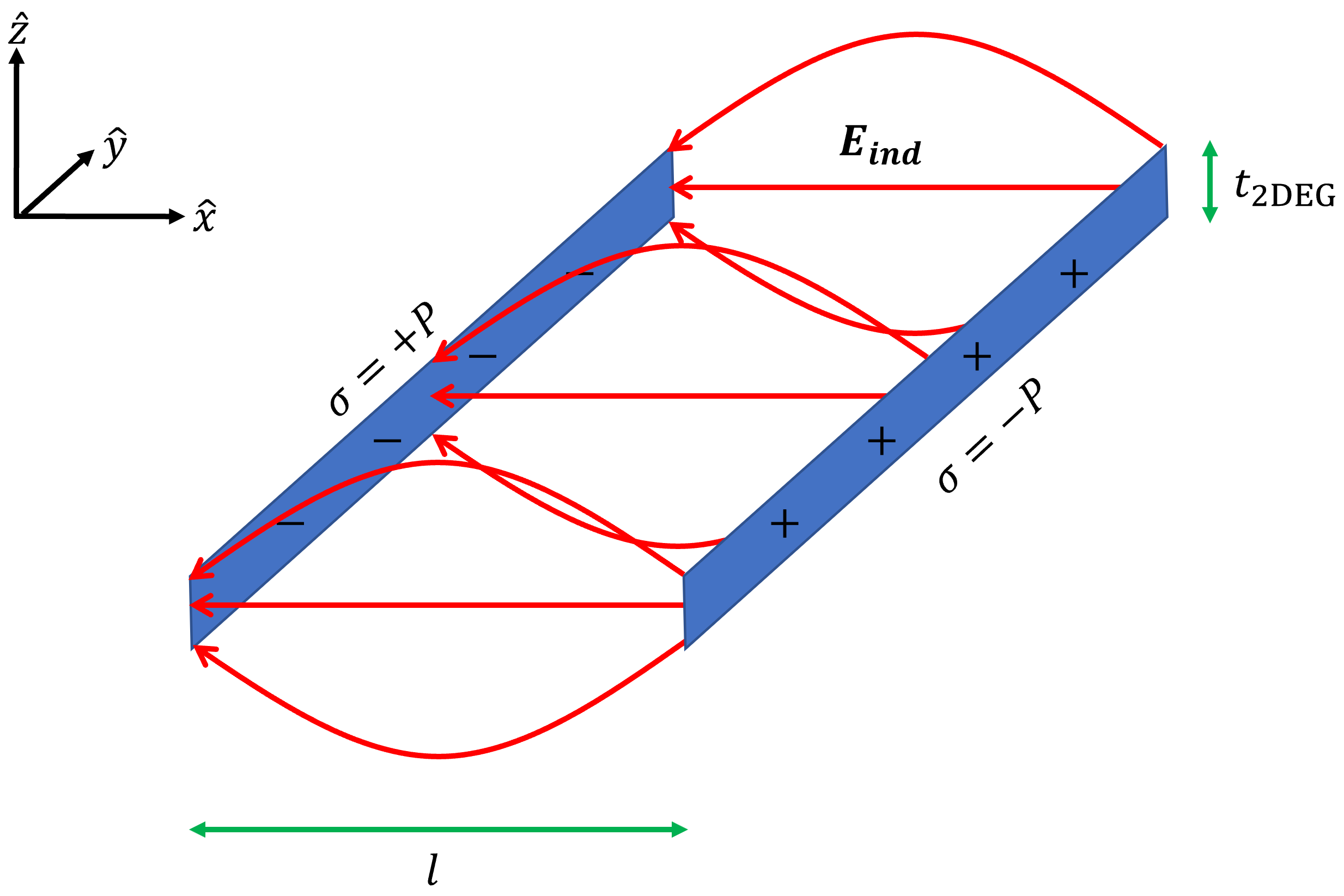}
	\caption{Depiction of induced electric field in the 2DEG due to charge polarization effectively giving rise to 2 plates of charge density $\pm P$ separated along the $x$-axis by the dipole length in the material, which we label as $l$.}
	\label{fig:inducedfield}
\end{figure}

We start with the definition of polarization: the number of dipole moments per unit volume multiplied by the amplitude of each moment. For now, we will express the size of each moment as $l$. Then, if we take a volumetric segment of the 2DEG with an $x$-axis length of $l$, the number of dipole moments in that segment becomes the following (see Fig.~\ref{fig:inducedfield}):
\begin{equation}
N_\mathrm{segment} = \frac{P l L_y L_z}{q_e l} = \frac{P L_y t_\mathrm{2DEG}}{q_e}.
\end{equation}
The charge density per unit of cross-sectional area is thus solved as follows:
\begin{equation}
\sigma_\mathrm{segment} = q_e \frac{N_\mathrm{segment}}{L_y t_\mathrm{2DEG}} = P.
\end{equation}
If we consider the dipoles in the 2DEG as a uniformly distributed cloud, then the induced electric field comes from two oppositely-charged plates, with area charge densities $\pm \sigma_\mathrm{segment} = \pm P$, separated by a distance $l$. For the case of a bulk material, the length of a plate along each cross-sectional axis far exceeds the dipole length $l$. Consequently, the electric field is equivalent to that between two infinite-area plates of charge density $\pm P$, i.e. $E_\mathrm{ind} = -P/\epsilon_0$. However, for the case of a thin-film material with thickness $t_\mathrm{2DEG}$, the average electric field between the plates is sharply attenuated if $t_\mathrm{2DEG} \lessapprox l$. We express this attenuation by defining a factor $a$, such that the average induced field becomes $\expect{E_\mathrm{ind}} = -aP/\epsilon_0$. This yields the following relationship between the external field $E_\mathrm{ext}$, overall field $E$, and polarization $P$:
\begin{equation}
\label{eq: field relationships thin film}
\epsilon_0 E_\mathrm{ext} = \epsilon_0 \Big(E - \expect{E_\mathrm{ind}}\Big) = \epsilon_0 E + aP.
\end{equation}
Note that if $k = 1$, the above expression reduces to the familiar relationship for bulk materials. However, in the thin-film limit, we need to solve for the induced electric field as a function of polarization by considering 2 \textit{finite} plates with charge density $\pm P$. The setup is depicted in Fig.~\ref{fig:inducedfield}.
Since $L_y \gg l$, the plates can be considered as essentially infinite along the $y$-axis. However, along the $z$-axis, each plate features a finite width $t_\mathrm{2DEG}$. The field lines therefore diverge from each plate before returning to the opposite plate, ensuring that while the electric field between the plates is uniform along the $y$-axis, it varies along the $z$- and especially $x$-axes. To calculate the induced electric field along the $x$-axis at a point $(x,z)$, we integrate the contributions from all of the infinitesimal charges along the 2 plates:
\begin{widetext}
\begin{align}
\begin{split}
&E_\mathrm{ind}(x,z) \\
&= -\frac{P}{4 \pi \epsilon_0} \int_{z - t_\mathrm{2DEG}/2}^{z + t_\mathrm{2DEG}/2} dz' \int_{-\infty}^{\infty} dy \bigg(\frac{x}{(x^2 + y^2 + z^2)^{3/2}} + \frac{l - x}{((l - x)^2 + y^2 + z^2)^{3/2}}\bigg) \\
&= -\frac{P}{4 \pi \epsilon_0} \int_{z - t_\mathrm{2DEG}/2}^{z + t_\mathrm{2DEG}/2} dz' \bigg(\frac{2x}{x^2 + z^2} + \frac{2(l - x)}{(l - x)^2 + z^2}\bigg) \\
&= -\frac{P}{4 \pi \epsilon_0} \bigg[2\tan^{-1}\bigg(\frac{z'}{x}\bigg) + 2\tan^{-1}\bigg(\frac{z'}{l - x}\bigg)\bigg]_{z' = z - t_\mathrm{2DEG}/2}^{z + t_\mathrm{2DEG}/2} \\
&= -\frac{P}{2 \pi \epsilon_0} \bigg(\tan^{-1}\bigg(\frac{t_\mathrm{2DEG}/2 + z}{x}\bigg) + \tan^{-1}\bigg(\frac{t_\mathrm{2DEG}/2 + z}{l - x}\bigg) \\
&\quad + \tan^{-1}\bigg(\frac{t_\mathrm{2DEG}/2 - z}{x}\bigg) + \tan^{-1}\bigg(\frac{t_\mathrm{2DEG}/2 - z}{l - x}\bigg)\bigg).
\end{split}
\end{align}
\end{widetext}
Next, we average this field over $x$ ranging from 0 to $l$ and $z$ ranging from $-t_\mathrm{2DEG}/2$ to $t_\mathrm{2DEG}/2$. Note that by symmetry, each of the terms features the same average value. We therefore pick one term to calculate the average induced field, yielding the following relationship between the induced field and the polarization:
\begin{widetext}
\begin{align}
\begin{split}
\expect{E_\mathrm{ind}} &= -4\frac{P}{2 \pi \epsilon_0} \expect{\tan^{-1}\bigg(\frac{t_\mathrm{2DEG}/2 + z}{x}\bigg)} \\
&= -\frac{2P}{\pi \epsilon_0 t_\mathrm{2DEG} l} \int_{-t_\mathrm{2DEG}/2}^{t_\mathrm{2DEG}/2} dz \int_0^l dx \tan^{-1}\bigg(\frac{t_\mathrm{2DEG}/2 + z}{x}\bigg) \\
&= -\frac{2P}{\pi \epsilon_0 t_\mathrm{2DEG} l} \int_{-t_\mathrm{2DEG}/2}^{t_\mathrm{2DEG}/2} dz \Bigg(\bigg(\frac{t_\mathrm{2DEG}/2 + z}{2}\bigg) \log{\bigg(1 + \frac{l^2}{(t_\mathrm{2DEG}/2 + z)^2}\bigg)} \\
&\quad + l \tan^{-1}\bigg(\frac{t_\mathrm{2DEG}/2 + z}{l}\bigg)\Bigg) \\
&\approx -\frac{2P}{\pi \epsilon_0 t_\mathrm{2DEG} l} \int_{-t_\mathrm{2DEG}/2}^{t_\mathrm{2DEG}/2} dz \bigg(\frac{t_\mathrm{2DEG}}{2} + z\bigg) \bigg(\log{\bigg(\frac{l}{t_\mathrm{2DEG}/2 + z}\bigg)} + 1\bigg) \\
&= -\frac{2P}{\pi \epsilon_0 t_\mathrm{2DEG} l} \bigg(\frac{t_\mathrm{2DEG}}{2}\bigg)^2 \bigg(2 \log{\bigg(\frac{l}{t_\mathrm{2DEG}}\bigg) + 3}\bigg) \\
&= -\frac{k P}{\epsilon_0},
\end{split}
\end{align}
\end{widetext}
where we have used the approximation $t_\mathrm{2DEG} \ll l$ in the fourth line. The attenuation factor $k$ takes the following form in terms of the dipole length and 2DEG thickness:
\begin{equation}
\label{eq: induced field attenuation factor}
a = \frac{t_\mathrm{2DEG}}{2 \pi l} \bigg(2 \log{\bigg(\frac{l}{t_\mathrm{2DEG}}\bigg) + 3}\bigg).
\end{equation}
The overall field $E$ inside the 2DEG can be solved by substituting the relationship $P = \epsilon_0 \chi^{(1)} E$ into Eq.~\eqref{eq: field relationships thin film}:
\begin{align}
\begin{split}
\epsilon_0 E_\mathrm{ext} &= \epsilon_0 E + a \epsilon_0 \chi^{(1)} E, \\
E &= \frac{E_\mathrm{ext}}{1 + a \chi^{(1)}} \approx \frac{E_\mathrm{ext}}{a \chi^{(1)}},
\end{split}
\end{align}
where in the last line, we have used the approximation $a \chi^{(1)} \gg 1$.

\bibliography{ref}

\end{document}